\documentclass[aps,prb,twocolumn,floatfix,longbibliography,superscriptaddress]{revtex4-2}
\usepackage[english]{babel}
\usepackage[utf8]{inputenc}
\usepackage[T1]{fontenc}
\usepackage{amsmath}
\usepackage{amssymb}
\usepackage{graphicx}
\usepackage{bm}
\usepackage{color}
\usepackage{hyperref}
\usepackage{upgreek}
\usepackage{scalerel}
\usepackage{pgffor}
\usepackage{pdfpages}
\usepackage{float}
\usepackage{appendix}
\usepackage{physics} 
\usepackage{lscape}   
\usepackage{silence}
\usepackage{soul,xcolor} 
\setstcolor{red} 
\usepackage{comment}  
\makeatletter
\AtBeginDocument{\let\LS@rot\@undefined}
\makeatother

\begin{document}
\title{Mitigating disorder-induced zero-energy states in weakly-coupled superconductor-semiconductor hybrid systems}
\author{Oladunjoye A. Awoga}
\email[  ]{oladunjoye.awoga@ftf.lth.se}
\author{Martin Leijnse}
\affiliation{Solid State Physics and NanoLund, Lund University, Box118, 22100 Lund, Sweden}
\author{Annica M. Black-Schaffer}
 \author{Jorge Cayao} 
 \email[ ]{ jorge.cayao@physics.uu.se}
\affiliation{Department of Physics and Astronomy, Uppsala University, Box 516, S-751 20 Uppsala, Sweden}

\date{\today}
\begin{abstract} 
Disorder has appeared as one of the main mechanisms to induce topologically trivial zero-energy states in superconductor-semiconductor systems, thereby challenging the detection of topological superconductivity and Majorana bound states. Here we demonstrate that, for disorder in any part of the system, the formation of disorder-induced trivial zero-energy states can to a large extent be mitigated by keeping the coupling between the semiconductor and superconductor weak. The only exception is strong disorder in the semiconductor, where instead the strong coupling regime is somewhat more robust against disorder. Furthermore, we  find that the topological phase in this weak coupling regime is robust against disorder, with a large and well-defined topological gap which is highly beneficial for topological protection. Our work  shows the advantages and disadvantages of weak and strong couplings under disorder, important for designing superconductor-semiconductor hybrid structures.
\end{abstract}\maketitle

\section{Introduction}
Superconductor-semiconductor (SC-SM) hybrid systems have received much attention in the last 10 years due to their potential for the realization of topological superconductivity and Majorana bound states (MBSs)~\cite{Leijnse2012Introduction,Aguadoreview17,Lutchyn2018Majorana,zhang2019next,frolov2020topological,Laubscher2021Majorana,Prada2020Adreev,flensberg2021engineered,marra2022majorana2}. MBSs have been predicted to appear under a strong applied magnetic field,  emerging at zero energy and located at the ends of the system~\cite{Oreg:PRL10,Alicea:PRB10,PhysRevLett.105.077001}. Although multiple detection schemes of MBSs have been reported,  quantized zero-bias conductance peaks have been one of the most pursued signatures~\cite{PhysRevLett.74.3451,PhysRevLett.98.237002,PhysRevLett.103.237001,PhysRevB.82.180516},  motivating many transport-based experiments to detect MBSs~\cite{Mourik:S12,Deng:NL12,Albrecht16,deng2016majorana,Suominen17,Nichele2017Scaling,vaitiekenas2021zero,Dvir2022Realization}. However, it is clear by now that zero-bias peaks do not necessarily represent evidence of MBSs because topologically trivial  zero-energy states (TZESs) can   produce similar signatures~\cite{PhysRevB.86.180503, Liu2012Zero, Liu2017QDot,PhysRevB.98.155314,Chen2019Ubiquitous,PhysRevLett.123.217003,10.21468/SciPostPhys.7.5.061,Zhang2020Transport,Prada2020Adreev,valentini2020nontopological,PhysRevLett.125.017701,PhysRevLett.125.116803,PhysRevB.104.134507,yu2020non,marra2022majorana,Hess2022trivial,chen2022topologically,sahu2022effect}.

Among the most relevant mechanisms known to cause the formation of TZESs are spatial inhomogeneities in the effective chemical potential profile of the SM~\cite{kells2021Near,PhysRevB.91.024514, 10.21468/SciPostPhys.7.5.061,Prada2020Adreev,PhysRevB.104.134507}. Such spatial variations can occur due to  distinct effects, such as gate voltages~\cite{deMoor2018Electric}, finite size effects of the SC \cite{Awoga2019Supercurrent,Reeg2018Zero,Awoga2022Robust,Escribano2022Proximity}, or random fluctuations due to charge inhomogeneities or generally scalar disorder ~\cite{PhysRevB.63.224204,PhysRevB.84.144526,Akhmerov2011Quantized,Potter2011Enginering,Bagrets:PRL12,PhysRevB.85.140513,Pientka2012Enhanced,Pikulin2012A,Sau:13,DasCole2016Proximity,PhysRevResearch.2.013377}. While the effect of gates can in principle be controlled~\cite{deMoor2018Electric,Thamm2022Machine}, the size of the SC or disorder are much harder to avoid in real samples. In fact, recent measurements of  zero-bias peaks have been interpreted in terms of disorder-induced TZESs ~\cite{Nichele2017Scaling,Zhang2021large,Ahn2021Estimating,DasSarma2021Disorder}, suggesting disorder is the main obstacle to realize MBSs \cite{PhysRevB.101.094506,aghaee2022inas1,sarma2022search}. Also, it has been shown that the coupling between the SC and SM considerably affects the low-energy properties. In particular, strong couplings, seen earlier as favorable due to the large induced gaps in the SM~\cite{Chang2015Hard}, induce detrimental effects such as  renormalization of the physical parameters of the SM, the need of large Zeeman fields to reach the topological phase, and also the formation of TZESs~\cite{Lee2017Scaling,Reeg2018Zero,Awoga2019Supercurrent,Awoga2022Robust}. As a result, the interplay of disorder and SC-SM coupling challenges the realization of topological superconductivity.   
\begin{figure}[!t]
	\centering
	\includegraphics[width=0.35\textwidth]{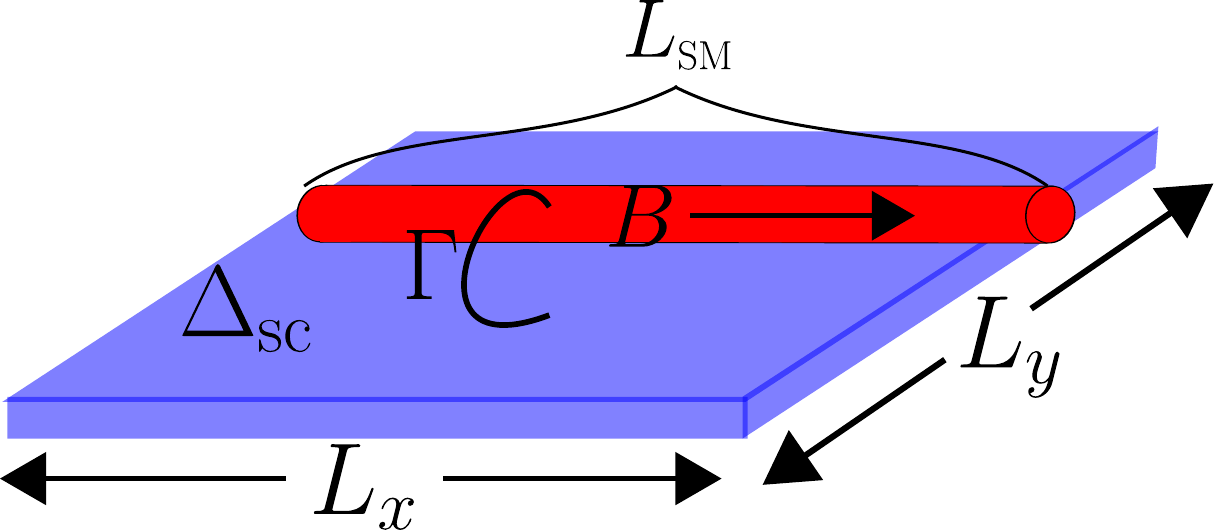} 
	\caption{Schematics of the system: SM (red) of length $L_{\scaleto{\rm SM}{4pt}}$ in a parallel magnetic field, $B$, coupled through $\Gamma$ to a thin film conventional SC of length $L_x$ and width $L_y$ (blue). Part of the SM is not coupled to the SC, thus remaining in the normal state, giving rise to a SN junction.}
	\label{fig1}
\end{figure}

In this work we study the influence of scalar disorder in SC-SM hybrid systems  and show that its impact can largely be suppressed in the weak SC-SM coupling regime. In particular, we discover that disorder in any part of the system does not induce TZESs, in stark contrast to the strong coupling regime. We find that this effect holds for a broad range of disorder strengths when present in the SM, SC, or in the coupling.   We find the only exception to be the case of very strong disorder in the SM, there the strong coupling regime shows  better resilience against disorder.  We also obtain that the topological phase in the weak coupling regime is robust against strong disorder, with a well-defined topological gap and the absence of disorder-induced in-gap states.  Our results thus allow tailoring the coupling strength depending on the dominant sources of disorder,  which is relevant when designing  efficient superconductor-semiconductor systems.

\section{SC-SM model} 
\label{sectionII}
We consider a disordered SC-SM system formed by coupling a 2D conventional spin-singlet $s$-wave SC and a 1D SM nanowire with strong spin-orbit coupling under  a Zeeman field $B$, as  shown in Fig.\,\ref{fig1}.  This coupled model goes beyond the usual 1D effective description and captures several realistic properties of SC-SM systems \cite{Awoga2019Supercurrent,Awoga2022Robust,Reeg2018Metallization,Reeg2017Finite,Reeg2018Zero,Stanescu2017Proximity}, such as the renormalization of all parameters in the SM and not just the inclusion of superconductivity, the hybridization between SM and SC, and the quality of the SC-SM interface, see also Ref.\,\cite{Leijnse2012Introduction,Aguadoreview17,Lutchyn2018Majorana,zhang2019next,frolov2020topological,Laubscher2021Majorana,Prada2020Adreev,flensberg2021engineered}.   
  Disorder is taken into account  in the form of non-magnetic scalar disorder \cite{PhysRev.109.1492}, as is it is likely the most unavoidable type of disorder present in all the regions of the SC-SM system \cite{Nichele2017Scaling,Zhang2021large,Ahn2021Estimating,DasSarma2021Disorder,PhysRevB.101.094506,aghaee2022inas1,sarma2022search,PhysRevB.104.094503,Mashkoori2023}.  For instance,  charge puddles or inhomogeneities lead to scalar disorder in the chemical potential of the SM and SC and, despite advances in the device fabrication, the interface between  SC and SM still suffers from imperfections leading to disorder in the coupling (or hybridization) strength $\Gamma$ between SC and SM.  
We thus model the total  SC-SM system by $H=H^{\rm c}+H^{\rm d}$, where $H^{\rm c(d)}=H_{\scaleto{\rm SM}{4pt}}^{\rm c(d)}+H_{\scaleto{\rm SC}{4pt}}^{\rm c(d)}+H_{\Gamma}^{\rm c(d)}$. Here, $H^{\rm c}$ describes the clean (c) system, while $H^{\rm d}$ models the disorder (d).  The clean system is   given by 
	\begin{equation}
	\begin{split}
		H_{\rm SM}^{\rm c} & =\sum_{r,r^\prime\beta,\beta^\prime} d_{r\beta}^{\dagger} \left[\left(\varepsilon_{\rm SM}^{} + B \sigma_{\beta\beta^\prime}^x \right)\delta_{rr^\prime} \right. \\
		&	\left. - \left(t_{\rm SM} - i\alpha_{\rm SM} \sigma_{\beta\beta^\prime}^y\right) \delta_{\langle r,r^\prime \rangle} \right] d_{r^\prime\beta^\prime} + \text{H.c.}\,,\\
		H_{\rm SC}^{\rm c} &=\sum_{j,j^\prime,\beta}c^{\dagger}_{j\beta} \left[\varepsilon_{\rm SC}^{} \delta_{jj^\prime}-t_{\rm SC}\delta_{\langle j,j^\prime \rangle}\right]c_{j^\prime\beta}\\
		& + \sum_{j,\beta,\beta^\prime}c^{\dagger}_{j\beta}\big[i\Delta_{\rm SC}\sigma_{\beta\beta^\prime}^y \big]c^{\dagger}_{j\beta^\prime}  + \text{H.c.}\,,\\
		H_{\Gamma}^{\rm c}&=-\sum_{r,j,\beta}  c^{\dagger}_{j\beta}\big[\Gamma\delta_{j_x,r}\delta_{j_y,\frac{L_y+1}{2}}\big]d_{r\beta} + \text{H.c}\,,
	\end{split}
	\label{eq:H}
\end{equation}
where $d_{r,\beta}$ destroys an electron with spin $\beta$ at site $r$ in the 1D SM of  length $L_{\rm SM}$,  $c_{m,\beta}$ destroys an electron with spin $\beta$ at site $j=(j_x,j_y)$ in the 2D SC,   $\delta_{\langle \dots \rangle} (\delta_{jj^\prime})$ enforces nearest-neighbor (on-site) terms only,  and $\sigma^n $ is the $n$-Pauli matrix in spin space. Also, $\varepsilon_{\rm SM}^{}=\left( 2t_{\rm SM}-\mu_{\rm SM}\right)$ is the SM onsite energy, $\varepsilon_{\rm SC}^{}=\left( 4t_{\rm SC}-\mu_{\rm SC}\right)$ is the SC onsite energy,  $\mu_{\rm SM (SC)}$  is the SM (SC) chemical potential, $t_{\rm SM (SC)}$ is the nearest neighbor hopping in the SM (SC), $\alpha_{\rm SM}$ is the Rashba spin-orbit  coupling in the SM, $\Delta_{\rm SC}$ is the onsite $s$-wave order parameter associated with conventional superconductivity, and $B$ is the external Zeeman field. 
To model scalar disorder, we consider random site-dependent fluctuations in the chemical potential of the SM and SC, as well as random fluctuations in the coupling strength $\Gamma$, given by 
\begin{equation}
	\begin{split}
		H_{\scaleto{\rm SM}{4pt}}^{\rm d} & =\sum_{r\beta} d_{r\beta}^{\dagger}\left[\delta \mu_{\scaleto{\rm SM}{4pt}}^{}\left(r\right)\right]  d_{r\beta}  + \text{H.c.}\,, \\
		H_{\scaleto{\rm SC}{4pt}}^{\rm d} & =\sum_{j\beta} c_{j\beta}^{\dagger} \left[\delta \mu_{\scaleto{\rm SC}{4pt}}^{}\left(j\right) \right] c_{j\beta}  + \text{H.c.}\,, \\
		H_{\Gamma}^{\rm d}&=-\sum_{r,j\beta}  c^{\dagger}_{j\beta}\big[\delta\Gamma\left(j\right)\delta_{j_x,r}\delta_{j_y,\frac{L_y+1}{2}}\big]d_{r\beta} + \text{H.c}.
	\end{split}
	\label{eq:Hdis}
\end{equation}
where $\delta Q \left(n\right) \in \left[-w_{Q},w_{Q}\right]$ describes the site-dependent random fluctuations in the quantity $Q=\lbrace\mu_{\scaleto{\rm SC}{4pt}},\,\mu_{\scaleto{\rm SM}{4pt}},\,\Gamma\rbrace$, with  $w_{Q}$ being the disorder strength. This approach ensures that $\langle\delta Q\rangle=0$.  To characterize the relative strength of the disorder, we generally compare the strength with respect to the quantity in the clean regime, $w_Q/Q$.
 
In terms of parameters, we consider realistic values  often used to describe SC-SM systems \cite{DasCole2016Proximity,Reeg2018Zero,Awoga2019Supercurrent,Awoga2022Robust}. In particular, for the SC we use  $|\Delta_{\rm SC}|=0.1t_{\rm SC}$, $\mu_{\rm SC}=0.38t_{\rm SC}$, and $L_y=11a$,   with  $a$  the discretization parameter.  In the SM,  we set $L_{\scaleto{\rm SM}{4pt}}=1000a$, $\mu_{\rm SM}=0.02t_{\scaleto{\rm SM}{4pt}}$,  $\alpha_{\scaleto{\rm SM}{4pt}}=0.05t_{\scaleto{\rm SM}{4pt}}$, and $t_{\scaleto{\rm SM}{4pt}}=4t_{\rm SC}$, which incorporates the mismatch in the lattice constants and effective masses of the SM and SC. We further leave a small part of the  SM uncovered by the SC in order to model a superconductor-normal (SN) junction, often used in transport experiments \cite{Prada2020Adreev}. To eliminate the presence of low-energy states and avoid severe gating effects in N, we keep  N very short with $L_{\rm N} = 2a$ \footnote{For a N region with a vanishing length, $L_{\rm N} = 0$, we have verified that our results do not change}. The physics of SC-SM systems also heavily depends on the coupling $\Gamma$ \cite{Awoga2022Robust,Awoga2019Supercurrent}.  In fact, the induced gap exhibits a linear dependence on the coupling strength at weak $\Gamma$, while it has a nonlinear behavior at stronger $\Gamma$, which eventually saturates at the values of the parent superconductor gap $\Delta_{\rm SC}$. This behavior of the induced gap enables us to identify two distinct regimes, which we refer to as the \textit{weak coupling regime}  with an induced gap linear in $\Gamma$ and the \textit{strong coupling regime} with an induced gap nonlinear with $\Gamma$. There is no sharp boundary between the weak and strong coupling regimes, but a smooth crossover that needs to be avoided when targeting either regime.  For the above parameters, we have identified that the  border between weak and strong coupling occurs for $\Gamma/t_{\rm SC} \sim 0.4$ \cite{Awoga2022Robust,Awoga2019Supercurrent}. Therefore, we  choose $\Gamma/t_{\rm SC} = 0.3$ and $\Gamma/t_{\rm SC} = 0.7$ as representative values of the weak and strong coupling regimes, respectively. We also note that, since  the induced gap reaches $\Delta_{\rm SC}$ at $\Gamma/t_{\rm SC}=1$, there is further no reason to consider $\Gamma/t_{\rm SC}>1$ for the strong coupling regime. Finally, for the considered parameters, we estimate that the disorder can be considered to be strong for $w_{\mu_{\scaleto{\rm SC}{4pt}}}/\mu_{\scaleto{\rm SC}{4pt}}\gtrsim 0.8$, $w_{\mu_{\scaleto{\rm SM}{4pt}}}/\mu_{\scaleto{\rm SM}{4pt}}\gtrsim 1$, and $w_{\Gamma}/\Gamma \gtrsim 0.3$, respectively,  see Appendix \ref{app:DisStrength} for details. 
Since we are interested in the role of disorder for inducing TZESs, we numerically solve  the full SC-SM Hamiltonian $H$ and focus on the low-energy spectrum. We employ the Arnoldi iteration method~\cite{arnoldi1951principle} because it efficiently allows us to address large systems.

\subsection{Clean regime}
Without disorder, the region of the SM that is in contact with the SC undergoes a topological phase transition (TPT) at a critical field $B_{\rm c}$ and enters into a topological phase where MBSs emerge at the end points of the SM \cite{Oreg:PRL10,Alicea:PRB10,PhysRevLett.105.077001,Maiani2021Topological}. Under ideal conditions, the topological properties can be understood from the two lowest energy levels, $E_{0,1}$. In the trivial phase at $B=0$, the induced gap is defined by $E_0$, which decreases as $B$ increases  until it reaches zero at the TPT  \footnote{The vanishing of the gap depends on system size: short SMs develop a sharp profile at the TPT but $E_{0}$ does not usually reach zero, while for long systems $E_{0}$ reaches zero.}. After the TPT,  the induced gap reopens but now it is defined by $E_{1}$, while  $E_{0}$ sticks to zero energy for long enough SMs, revealing the MBSs. The induced gap in the topological phase (the topological gap)  isolates the MBSs from the quasi-continuum and thus provides protection \cite{Sarma:16}.

%
\begin{figure}[!t]
	\centering
	\includegraphics[width=.48\textwidth]{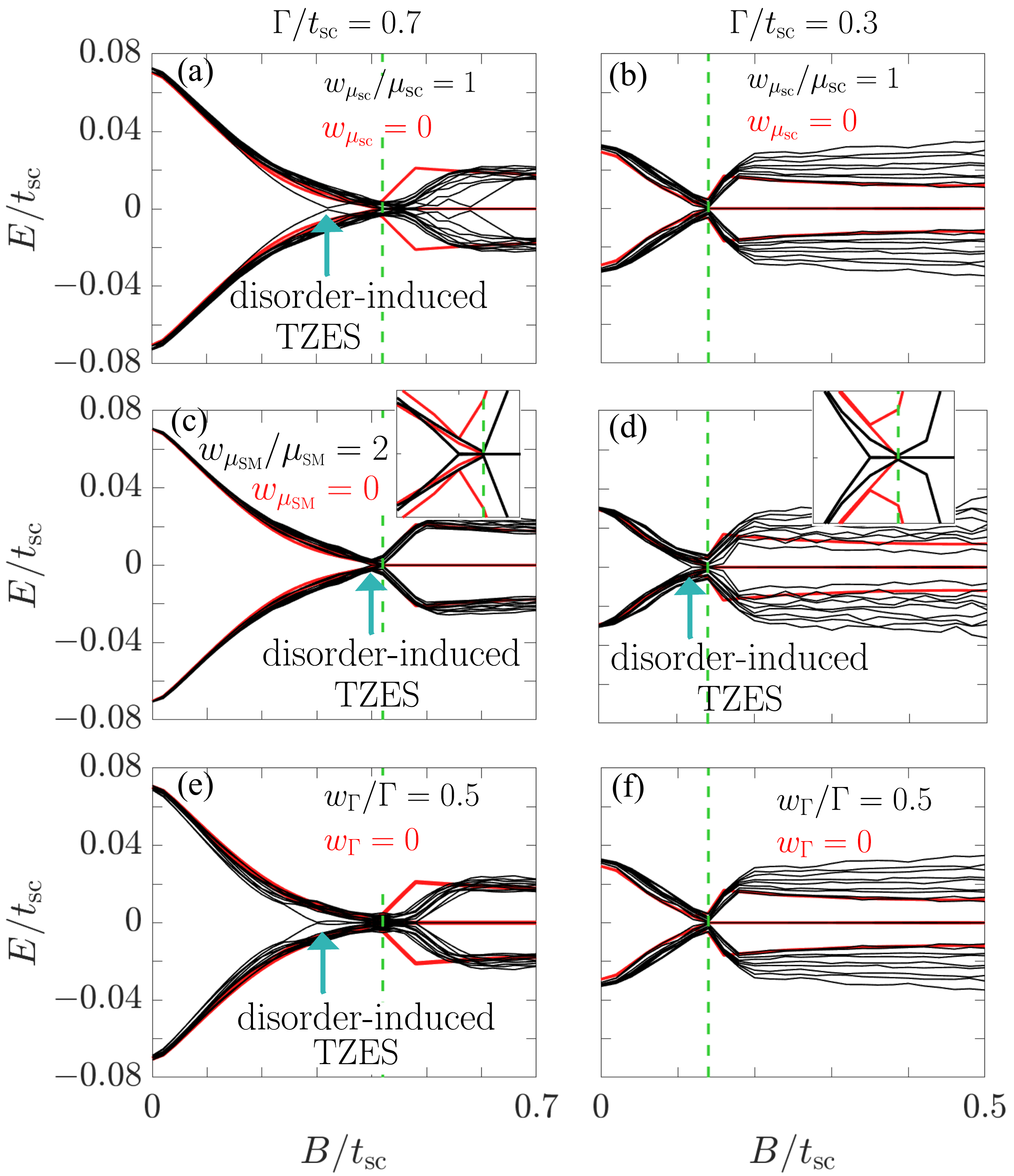} 
	\caption{Low-energy spectrum as a function of the Zeeman field $B$ for a single realization  (20 lowest levels, black) for disorder in the SC (a,b), SM (c,d), and at the interface (e,f) for strong (left) and weak (right) coupling $\Gamma$. The chosen values of disorder correspond to strong disorder, see text.  Insets show zooms around the TPT (dashed green). Clean system spectrum  (four lowest levels, red). 
	}
	\label{Fig2}
\end{figure}
\section{Disorder-induced TZESs} 
We next analyze the SC-SM system under disorder in the SC, SM, and SC-SM interface, as described by Eq.~\eqref{eq:Hdis} and separate the behavior for weak and strong coupling $\Gamma$. In Fig.\,\ref{Fig2} we show the low-energy spectrum as a function of $B$ under a single but representative \footnote{We have checked that other random disorder realizations support our claims.} realization of strong disorder (black) in the SC (a,b), SM (d,e), and at the interface (e,f) for  strong (left) and weak (right) coupling. To contrast the disordered results, we also show the four lowest  energy levels of the clean system (red), which display a clear TPT (dashed green) and no TZESs.

For strongly coupled SC-SM systems ($\Gamma/t_{\rm SC}=0.7$), the  immediate observation is that finite disorder in the SC  or at the interface induces TZESs well before the TPT [Fig. 2(a,g)].  The high impact of disorder in this regime occurs due to the large  renormalization induced by the SC on the SM parameters. Specifically, the chemical potential of the SM acquires a highly inhomogeneous spatial profile that can confine zero-energy states \cite{Awoga2019Supercurrent}. Disorder in the SM seems to be less detrimental, as it only affects the trivial phase just before the TPT,  but there also technically induces TZESs [Fig. 2(c)].   We find that these TZESs are  largely spatially located throughout the SC and  also remain if we remove the SN junction \footnote{To be precise, in the strong coupling regime, while most of the wavefunctions of the TZESs are mostly located in the SC,  part of them are also in the SM but that is much smaller.}.  We note here that the presence of TZESs is not violating the Anderson's theorem  about disorder-robustness in $s$-wave superconductors \cite{ANDERSON195926}, since they occur in a regime where time-reversal symmetry is broken and effective  $p$-wave spin-triplet superconducting correlations form \cite{PhysRevLett.87.037004, RevModPhys.77.1321, linder2015superconducting, PhysRevB.92.014513,solbook, PhysRevB.98.075425, PhysRevB.99.184512,PhysRevB.100.115433, cayao2019odd}.
Furthermore, we  note that for disorder in the SC or at the interface [Fig.\,\ref{Fig2}(a,e)], it is hard to discern the TPT and gap closing in the strong coupling regime due to an accumulation of low-energy states on both sides of the TPT  \cite{PhysRevResearch.2.013377,Ahn2021Estimating,DasSarma2021Disorder,DasCole2016Proximity}. This reduces the topological gap and even leads to the appearance of low-energy levels that coexist with MBSs above the TPT.    Therefore, disorder in the SC and interface is highly detrimental for MBSs in the strong coupling regime.
 
\begin{figure}[!t]
	\centering
	\includegraphics[width=.48\textwidth]{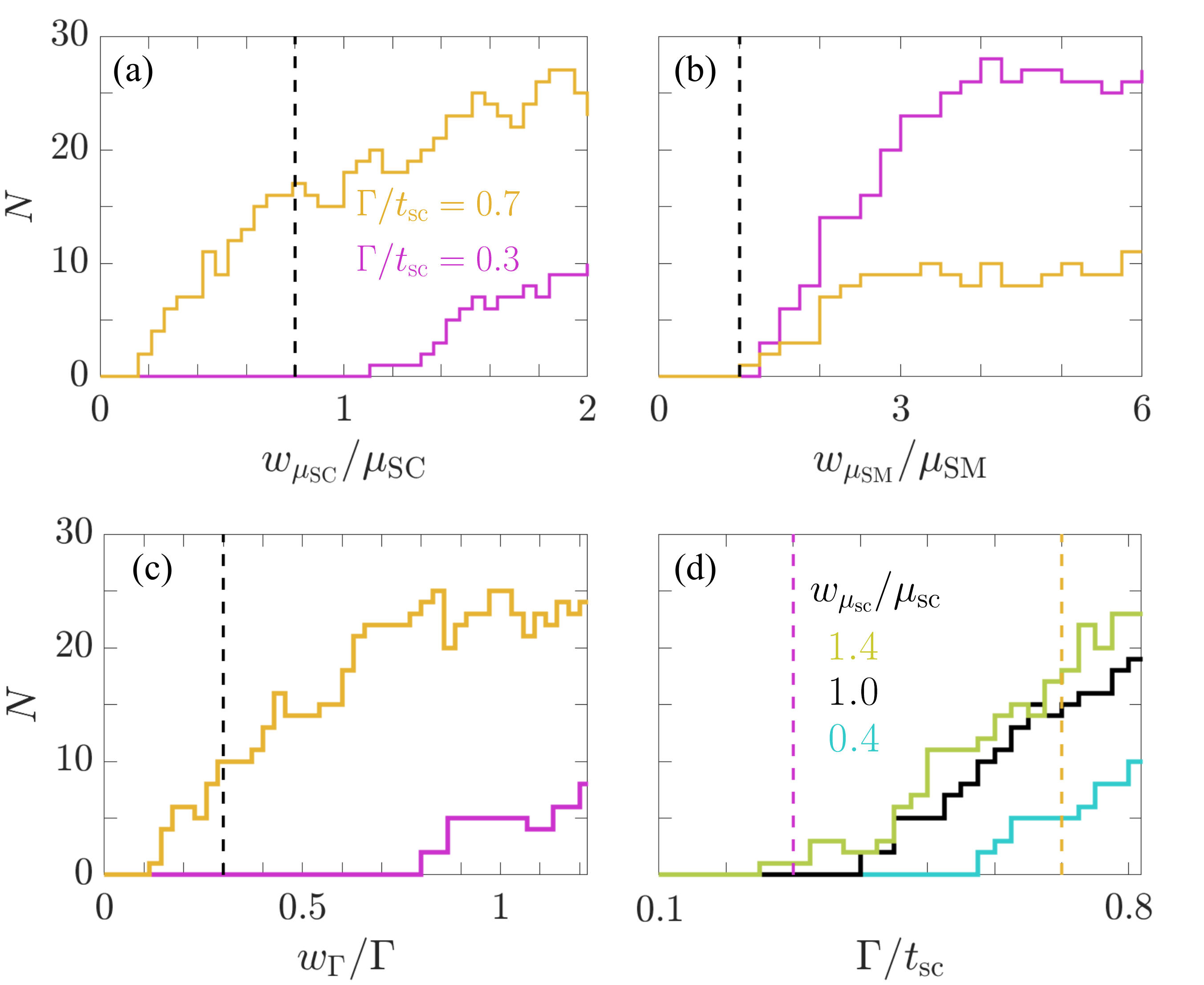} 
	\caption{Number of disorder configurations $N$ resulting in disorder-induced TZES (out of 30 random realizations) as a function of disorder strength at weak (yellow) and strong (purple) coupling $\Gamma$ for disorder in the SC (a), SM (b), and at the interface (c). Vertical black dashed lines indicates the boundary between weak (left side) and strong (right side) disorder.  (d) $N$ as a function of coupling strength, for disorder in the SC at different disorder strengths. Vertical colored lines denote $\Gamma/t_{\rm SC}=0.3, 0.7$.}
	\label{Fig3}
\end{figure}

For weakly coupled SC-SM systems  ($\Gamma/t_{\rm SC}=0.3$) we find that the impact of disorder on the low-energy spectrum is surprisingly largely absent, provided the strengths of disorder do not considerably surpass the values of their respective quantities in the clean regime [Fig.~\ref{Fig2}(b,d,f)]. In particular, the low-energy spectrum exhibits a  behavior similar to that  in the clean regime with no induced TZESs,   thus revealing an important advantage over the strong coupling regime \cite{PhysRevResearch.2.013377,Ahn2021Estimating,DasSarma2021Disorder,DasCole2016Proximity}. 
While this behavior practically occurs for disorder in all regions, it is fair to notice that the strong coupling reflects more resilience for disorder in the SM [Fig.~\ref{Fig2}(c,d)]. As no disorder-induced TZESs appear in the weak coupling regime, the TPT is also easily identified by naked eye, unlike in the strong coupling regime.  Furthermore, in the topological phase above the TPT, the low-energy spectrum reveals the emergence of MBSs and finite topological gap, both highly robust against disorder for weak coupling.  We note that the stability of the topological gap against disorder in the SC for weak $\Gamma$ is in line with Ref.\,\cite{PhysRevB.85.140513}, which showed that the impurity scattering rate in the SC  involves higher-order tunneling processes and is suppressed due to the destructive quantum interference of quasi-particle and quasi-hole trajectories. In a broader perspective, the robustness against disorder in both the trivial and topological phases in the weak coupling regime originates from the vanishing  renormalization of the effective  chemical potential in the SM for weak  $\Gamma$ \footnote{Apart from the chemical potential, other parameters such as spin-orbit coupling and $g$-factor also suffer a renormalization in the strong coupling regime}. To further explore the impact of disorder, we verify that the weak coupling regime remains robust  in a realistic scenario with disorder in all parts of the system, see Appendix \ref{App:DisAll}. 
These results clearly demonstrate the advantages and disadvantages of weak and strong couplings in disordered SC-SM systems. 

In order to obtain further understanding of the role of disorder, we show in Figs.\,\ref{Fig3}(a-c) the number of disorder configurations that produce disorder-induced TZESs ($N$), obtained by analyzing 30 random disorder realizations, as a function of disorder strength for all three types of disorder considered. We here only count disorder configurations that induce clearly isolated TZESs, like those seen in Fig.~\ref{Fig2},  in the Zeeman field range $B\in [0, B_{\rm c}]$. As the disorder strength increases,  $N$ also increases but with a very different behavior for weak (purple) and strong  (yellow) coupling $\Gamma$. Notably, for both disorder in the SC and at the interface, disorder-induced TZESs already form for very weak disorder at strong coupling $\Gamma$, while very strong disorder is required for TZESs to appear in the weak coupling regime [Fig.~\ref{Fig3}(a,c)]. In fact, in the latter situation, the disorder is often strong enough to also destroy superconductivity within a self-consistent calculation~\cite{Awoga2017disorder}.  On the other hand, for  disorder in the SM, disorder-induced TZESs only form for very strong disorder in both the weak and strong coupling regimes [Fig.~\ref{Fig3}(b)]. Thus, while it is always possible to avoid disorder-induced TZESs in weakly coupled SC-SM systems, disorder-induced TZES can  be avoided in strongly coupled SC-SM systems only for disorder in the SM.  To further inspect the robustness of the weak coupling regime, we plot in Fig.~\ref{Fig3}(d) $N$  as a function of $\Gamma$ for different strengths of disorder in the SC. The overall observation is that there exists always a regime at weak $\Gamma$ where no disorder-induced TZESs appear, with weaker coupling requiring ever stronger disorder  to form TZESs. We find similar results for disorder in the SM and at the interface. Thus, we find that it is possible to avoid disorder-induced TZESs in SC-SM systems even beyond the strong disorder limit as long the SC-SM coupling is kept weak. 

 
\begin{figure}[!t]
	\centering
	\includegraphics[width=0.48\textwidth]{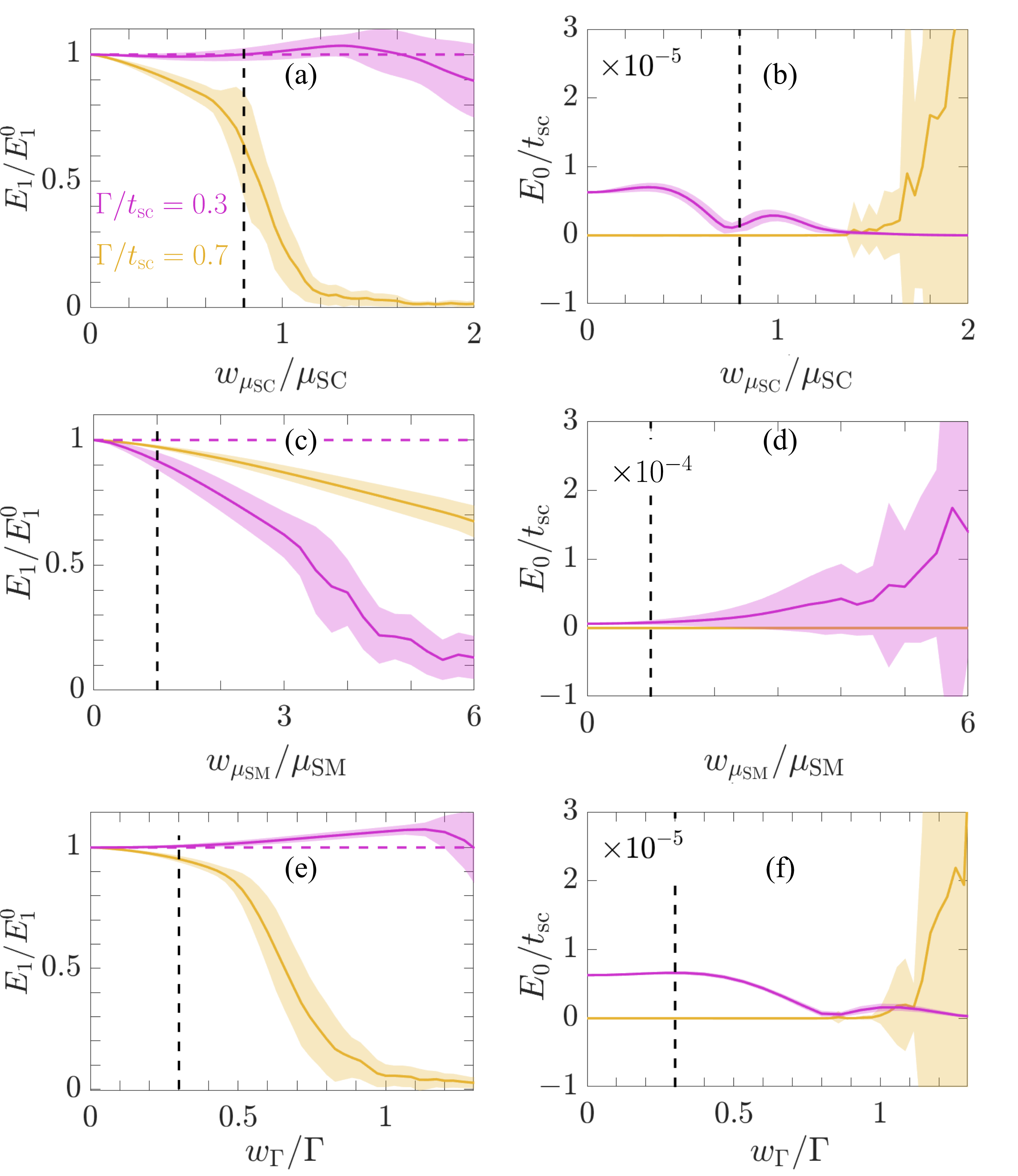} 
\caption{Disorder-averaged calculations (30 random configurations) in the topological phase at $B/B_{\rm c}=1.5$ showing $E_1/E_1^0$ (left) and $E_0/t_{\rm SC}$ (right) as functions of disorder strength for disorder in the SC (a,b), SM (c,d) and at the interface (e,f) for weak (purple) and strong (yellow) coupling $\Gamma$. Shaded regions represent one standard deviation. Vertical black dashed lines indicates the boundary between weak (left side) and strong (right side) disorder.}
	\label{Fig4}
\end{figure}

\section{Robustness of topological phase}
Finally, we explore the role of disorder in the topological phase, focusing on the two lowest   levels $E_{0(1)}$, because they determine the presence of MBSs (topological gap). In Fig.~\ref{Fig4} we plot the disorder-averaged $E_{0}$  (left) and $E_{1}$ (right) at fixed $B/B_c$ beyond the TPT as functions of the disorder strength for disorder in the SC (top), SM (middle), and at the interface  (bottom) for both the weak (purple) and strong (yellow) coupling regimes. The calculations correspond to  30 random disorder configurations with the standard deviation indicated (shaded regions), but we find that the results remain unchanged for increased number of configurations.  By direct inspection we note that $E_0$ and $E_1$  strongly depend on the strength of disorder, with non-trivial behavior for weak and strong $\Gamma$.
First, focusing on the topological gap $E_{1}$, we find that as disorder in the SC or at the interface increases, $E_{1}$ undergoes a notably fast reduction  in the strong coupling regime, acquiring a vanishing value just beyond what we characterize as the strong disorder regime Fig.~\ref{Fig4}(a,e).  In contrast, $E_1$ remains robust  for weak coupling and becomes even larger than the clean reference value $E_1^0$ (dashed line) for large and even strong disorder; it only decreases for very strong disorder, at first with an increased spread,  purple in Fig.~\ref{Fig4}(a,e).  While the robustness of the topological phase under weak disorder was reported earlier~\cite{Awoga2017disorder,Haim2019Benefits}, we emphasize that the robustness of the weak coupling regime seen here occurs even under strong disorder strengths,  for disorder either in the SC or at the interface.
For disorder in the SM, however, we find a faster reduction of the topological gap in the weak coupling regime than for strong coupling, but the reduction is  significant only for very strong disorder, such that the gap is largely preserved in both coupling regimes for weak to strong disorder.
 
For the lowest energy level $E_0$, representing the MBSs, we find a similar disorder robustness for disorder  both in the SC and at the interface for weak coupling  (purple in Figs.~\ref{Fig4}(b,f)). In particular, while a weak SC-SM coupling results in the $E_0$ level not being at zero energy in the clean system due to a finite Majorana localization length, disorder actually pushes this $E_0$ level towards zero by promoting smaller localization lengths,  consistent with the increased topological gap. Thus, the MBSs become  more localized for finite disorder in the SC and at the interface in the weak coupling regime.
In contrast, for strong coupling, $E_0$ clearly starts at zero energy in the clean regime, but then with increasing disorder becomes finite and also heavily dependent on the specific disorder configuration, as seen by the large spread.
The behavior is again different for disorder in the SM, where $E_{0}$ instead increases more with disorder at weak coupling than at strong coupling, but, again, the effect is not very noticeable until very strong disorder is considered. Taken together, the reduction in $E_0$ and small increase in $E_1$ with increasing disorder strength for both disorder in the SC and at the interface, together with an overall stability against disorder in the SM, implies that the topological phase in weakly-coupled SC-SM systems is very robust in the full range of weak to moderately strong disorder. In contrast, strongly coupled SC-SM systems show a pronounced fragility towards disorder both in the SC and at the interface, which stems from the strong renormalization caused in the SM parameters. In all of the above results we consider systems in which there are no TZESs in the clean regime, but we have verified that the results remain even if the clean system already hosts TZESs, e.g.~due to an interplay of the finite size of the SC and strong coupling, see Appendix \ref{App:DisTZES}.

\section{Conclusions}
To summarize, the weak coupling regime in SC-SM systems is robust against weak to moderately strong disorder and offers a powerful way to mitigate the formation of disorder-induced TZESs, in contrast to what has been reported for strong coupling \cite{PhysRevResearch.2.013377,Ahn2021Estimating,DasSarma2021Disorder,DasCole2016Proximity}. Furthermore, disorder in the SC or at the interface can even help generate more stable MBSs and a larger topological gap, both beneficial properties for designing future SC-SM structures with enhanced topological protection. At the same time, strongly coupled systems are more stable against very strong disorder in the SM.  After having identified how to mitigate the detrimental impact of disorder in SC-SM systems, the next step towards modeling realistic Majorana devices may include electrostatic effects \cite{dominguez2017zero,PhysRevB.98.035428,escribano2018interaction,PhysRevX.8.031040,PhysRevX.8.031041} and multichannel SMs \cite{PhysRevLett.106.127001,Stanescu2011Majorana,PhysRevLett.112.137001,PhysRevB.86.121103,PhysRevLett.105.227003}.
Our work thus illustrates the advantages and disadvantages of weakly  and strongly coupled SC-SM systems under scalar disorder, with potential use  for simulating and understanding actual Majorana experiments.

\begin{acknowledgements}
O.A.~and M.L.~acknowledge funding from NanoLund, the Swedish Research Council (VR) and the European Research Council (ERC) under the European Union's Horizon 2020 research and innovation programme under the Grant Agreement No.~856526. J.C.~acknowledges financial support from the Swedish Research Council  (Vetenskapsr\aa det Grant  No.~2021-04121), the G\"{o}ran Gustafsson Foundation (Grant No. 2216), and the Carl Trygger's Foundation (Grant No.~22: 2093).  A.B.S.~acknowledges financial support from the Swedish Research Council  (Vetenskapsr\aa det Grant No.~2018-03488) and the Knut and Alice Wallenberg Foundation through the Wallenberg Academy Fellows program. Simulations were enabled by resources provided by the Swedish National Infrastructure for Computing (SNIC) at the High Performance Computing Center North (HPC2N), Kebnekaise cluster, partially funded by the Swedish Research Council through Grant No.~2018-05973.
\end{acknowledgements}

%
\appendix
\section{Estimation of disorder strength}
\label{app:DisStrength}
In the main text we stated that disorder is strong when $w_{\mu_{\scaleto{\rm SC}{4pt}}}/\mu_{\scaleto{\rm SC}{4pt}}\gtrsim 0.8$, $w_{\mu_{\scaleto{\rm SM}{4pt}}}/\mu_{\scaleto{\rm SM}{4pt}}\gtrsim 1$,  and $w_{\Gamma}/\Gamma \gtrsim 0.3$, for disorder in the SC, SM, and at the interface, respectively. In this  Appendix, we provide details of this estimation.

We characterize the regimes of disorder in the SC and SM using the ratio $\xi/l$, where  $l$ the mean free path and $\xi$ is the superconducting coherence length \cite{Datta:97}. When using this ratio $\xi/l$ it is standard to classify the regime with $\xi/l\ll1$ as a weak disordered regime, while  $\xi/l \gtrsim 1$ denotes strong disorder. 
Here the mean free path is given by $l=v_{\rm F}/\pi N(0)w_{\scaleto{\rm SC,SM}{4pt}}^2$, where $v_{\rm F}$ and $N(0)$ are the Fermi velocity and normal state density of states, respectively and given by separate values  the SC and SM~\cite{Potter2011Enginering,DasCole2016Proximity}.
Also the superconducting coherence length is given by $\xi =v_{\rm F}/\Delta$, where $\Delta$ is the superconducting order parameter, or equivalently gap, in the SC, while it is the proximity-induced (superconducting) gap, $\Delta_{\rm ind}$, in the SM.

To estimate the disorder strength required to be in strong disorder regime, we treat the SC and SM separately. We consider the parameters listed in the Sec.\,\ref{sectionII}.
For the SC, we use the density of states for a 2D SC which is $N(0) = m^*/(\pi \hbar^2)=1/(2\pi t)$. We then obtain $\xi/l= w_{\mu_{\scaleto{\rm SC}{4pt}}}^2/(2\Delta t)$, which for our parameters gives $\xi/l \gtrsim 1$ when $w_{\mu_{\scaleto{\rm SC}{4pt}}} \gtrsim 0.4t_{\scaleto{\rm SC}{4pt}}$. This results in the SC being in the strong disorder regime for $w_{\mu_{\scaleto{\rm SC}{4pt}}}/\mu_{\scaleto{\rm SC}{4pt}} \gtrsim 0.8$, as stated in the main text. 
Similarly, for the 1D SM, we have $N(0)=2/(\pi \hbar v_{\rm F})$.  Thus, $\xi/l=2 w_{\mu_{\scaleto{\rm SM}{4pt}}}^2/(\Delta_{\rm ind} v_{\rm F})$ for the SM. For our parameter choices we find $\Delta_{\rm ind} \approx 0.04t_{\scaleto{\rm SC}{4pt}}\, \left(\Delta_{\rm ind} \approx 0.07t_{\scaleto{\rm SC}{4pt}}\right)$ for $\Gamma= 0.3t_{\scaleto{\rm SC}{4pt}}\, \left(\Gamma= 0.7t_{\scaleto{\rm SC}{4pt}}\right)$. We then obtain that the disorder strength for the strong disorder regime must satisfy $w_{\mu_{\scaleto{\rm SM}{4pt}}} \gtrsim 0.02 t_{\scaleto{\rm SM}{4pt}}, \left(w_{\mu_{\scaleto{\rm SM}{4pt}}} \gtrsim  0.03 t_{\scaleto{\rm SM}{4pt}}\right)$ for $\Gamma= 0.3t_{\scaleto{\rm SC}{4pt}}\, \left(\Gamma= 0.7t_{\scaleto{\rm SC}{4pt}}\right)$. This implies that $w_{\mu_{\scaleto{\rm SM}{4pt}}}/\mu_{\scaleto{\rm SM}{4pt}}\gtrsim 1$  characterizes the strong disorder regime, as stated in the main text. 

In the case of disorder at the interface, modeled as disorder in the coupling $\Gamma$, we follow a previous study \cite{Stanescu2011Majorana} and consider fluctuations larger than $30\%$ of the coupling strength to be in the strong disorder regime. Thus, strong disorder here occurs for  $w_{\Gamma}/\Gamma \gtrsim 0.3$, as stated in the main text.

%
\section{Disorder in all components}
\label{App:DisAll}
\begin{figure}[!t]
	\centering
	\includegraphics[width=1\linewidth]{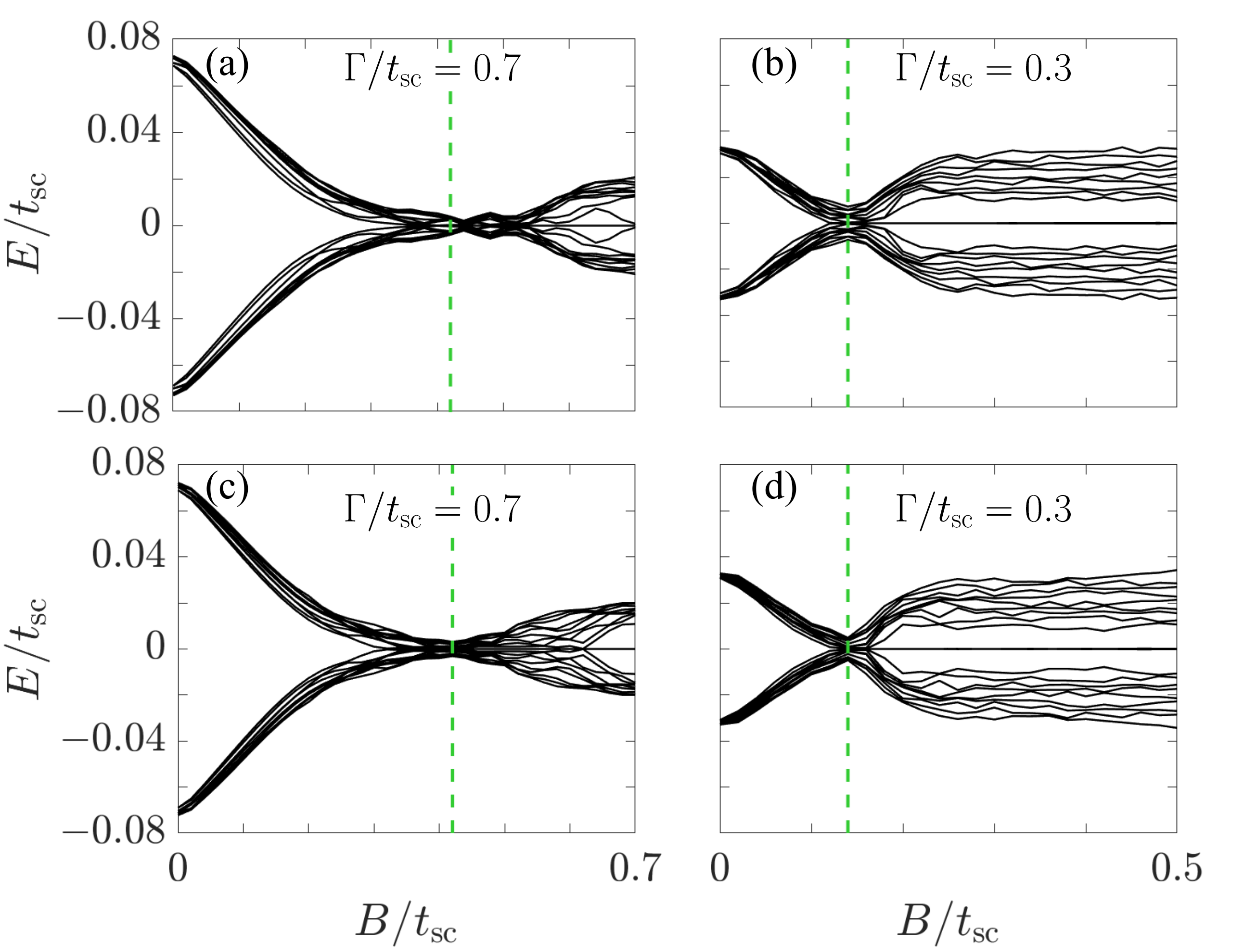}
	\caption{Low-energy spectrum as a function of Zeeman field $B/t_{\rm sc}$ for a single realization of disorder present in all the components: SC, SM, and at the interface. Top and bottom rows correspond to different disorder realizations. The left (right) column  corresponds to strong (weak) coupling. Dashed green line mark the topological phase transition point in the clean limit. All parameters are the same as Fig.~2 in the main text with disorder strengths $w_{\rm sc}/\mu_{\rm sc}=1$, $w_{\rm SM}/\mu_{\rm SM}=2$, and $w_{\Gamma}/\Gamma = 0.5$.}
	\label{FigS3}
\end{figure}
In the main text we consider disorder in only one component of the SC-SM system at a time, in order to identify  which part is more or less detrimental and how it can be mitigated in the weak coupling regime. In real samples, however, disorder is  likely present in all parts of the system at the same time. In this appendix we consider this situation and perform calculations  for scalar disorder  in all components simultaneously, i.e.~in the SC, SM, and at the interface. We plot the low-energy spectrum as a function of the Zeeman field in  Fig.~\ref{FigS3} for a single disorder realization at  strong (left column) and weak coupling (right column). We observe that, for strong couplings, there is an avalanche of disorder-induced zero-energy states in both the trivial and topological regimes, which ruins the topological gap, any possible identification of the topological phase transition,  and even a clear detection of  MBSs. In contrast, all these issues of the strong coupling regime are mitigated at weak coupling: the formation of trivial zero-energy states is clearly suppressed allowing to identify a clear topological phase transition, while the absence of ingap states in the topological phase generates a robust topological gap and only MBSs at zero energy. Thus, even when disorder is present in all components of the SC-SM system, the weak coupling regime proves to be beneficial for mitigating the formation of TZESs and detection of MBSs, thus supporting our  findings in the main text.

\section{Influence of disorder on TZES appearing in the clean limit}
\label{App:DisTZES}
In the main text we consider only systems that  do not host any TZES in the clean limit, such that the TZES in the main text are all induced by disorder. For different parameter choices it is however possible to also have TZESs appearing in the N region of the SM in the clean regime, which we here simply refer to as clean-limit TZESs. These clean-limit TZESs occur due to the renormalization of the effective chemical potential in the SM, which, due to the uncovered N region in our case, acquires an inhomogeneous profile, see Refs.\,\cite{Reeg2018Zero,Awoga2019Supercurrent,Awoga2022Robust}. In this appendix we provide additional data showing that our conclusions in the main text is not dependent on whether the clean-limit hosts TZESs already or not. To achieve TZESs in the clean limit we change the parameters slightly from the main text by setting $\mu_{\scaleto{\rm SC}{4pt}}=0.5t_{\scaleto{\rm SC}{4pt}}$, but keep all other parameters the same. This produces clean-limit TZESs in the N region in the strong coupling regime.
 
\begin{figure}[!t]
\centering
\includegraphics[width=0.48\textwidth]{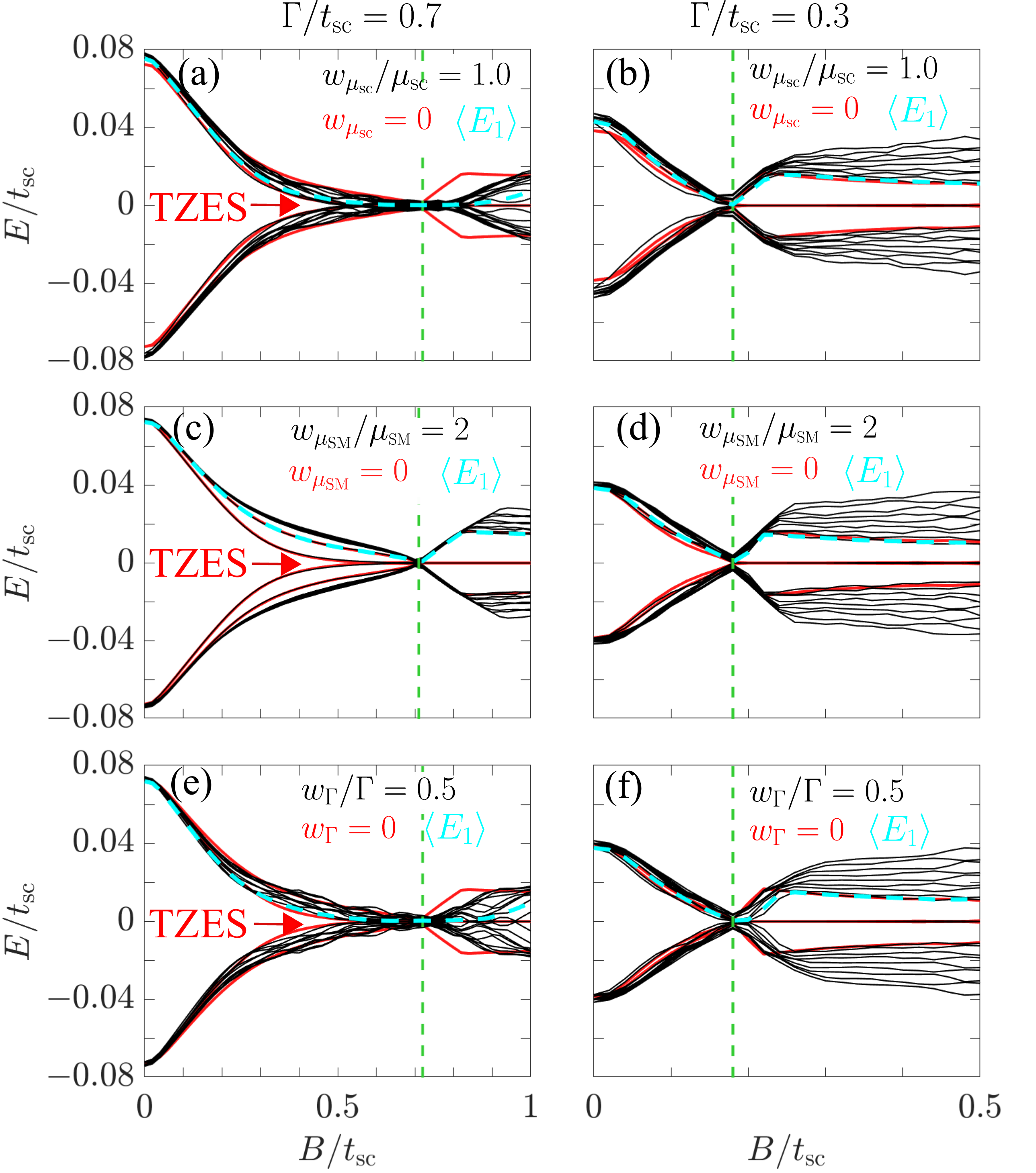} 
\caption{Low-energy spectrum as a function of the Zeeman field, $B$ for a single disorder configuration (20 lowest levels, black), for disorder in the SC (a,b), SM (c,d) and at the interface (e,f), strong (left) and weak (right) coupling $\Gamma$. 
Clean system spectrum (4 lowest levels only, red) and disorder-averaged $E_1$ (30 configurations) (cyan). Dashed green lines denote TPT. Compare with Fig.~2 in the main text.}
\label{FigS1}
\end{figure}

In Fig.~\ref{FigS1} we plot the low-energy spectrum as a function of the Zeeman field $B$ for both the strong (left) and weak (right) coupling regimes for strong disorder in the SC (top), SM (middle), and at the interface (bottom).  We compare these single disorder results (black) with results for the clean system (red) and also the disorder-averaged $E_1$ (cyan). Thus Fig.~\ref{FigS1} is the same as Fig.~2 in the main text, except that in Fig.~\ref{FigS1} we have also TZES in the clean limit for strong coupling. Comparing the two figures it is clear that the influence of disorder is very similar, namely, for disorder in the SC or at the interface, we find that at weak coupling there are no disorder-induced TZESs, while for strong coupling we clearly find additional disorder-induced TZESs close to the topological phase transition. For disorder in the SM, weak or strong coupling matters less, and the TZES in the clean limit for strong coupling is the only TZES in the system. 
These results are further solidified by studying the disorder-averaged $E_1$, which,  for both disorder in the SC and at the interface, it comes close to zero in the strong coupling regime while it remains almost same as  in the clean system value in the weak coupling regime. 

\begin{figure}[!t]
	\centering
	\includegraphics[width=0.48\textwidth]{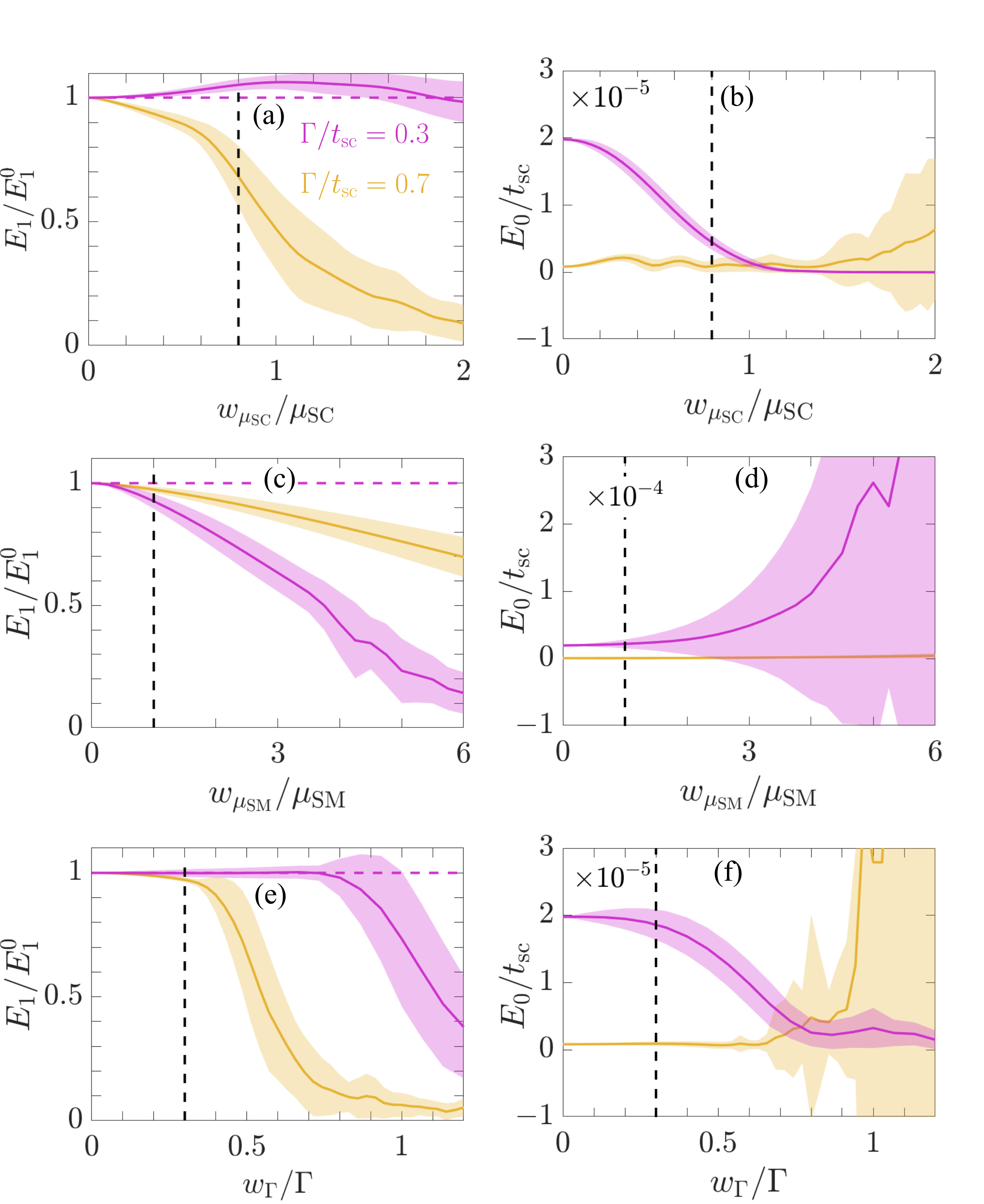} 
	\caption{Disorder-averaged calculations (30 random configurations) in the topological phase at $B/B_{\rm c}=1.5$ showing $E_0/t_{\scaleto{\rm SC}{4pt}}$ (left) and $E_1/E_1^0$ (right) as functions of disorder strength for disorder in the SC (a,b), SM (c,d) and at the interface (e,f) for weak (purple) and strong (yellow) coupling $\Gamma$. Shaded regions represent one standard deviation. Compare with Fig.~4 in the main text.  Vertical black dashed lines indicates the boundary between weak (left side) and strong (right side) disorder.}
	\label{FigS2}
\end{figure}

In Fig.~\ref{FigS2}  we plot disorder-averaged $E_1$ (left) and $E_0$ (right) in the topological phase, for disorder in the SC (top), SM (middle) and at the interface (bottom) for both strong (yellow) and weak (purple) SC-SM coupling. This is thus the equivalent of Fig.~4 in the main text, but now for a system with TZES in the clean limit for strong coupling. By direct comparison we see that there are no qualitative change of the results between Fig.~\ref{FigS2} and Fig.~4 in the main text. To summarize, the above results supports the conclusion of the main text, in that the weak coupling regime is more robust to disorder than the strong coupling regime, irrespective of whether TZESs are present in clean limit with strong coupling or not.

%
\bibliography{biblio}

\begin{thebibliography}{105}%
\makeatletter
\providecommand \@ifxundefined [1]{%
 \@ifx{#1\undefined}
}%
\providecommand \@ifnum [1]{%
 \ifnum #1\expandafter \@firstoftwo
 \else \expandafter \@secondoftwo
 \fi
}%
\providecommand \@ifx [1]{%
 \ifx #1\expandafter \@firstoftwo
 \else \expandafter \@secondoftwo
 \fi
}%
\providecommand \natexlab [1]{#1}%
\providecommand \enquote  [1]{``#1''}%
\providecommand \bibnamefont  [1]{#1}%
\providecommand \bibfnamefont [1]{#1}%
\providecommand \citenamefont [1]{#1}%
\providecommand \href@noop [0]{\@secondoftwo}%
\providecommand \href [0]{\begingroup \@sanitize@url \@href}%
\providecommand \@href[1]{\@@startlink{#1}\@@href}%
\providecommand \@@href[1]{\endgroup#1\@@endlink}%
\providecommand \@sanitize@url [0]{\catcode `\\12\catcode `\$12\catcode
  `\&12\catcode `\#12\catcode `\^12\catcode `\_12\catcode `\%12\relax}%
\providecommand \@@startlink[1]{}%
\providecommand \@@endlink[0]{}%
\providecommand \url  [0]{\begingroup\@sanitize@url \@url }%
\providecommand \@url [1]{\endgroup\@href {#1}{\urlprefix }}%
\providecommand \urlprefix  [0]{URL }%
\providecommand \Eprint [0]{\href }%
\providecommand \doibase [0]{https://doi.org/}%
\providecommand \selectlanguage [0]{\@gobble}%
\providecommand \bibinfo  [0]{\@secondoftwo}%
\providecommand \bibfield  [0]{\@secondoftwo}%
\providecommand \translation [1]{[#1]}%
\providecommand \BibitemOpen [0]{}%
\providecommand \bibitemStop [0]{}%
\providecommand \bibitemNoStop [0]{.\EOS\space}%
\providecommand \EOS [0]{\spacefactor3000\relax}%
\providecommand \BibitemShut  [1]{\csname bibitem#1\endcsname}%
\let\auto@bib@innerbib\@empty
\bibitem [{\citenamefont {Leijnse}\ and\ \citenamefont
  {Flensberg}(2012)}]{Leijnse2012Introduction}%
  \BibitemOpen
  \bibfield  {author} {\bibinfo {author} {\bibfnamefont {M.}~\bibnamefont
  {Leijnse}}\ and\ \bibinfo {author} {\bibfnamefont {K.}~\bibnamefont
  {Flensberg}},\ }\bibfield  {title} {\bibinfo {title} {Introduction to
  topological superconductivity and {M}ajorana fermions},\ }\href
  {https://doi.org/10.1088/0268-1242/27/12/124003} {\bibfield  {journal}
  {\bibinfo  {journal} {Semicond. Sci. Technol.}\ }\textbf {\bibinfo {volume}
  {27}},\ \bibinfo {pages} {124003} (\bibinfo {year} {2012})}\BibitemShut
  {NoStop}%
\bibitem [{\citenamefont {Aguado}(2017)}]{Aguadoreview17}%
  \BibitemOpen
  \bibfield  {author} {\bibinfo {author} {\bibfnamefont {R.}~\bibnamefont
  {Aguado}},\ }\bibfield  {title} {\bibinfo {title} {Majorana quasiparticles in
  condensed matter},\ }\href {https://doi.org/10.1393/ncr/i2017-10141-9}
  {\bibfield  {journal} {\bibinfo  {journal} {Riv. Nuovo Cimento}\ }\textbf
  {\bibinfo {volume} {40}},\ \bibinfo {pages} {523} (\bibinfo {year}
  {2017})}\BibitemShut {NoStop}%
\bibitem [{\citenamefont {Lutchyn}\ \emph {et~al.}(2018)\citenamefont
  {Lutchyn}, \citenamefont {Bakkers}, \citenamefont {Kouwenhoven},
  \citenamefont {Krogstrup}, \citenamefont {Marcus},\ and\ \citenamefont
  {Oreg}}]{Lutchyn2018Majorana}%
  \BibitemOpen
  \bibfield  {author} {\bibinfo {author} {\bibfnamefont {R.~M.}\ \bibnamefont
  {Lutchyn}}, \bibinfo {author} {\bibfnamefont {E.~P. A.~M.}\ \bibnamefont
  {Bakkers}}, \bibinfo {author} {\bibfnamefont {L.~P.}\ \bibnamefont
  {Kouwenhoven}}, \bibinfo {author} {\bibfnamefont {P.}~\bibnamefont
  {Krogstrup}}, \bibinfo {author} {\bibfnamefont {C.~M.}\ \bibnamefont
  {Marcus}},\ and\ \bibinfo {author} {\bibfnamefont {Y.}~\bibnamefont {Oreg}},\
  }\bibfield  {title} {\bibinfo {title} {Majorana zero modes in
  superconductor-semiconductor heterostructures},\ }\href
  {https://doi.org/10.1038/s41578-018-0003-1} {\bibfield  {journal} {\bibinfo
  {journal} {Nat. Rev. Mater.}\ }\textbf {\bibinfo {volume} {3}},\ \bibinfo
  {pages} {52} (\bibinfo {year} {2018})}\BibitemShut {NoStop}%
\bibitem [{\citenamefont {Zhang}\ \emph {et~al.}(2019)\citenamefont {Zhang},
  \citenamefont {Liu}, \citenamefont {Wimmer},\ and\ \citenamefont
  {Kouwenhoven}}]{zhang2019next}%
  \BibitemOpen
  \bibfield  {author} {\bibinfo {author} {\bibfnamefont {H.}~\bibnamefont
  {Zhang}}, \bibinfo {author} {\bibfnamefont {D.~E.}\ \bibnamefont {Liu}},
  \bibinfo {author} {\bibfnamefont {M.}~\bibnamefont {Wimmer}},\ and\ \bibinfo
  {author} {\bibfnamefont {L.~P.}\ \bibnamefont {Kouwenhoven}},\ }\bibfield
  {title} {\bibinfo {title} {Next steps of quantum transport in majorana
  nanowire devices},\ }\href@noop {} {\bibfield  {journal} {\bibinfo  {journal}
  {Nat. Commun.}\ }\textbf {\bibinfo {volume} {10}},\ \bibinfo {pages} {5128}
  (\bibinfo {year} {2019})}\BibitemShut {NoStop}%
\bibitem [{\citenamefont {Frolov}\ \emph {et~al.}(2020)\citenamefont {Frolov},
  \citenamefont {Manfra},\ and\ \citenamefont {Sau}}]{frolov2020topological}%
  \BibitemOpen
  \bibfield  {author} {\bibinfo {author} {\bibfnamefont {S.~M.}\ \bibnamefont
  {Frolov}}, \bibinfo {author} {\bibfnamefont {M.~J.}\ \bibnamefont {Manfra}},\
  and\ \bibinfo {author} {\bibfnamefont {J.~D.}\ \bibnamefont {Sau}},\
  }\bibfield  {title} {\bibinfo {title} {Topological superconductivity in
  hybrid devices},\ }\href@noop {} {\bibfield  {journal} {\bibinfo  {journal}
  {Nat. Phys.}\ }\textbf {\bibinfo {volume} {16}},\ \bibinfo {pages} {718}
  (\bibinfo {year} {2020})}\BibitemShut {NoStop}%
\bibitem [{\citenamefont {Laubscher}\ and\ \citenamefont
  {Klinovaja}(2021)}]{Laubscher2021Majorana}%
  \BibitemOpen
  \bibfield  {author} {\bibinfo {author} {\bibfnamefont {K.}~\bibnamefont
  {Laubscher}}\ and\ \bibinfo {author} {\bibfnamefont {J.}~\bibnamefont
  {Klinovaja}},\ }\bibfield  {title} {\bibinfo {title} {Majorana bound states
  in semiconducting nanostructures},\ }\href
  {https://doi.org/10.1063/5.0055997} {\bibfield  {journal} {\bibinfo
  {journal} {J. Appl. Phys.}\ }\textbf {\bibinfo {volume} {130}},\ \bibinfo
  {pages} {081101} (\bibinfo {year} {2021})}\BibitemShut {NoStop}%
\bibitem [{\citenamefont {Prada}\ \emph {et~al.}(2020)\citenamefont {Prada},
  \citenamefont {San-Jose}, \citenamefont {de~Moor}, \citenamefont {Geresdi},
  \citenamefont {Lee}, \citenamefont {Klinovaja}, \citenamefont {Loss},
  \citenamefont {Nyg{\aa}rd}, \citenamefont {Aguado},\ and\ \citenamefont
  {Kouwenhoven}}]{Prada2020Adreev}%
  \BibitemOpen
  \bibfield  {author} {\bibinfo {author} {\bibfnamefont {E.}~\bibnamefont
  {Prada}}, \bibinfo {author} {\bibfnamefont {P.}~\bibnamefont {San-Jose}},
  \bibinfo {author} {\bibfnamefont {M.~W.}\ \bibnamefont {de~Moor}}, \bibinfo
  {author} {\bibfnamefont {A.}~\bibnamefont {Geresdi}}, \bibinfo {author}
  {\bibfnamefont {E.~J.}\ \bibnamefont {Lee}}, \bibinfo {author} {\bibfnamefont
  {J.}~\bibnamefont {Klinovaja}}, \bibinfo {author} {\bibfnamefont
  {D.}~\bibnamefont {Loss}}, \bibinfo {author} {\bibfnamefont {J.}~\bibnamefont
  {Nyg{\aa}rd}}, \bibinfo {author} {\bibfnamefont {R.}~\bibnamefont {Aguado}},\
  and\ \bibinfo {author} {\bibfnamefont {L.~P.}\ \bibnamefont {Kouwenhoven}},\
  }\bibfield  {title} {\bibinfo {title} {From {A}ndreev to {M}ajorana bound
  states in hybrid superconductor--semiconductor nanowires},\ }\href
  {https://doi.org/10.1038/s42254-020-0228-y} {\bibfield  {journal} {\bibinfo
  {journal} {Nat. Rev. Phys}\ }\textbf {\bibinfo {volume} {2}},\ \bibinfo
  {pages} {575} (\bibinfo {year} {2020})}\BibitemShut {NoStop}%
\bibitem [{\citenamefont {Flensberg}\ \emph {et~al.}(2021)\citenamefont
  {Flensberg}, \citenamefont {von Oppen},\ and\ \citenamefont
  {Stern}}]{flensberg2021engineered}%
  \BibitemOpen
  \bibfield  {author} {\bibinfo {author} {\bibfnamefont {K.}~\bibnamefont
  {Flensberg}}, \bibinfo {author} {\bibfnamefont {F.}~\bibnamefont {von
  Oppen}},\ and\ \bibinfo {author} {\bibfnamefont {A.}~\bibnamefont {Stern}},\
  }\bibfield  {title} {\bibinfo {title} {Engineered platforms for topological
  superconductivity and majorana zero modes},\ }\href@noop {} {\bibfield
  {journal} {\bibinfo  {journal} {Nat. Rev. Mater.}\ }\textbf {\bibinfo
  {volume} {6}},\ \bibinfo {pages} {944} (\bibinfo {year} {2021})}\BibitemShut
  {NoStop}%
\bibitem [{\citenamefont {Marra}(2022)}]{marra2022majorana2}%
  \BibitemOpen
  \bibfield  {author} {\bibinfo {author} {\bibfnamefont {P.}~\bibnamefont
  {Marra}},\ }\bibfield  {title} {\bibinfo {title} {Majorana nanowires for
  topological quantum computation},\ }\href {https://doi.org/10.1063/5.0102999}
  {\bibfield  {journal} {\bibinfo  {journal} {J. Appl. Phys.}\ }\textbf
  {\bibinfo {volume} {132}},\ \bibinfo {pages} {231101} (\bibinfo {year}
  {2022})}\BibitemShut {NoStop}%
\bibitem [{\citenamefont {Oreg}\ \emph {et~al.}(2010)\citenamefont {Oreg},
  \citenamefont {Refael},\ and\ \citenamefont {von Oppen}}]{Oreg:PRL10}%
  \BibitemOpen
  \bibfield  {author} {\bibinfo {author} {\bibfnamefont {Y.}~\bibnamefont
  {Oreg}}, \bibinfo {author} {\bibfnamefont {G.}~\bibnamefont {Refael}},\ and\
  \bibinfo {author} {\bibfnamefont {F.}~\bibnamefont {von Oppen}},\ }\bibfield
  {title} {\bibinfo {title} {Helical liquids and majorana bound states in
  quantum wires},\ }\href@noop {} {\bibfield  {journal} {\bibinfo  {journal}
  {Phys. Rev. Lett.}\ }\textbf {\bibinfo {volume} {105}},\ \bibinfo {pages}
  {177002} (\bibinfo {year} {2010})}\BibitemShut {NoStop}%
\bibitem [{\citenamefont {Alicea}(2010)}]{Alicea:PRB10}%
  \BibitemOpen
  \bibfield  {author} {\bibinfo {author} {\bibfnamefont {J.}~\bibnamefont
  {Alicea}},\ }\bibfield  {title} {\bibinfo {title} {Majorana fermions in a
  tunable semiconductor device},\ }\href
  {https://link.aps.org/doi/10.1103/PhysRevB.81.125318} {\bibfield  {journal}
  {\bibinfo  {journal} {Phys. Rev. B}\ }\textbf {\bibinfo {volume} {81}},\
  \bibinfo {pages} {125318} (\bibinfo {year} {2010})}\BibitemShut {NoStop}%
\bibitem [{\citenamefont {Lutchyn}\ \emph {et~al.}(2010)\citenamefont
  {Lutchyn}, \citenamefont {Sau},\ and\ \citenamefont
  {Das~Sarma}}]{PhysRevLett.105.077001}%
  \BibitemOpen
  \bibfield  {author} {\bibinfo {author} {\bibfnamefont {R.~M.}\ \bibnamefont
  {Lutchyn}}, \bibinfo {author} {\bibfnamefont {J.~D.}\ \bibnamefont {Sau}},\
  and\ \bibinfo {author} {\bibfnamefont {S.}~\bibnamefont {Das~Sarma}},\
  }\bibfield  {title} {\bibinfo {title} {Majorana fermions and a topological
  phase transition in semiconductor-superconductor heterostructures},\ }\href
  {https://doi.org/10.1103/PhysRevLett.105.077001} {\bibfield  {journal}
  {\bibinfo  {journal} {Phys. Rev. Lett.}\ }\textbf {\bibinfo {volume} {105}},\
  \bibinfo {pages} {077001} (\bibinfo {year} {2010})}\BibitemShut {NoStop}%
\bibitem [{\citenamefont {Tanaka}\ and\ \citenamefont
  {Kashiwaya}(1995)}]{PhysRevLett.74.3451}%
  \BibitemOpen
  \bibfield  {author} {\bibinfo {author} {\bibfnamefont {Y.}~\bibnamefont
  {Tanaka}}\ and\ \bibinfo {author} {\bibfnamefont {S.}~\bibnamefont
  {Kashiwaya}},\ }\bibfield  {title} {\bibinfo {title} {Theory of tunneling
  spectroscopy of $\mathit{d}$-wave superconductors},\ }\href
  {https://doi.org/10.1103/PhysRevLett.74.3451} {\bibfield  {journal} {\bibinfo
   {journal} {Phys. Rev. Lett.}\ }\textbf {\bibinfo {volume} {74}},\ \bibinfo
  {pages} {3451} (\bibinfo {year} {1995})}\BibitemShut {NoStop}%
\bibitem [{\citenamefont {Bolech}\ and\ \citenamefont
  {Demler}(2007)}]{PhysRevLett.98.237002}%
  \BibitemOpen
  \bibfield  {author} {\bibinfo {author} {\bibfnamefont {C.~J.}\ \bibnamefont
  {Bolech}}\ and\ \bibinfo {author} {\bibfnamefont {E.}~\bibnamefont
  {Demler}},\ }\bibfield  {title} {\bibinfo {title} {Observing {M}ajorana bound
  states in $p$-wave superconductors using noise measurements in tunneling
  experiments},\ }\href {https://doi.org/10.1103/PhysRevLett.98.237002}
  {\bibfield  {journal} {\bibinfo  {journal} {Phys. Rev. Lett.}\ }\textbf
  {\bibinfo {volume} {98}},\ \bibinfo {pages} {237002} (\bibinfo {year}
  {2007})}\BibitemShut {NoStop}%
\bibitem [{\citenamefont {Law}\ \emph {et~al.}(2009)\citenamefont {Law},
  \citenamefont {Lee},\ and\ \citenamefont {Ng}}]{PhysRevLett.103.237001}%
  \BibitemOpen
  \bibfield  {author} {\bibinfo {author} {\bibfnamefont {K.~T.}\ \bibnamefont
  {Law}}, \bibinfo {author} {\bibfnamefont {P.~A.}\ \bibnamefont {Lee}},\ and\
  \bibinfo {author} {\bibfnamefont {T.~K.}\ \bibnamefont {Ng}},\ }\bibfield
  {title} {\bibinfo {title} {Majorana fermion induced resonant {A}ndreev
  reflection},\ }\href {https://doi.org/10.1103/PhysRevLett.103.237001}
  {\bibfield  {journal} {\bibinfo  {journal} {Phys. Rev. Lett.}\ }\textbf
  {\bibinfo {volume} {103}},\ \bibinfo {pages} {237001} (\bibinfo {year}
  {2009})}\BibitemShut {NoStop}%
\bibitem [{\citenamefont {Flensberg}(2010)}]{PhysRevB.82.180516}%
  \BibitemOpen
  \bibfield  {author} {\bibinfo {author} {\bibfnamefont {K.}~\bibnamefont
  {Flensberg}},\ }\bibfield  {title} {\bibinfo {title} {Tunneling
  characteristics of a chain of {M}ajorana bound states},\ }\href
  {https://doi.org/10.1103/PhysRevB.82.180516} {\bibfield  {journal} {\bibinfo
  {journal} {Phys. Rev. B}\ }\textbf {\bibinfo {volume} {82}},\ \bibinfo
  {pages} {180516} (\bibinfo {year} {2010})}\BibitemShut {NoStop}%
\bibitem [{\citenamefont {Mourik}\ \emph {et~al.}(2012)\citenamefont {Mourik},
  \citenamefont {Zuo}, \citenamefont {Frolov}, \citenamefont {Plissard},
  \citenamefont {Bakkers},\ and\ \citenamefont {Kouwenhoven}}]{Mourik:S12}%
  \BibitemOpen
  \bibfield  {author} {\bibinfo {author} {\bibfnamefont {V.}~\bibnamefont
  {Mourik}}, \bibinfo {author} {\bibfnamefont {K.}~\bibnamefont {Zuo}},
  \bibinfo {author} {\bibfnamefont {S.}~\bibnamefont {Frolov}}, \bibinfo
  {author} {\bibfnamefont {S.}~\bibnamefont {Plissard}}, \bibinfo {author}
  {\bibfnamefont {E.}~\bibnamefont {Bakkers}},\ and\ \bibinfo {author}
  {\bibfnamefont {L.}~\bibnamefont {Kouwenhoven}},\ }\bibfield  {title}
  {\bibinfo {title} {Signatures of {M}ajorana fermions in hybrid
  superconductor-semiconductor nanowire devices},\ }\href
  {http://www.sciencemag.org/content/early/2012/04/11/science.1222360.abstract}
  {\bibfield  {journal} {\bibinfo  {journal} {Science}\ }\textbf {\bibinfo
  {volume} {336}},\ \bibinfo {pages} {1003} (\bibinfo {year}
  {2012})}\BibitemShut {NoStop}%
\bibitem [{\citenamefont {Deng}\ \emph {et~al.}(2012)\citenamefont {Deng},
  \citenamefont {Yu}, \citenamefont {Huang}, \citenamefont {Larsson},
  \citenamefont {Caroff},\ and\ \citenamefont {Xu}}]{Deng:NL12}%
  \BibitemOpen
  \bibfield  {author} {\bibinfo {author} {\bibfnamefont {M.~T.}\ \bibnamefont
  {Deng}}, \bibinfo {author} {\bibfnamefont {C.~L.}\ \bibnamefont {Yu}},
  \bibinfo {author} {\bibfnamefont {G.~Y.}\ \bibnamefont {Huang}}, \bibinfo
  {author} {\bibfnamefont {M.}~\bibnamefont {Larsson}}, \bibinfo {author}
  {\bibfnamefont {P.}~\bibnamefont {Caroff}},\ and\ \bibinfo {author}
  {\bibfnamefont {H.~Q.}\ \bibnamefont {Xu}},\ }\bibfield  {title} {\bibinfo
  {title} {Anomalous zero-bias conductance peak in a{ Nb--InSb} nanowire--{Nb}
  hybrid device},\ }\href {https://doi.org/10.1021/nl303758w} {\bibfield
  {journal} {\bibinfo  {journal} {Nano Lett.}\ }\textbf {\bibinfo {volume}
  {12}},\ \bibinfo {pages} {6414} (\bibinfo {year} {2012})}\BibitemShut
  {NoStop}%
\bibitem [{\citenamefont {Albrecht}\ \emph {et~al.}(2016)\citenamefont
  {Albrecht}, \citenamefont {Higginbotham}, \citenamefont {Madsen},
  \citenamefont {Kuemmeth}, \citenamefont {Jespersen}, \citenamefont
  {Nyg{\aa}rd}, \citenamefont {Krogstrup},\ and\ \citenamefont
  {Marcus}}]{Albrecht16}%
  \BibitemOpen
  \bibfield  {author} {\bibinfo {author} {\bibfnamefont {S.~M.}\ \bibnamefont
  {Albrecht}}, \bibinfo {author} {\bibfnamefont {A.~P.}\ \bibnamefont
  {Higginbotham}}, \bibinfo {author} {\bibfnamefont {M.}~\bibnamefont
  {Madsen}}, \bibinfo {author} {\bibfnamefont {F.}~\bibnamefont {Kuemmeth}},
  \bibinfo {author} {\bibfnamefont {T.~S.}\ \bibnamefont {Jespersen}}, \bibinfo
  {author} {\bibfnamefont {J.}~\bibnamefont {Nyg{\aa}rd}}, \bibinfo {author}
  {\bibfnamefont {P.}~\bibnamefont {Krogstrup}},\ and\ \bibinfo {author}
  {\bibfnamefont {C.~M.}\ \bibnamefont {Marcus}},\ }\bibfield  {title}
  {\bibinfo {title} {Exponential protection of zero modes in {M}ajorana
  islands},\ }\href {https://doi.org/10.1038/nature17162} {\bibfield  {journal}
  {\bibinfo  {journal} {Nature}\ }\textbf {\bibinfo {volume} {531}},\ \bibinfo
  {pages} {206} (\bibinfo {year} {2016})}\BibitemShut {NoStop}%
\bibitem [{\citenamefont {Deng}\ \emph {et~al.}(2016)\citenamefont {Deng},
  \citenamefont {Vaitiek{\.e}nas}, \citenamefont {Hansen}, \citenamefont
  {Danon}, \citenamefont {Leijnse}, \citenamefont {Flensberg}, \citenamefont
  {Nyg{\aa}rd}, \citenamefont {Krogstrup},\ and\ \citenamefont
  {Marcus}}]{deng2016majorana}%
  \BibitemOpen
  \bibfield  {author} {\bibinfo {author} {\bibfnamefont {M.~T.}\ \bibnamefont
  {Deng}}, \bibinfo {author} {\bibfnamefont {S.}~\bibnamefont
  {Vaitiek{\.e}nas}}, \bibinfo {author} {\bibfnamefont {E.~B.}\ \bibnamefont
  {Hansen}}, \bibinfo {author} {\bibfnamefont {J.}~\bibnamefont {Danon}},
  \bibinfo {author} {\bibfnamefont {M.}~\bibnamefont {Leijnse}}, \bibinfo
  {author} {\bibfnamefont {K.}~\bibnamefont {Flensberg}}, \bibinfo {author}
  {\bibfnamefont {J.}~\bibnamefont {Nyg{\aa}rd}}, \bibinfo {author}
  {\bibfnamefont {P.}~\bibnamefont {Krogstrup}},\ and\ \bibinfo {author}
  {\bibfnamefont {C.~M.}\ \bibnamefont {Marcus}},\ }\bibfield  {title}
  {\bibinfo {title} {Majorana bound state in a coupled quantum-dot
  hybrid-nanowire system},\ }\href
  {https://www.science.org/doi/10.1126/science.aaf3961} {\bibfield  {journal}
  {\bibinfo  {journal} {Science}\ }\textbf {\bibinfo {volume} {354}},\ \bibinfo
  {pages} {1557} (\bibinfo {year} {2016})}\BibitemShut {NoStop}%
\bibitem [{\citenamefont {Suominen}\ \emph {et~al.}(2017)\citenamefont
  {Suominen}, \citenamefont {Kjaergaard}, \citenamefont {Hamilton},
  \citenamefont {Shabani}, \citenamefont {Palmstr\o{}m}, \citenamefont
  {Marcus},\ and\ \citenamefont {Nichele}}]{Suominen17}%
  \BibitemOpen
  \bibfield  {author} {\bibinfo {author} {\bibfnamefont {H.~J.}\ \bibnamefont
  {Suominen}}, \bibinfo {author} {\bibfnamefont {M.}~\bibnamefont
  {Kjaergaard}}, \bibinfo {author} {\bibfnamefont {A.~R.}\ \bibnamefont
  {Hamilton}}, \bibinfo {author} {\bibfnamefont {J.}~\bibnamefont {Shabani}},
  \bibinfo {author} {\bibfnamefont {C.~J.}\ \bibnamefont {Palmstr\o{}m}},
  \bibinfo {author} {\bibfnamefont {C.~M.}\ \bibnamefont {Marcus}},\ and\
  \bibinfo {author} {\bibfnamefont {F.}~\bibnamefont {Nichele}},\ }\bibfield
  {title} {\bibinfo {title} {Zero-energy modes from coalescing {A}ndreev states
  in a two-dimensional semiconductor-superconductor hybrid platform},\ }\href
  {https://link.aps.org/doi/10.1103/PhysRevLett.119.176805} {\bibfield
  {journal} {\bibinfo  {journal} {Phys. Rev. Lett.}\ }\textbf {\bibinfo
  {volume} {119}},\ \bibinfo {pages} {176805} (\bibinfo {year}
  {2017})}\BibitemShut {NoStop}%
\bibitem [{\citenamefont {Nichele}\ \emph {et~al.}(2017)\citenamefont
  {Nichele}, \citenamefont {Drachmann}, \citenamefont {Whiticar}, \citenamefont
  {O'Farrell}, \citenamefont {Suominen}, \citenamefont {Fornieri},
  \citenamefont {Wang}, \citenamefont {Gardner}, \citenamefont {Thomas},
  \citenamefont {Hatke}, \citenamefont {Krogstrup}, \citenamefont {Manfra},
  \citenamefont {Flensberg},\ and\ \citenamefont
  {Marcus}}]{Nichele2017Scaling}%
  \BibitemOpen
  \bibfield  {author} {\bibinfo {author} {\bibfnamefont {F.}~\bibnamefont
  {Nichele}}, \bibinfo {author} {\bibfnamefont {A.~C.~C.}\ \bibnamefont
  {Drachmann}}, \bibinfo {author} {\bibfnamefont {A.~M.}\ \bibnamefont
  {Whiticar}}, \bibinfo {author} {\bibfnamefont {E.~C.~T.}\ \bibnamefont
  {O'Farrell}}, \bibinfo {author} {\bibfnamefont {H.~J.}\ \bibnamefont
  {Suominen}}, \bibinfo {author} {\bibfnamefont {A.}~\bibnamefont {Fornieri}},
  \bibinfo {author} {\bibfnamefont {T.}~\bibnamefont {Wang}}, \bibinfo {author}
  {\bibfnamefont {G.~C.}\ \bibnamefont {Gardner}}, \bibinfo {author}
  {\bibfnamefont {C.}~\bibnamefont {Thomas}}, \bibinfo {author} {\bibfnamefont
  {A.~T.}\ \bibnamefont {Hatke}}, \bibinfo {author} {\bibfnamefont
  {P.}~\bibnamefont {Krogstrup}}, \bibinfo {author} {\bibfnamefont {M.~J.}\
  \bibnamefont {Manfra}}, \bibinfo {author} {\bibfnamefont {K.}~\bibnamefont
  {Flensberg}},\ and\ \bibinfo {author} {\bibfnamefont {C.~M.}\ \bibnamefont
  {Marcus}},\ }\bibfield  {title} {\bibinfo {title} {Scaling of {M}ajorana
  zero-bias conductance peaks},\ }\href
  {https://link.aps.org/doi/10.1103/PhysRevLett.119.136803} {\bibfield
  {journal} {\bibinfo  {journal} {Phys. Rev. Lett.}\ }\textbf {\bibinfo
  {volume} {119}},\ \bibinfo {pages} {136803} (\bibinfo {year}
  {2017})}\BibitemShut {NoStop}%
\bibitem [{\citenamefont {Vaitiek{\.e}nas}\ \emph {et~al.}(2021)\citenamefont
  {Vaitiek{\.e}nas}, \citenamefont {Liu}, \citenamefont {Krogstrup},\ and\
  \citenamefont {Marcus}}]{vaitiekenas2021zero}%
  \BibitemOpen
  \bibfield  {author} {\bibinfo {author} {\bibfnamefont {S.}~\bibnamefont
  {Vaitiek{\.e}nas}}, \bibinfo {author} {\bibfnamefont {Y.}~\bibnamefont
  {Liu}}, \bibinfo {author} {\bibfnamefont {P.}~\bibnamefont {Krogstrup}},\
  and\ \bibinfo {author} {\bibfnamefont {C.}~\bibnamefont {Marcus}},\
  }\bibfield  {title} {\bibinfo {title} {Zero-bias peaks at zero magnetic field
  in ferromagnetic hybrid nanowires},\ }\href@noop {} {\bibfield  {journal}
  {\bibinfo  {journal} {Nat. Phys.}\ }\textbf {\bibinfo {volume} {17}},\
  \bibinfo {pages} {43} (\bibinfo {year} {2021})}\BibitemShut {NoStop}%
\bibitem [{\citenamefont {Dvir}\ \emph {et~al.}(2022)\citenamefont {Dvir},
  \citenamefont {Wang}, \citenamefont {van Loo}, \citenamefont {Liu},
  \citenamefont {Mazur}, \citenamefont {Bordin}, \citenamefont {Haaf},
  \citenamefont {Wang}, \citenamefont {van Driel}, \citenamefont {Zatelli},
  \citenamefont {Li}, \citenamefont {Malinowski}, \citenamefont {Gazibegovic},
  \citenamefont {Badawy}, \citenamefont {Bakkers}, \citenamefont {Wimmer},\
  and\ \citenamefont {Kouwenhoven}}]{Dvir2022Realization}%
  \BibitemOpen
  \bibfield  {author} {\bibinfo {author} {\bibfnamefont {T.}~\bibnamefont
  {Dvir}}, \bibinfo {author} {\bibfnamefont {G.}~\bibnamefont {Wang}}, \bibinfo
  {author} {\bibfnamefont {N.}~\bibnamefont {van Loo}}, \bibinfo {author}
  {\bibfnamefont {C.-X.}\ \bibnamefont {Liu}}, \bibinfo {author} {\bibfnamefont
  {G.~P.}\ \bibnamefont {Mazur}}, \bibinfo {author} {\bibfnamefont
  {A.}~\bibnamefont {Bordin}}, \bibinfo {author} {\bibfnamefont {S.~L. D.~t.}\
  \bibnamefont {Haaf}}, \bibinfo {author} {\bibfnamefont {J.-Y.}\ \bibnamefont
  {Wang}}, \bibinfo {author} {\bibfnamefont {D.}~\bibnamefont {van Driel}},
  \bibinfo {author} {\bibfnamefont {F.}~\bibnamefont {Zatelli}}, \bibinfo
  {author} {\bibfnamefont {X.}~\bibnamefont {Li}}, \bibinfo {author}
  {\bibfnamefont {F.~K.}\ \bibnamefont {Malinowski}}, \bibinfo {author}
  {\bibfnamefont {S.}~\bibnamefont {Gazibegovic}}, \bibinfo {author}
  {\bibfnamefont {G.}~\bibnamefont {Badawy}}, \bibinfo {author} {\bibfnamefont
  {E.~P. A.~M.}\ \bibnamefont {Bakkers}}, \bibinfo {author} {\bibfnamefont
  {M.}~\bibnamefont {Wimmer}},\ and\ \bibinfo {author} {\bibfnamefont {L.~P.}\
  \bibnamefont {Kouwenhoven}},\ }\bibfield  {title} {\bibinfo {title}
  {Realization of a minimal {K}itaev chain in coupled quantum dots},\ }\href
  {https://arxiv.org/abs/2206.08045} {\bibfield  {journal} {\bibinfo  {journal}
  {arXiv:2206.08045}\ } (\bibinfo {year} {2022})}\BibitemShut {NoStop}%
\bibitem [{\citenamefont {Prada}\ \emph {et~al.}(2012)\citenamefont {Prada},
  \citenamefont {San-Jose},\ and\ \citenamefont {Aguado}}]{PhysRevB.86.180503}%
  \BibitemOpen
  \bibfield  {author} {\bibinfo {author} {\bibfnamefont {E.}~\bibnamefont
  {Prada}}, \bibinfo {author} {\bibfnamefont {P.}~\bibnamefont {San-Jose}},\
  and\ \bibinfo {author} {\bibfnamefont {R.}~\bibnamefont {Aguado}},\
  }\bibfield  {title} {\bibinfo {title} {Transport spectroscopy of {NS}
  nanowire junctions with {M}ajorana fermions},\ }\href
  {https://doi.org/10.1103/PhysRevB.86.180503} {\bibfield  {journal} {\bibinfo
  {journal} {Phys. Rev. B}\ }\textbf {\bibinfo {volume} {86}},\ \bibinfo
  {pages} {180503} (\bibinfo {year} {2012})}\BibitemShut {NoStop}%
\bibitem [{\citenamefont {Liu}\ \emph {et~al.}(2012)\citenamefont {Liu},
  \citenamefont {Potter}, \citenamefont {Law},\ and\ \citenamefont
  {Lee}}]{Liu2012Zero}%
  \BibitemOpen
  \bibfield  {author} {\bibinfo {author} {\bibfnamefont {J.}~\bibnamefont
  {Liu}}, \bibinfo {author} {\bibfnamefont {A.~C.}\ \bibnamefont {Potter}},
  \bibinfo {author} {\bibfnamefont {K.~T.}\ \bibnamefont {Law}},\ and\ \bibinfo
  {author} {\bibfnamefont {P.~A.}\ \bibnamefont {Lee}},\ }\bibfield  {title}
  {\bibinfo {title} {Zero-bias peaks in the tunneling conductance of
  spin-orbit-coupled superconducting wires with and without {M}ajorana
  end-states},\ }\href {https://doi.org/10.1103/PhysRevLett.109.267002}
  {\bibfield  {journal} {\bibinfo  {journal} {Phys. Rev. Lett.}\ }\textbf
  {\bibinfo {volume} {109}},\ \bibinfo {pages} {267002} (\bibinfo {year}
  {2012})}\BibitemShut {NoStop}%
\bibitem [{\citenamefont {Liu}\ \emph {et~al.}(2017)\citenamefont {Liu},
  \citenamefont {Sau}, \citenamefont {Stanescu},\ and\ \citenamefont
  {Das~Sarma}}]{Liu2017QDot}%
  \BibitemOpen
  \bibfield  {author} {\bibinfo {author} {\bibfnamefont {C.-X.}\ \bibnamefont
  {Liu}}, \bibinfo {author} {\bibfnamefont {J.~D.}\ \bibnamefont {Sau}},
  \bibinfo {author} {\bibfnamefont {T.~D.}\ \bibnamefont {Stanescu}},\ and\
  \bibinfo {author} {\bibfnamefont {S.}~\bibnamefont {Das~Sarma}},\ }\bibfield
  {title} {\bibinfo {title} {Andreev bound states versus {M}ajorana bound
  states in quantum dot-nanowire-superconductor hybrid structures: {T}rivial
  versus topological zero-bias conductance peaks},\ }\href
  {https://doi.org/10.1103/PhysRevB.96.075161} {\bibfield  {journal} {\bibinfo
  {journal} {Phys. Rev. B}\ }\textbf {\bibinfo {volume} {96}},\ \bibinfo
  {pages} {075161} (\bibinfo {year} {2017})}\BibitemShut {NoStop}%
\bibitem [{\citenamefont {Moore}\ \emph {et~al.}(2018)\citenamefont {Moore},
  \citenamefont {Zeng}, \citenamefont {Stanescu},\ and\ \citenamefont
  {Tewari}}]{PhysRevB.98.155314}%
  \BibitemOpen
  \bibfield  {author} {\bibinfo {author} {\bibfnamefont {C.}~\bibnamefont
  {Moore}}, \bibinfo {author} {\bibfnamefont {C.}~\bibnamefont {Zeng}},
  \bibinfo {author} {\bibfnamefont {T.~D.}\ \bibnamefont {Stanescu}},\ and\
  \bibinfo {author} {\bibfnamefont {S.}~\bibnamefont {Tewari}},\ }\bibfield
  {title} {\bibinfo {title} {Quantized zero-bias conductance plateau in
  semiconductor-superconductor heterostructures without topological majorana
  zero modes},\ }\href {https://doi.org/10.1103/PhysRevB.98.155314} {\bibfield
  {journal} {\bibinfo  {journal} {Phys. Rev. B}\ }\textbf {\bibinfo {volume}
  {98}},\ \bibinfo {pages} {155314} (\bibinfo {year} {2018})}\BibitemShut
  {NoStop}%
\bibitem [{\citenamefont {Chen}\ \emph {et~al.}(2019)\citenamefont {Chen},
  \citenamefont {Woods}, \citenamefont {Yu}, \citenamefont {Hocevar},
  \citenamefont {Car}, \citenamefont {Plissard}, \citenamefont {Bakkers},
  \citenamefont {Stanescu},\ and\ \citenamefont {Frolov}}]{Chen2019Ubiquitous}%
  \BibitemOpen
  \bibfield  {author} {\bibinfo {author} {\bibfnamefont {J.}~\bibnamefont
  {Chen}}, \bibinfo {author} {\bibfnamefont {B.~D.}\ \bibnamefont {Woods}},
  \bibinfo {author} {\bibfnamefont {P.}~\bibnamefont {Yu}}, \bibinfo {author}
  {\bibfnamefont {M.}~\bibnamefont {Hocevar}}, \bibinfo {author} {\bibfnamefont
  {D.}~\bibnamefont {Car}}, \bibinfo {author} {\bibfnamefont {S.~R.}\
  \bibnamefont {Plissard}}, \bibinfo {author} {\bibfnamefont {E.~P. A.~M.}\
  \bibnamefont {Bakkers}}, \bibinfo {author} {\bibfnamefont {T.~D.}\
  \bibnamefont {Stanescu}},\ and\ \bibinfo {author} {\bibfnamefont {S.~M.}\
  \bibnamefont {Frolov}},\ }\bibfield  {title} {\bibinfo {title} {Ubiquitous
  non-{M}ajorana zero-bias conductance peaks in nanowire devices},\ }\href
  {https://doi.org/10.1103/PhysRevLett.123.107703} {\bibfield  {journal}
  {\bibinfo  {journal} {Phys. Rev. Lett.}\ }\textbf {\bibinfo {volume} {123}},\
  \bibinfo {pages} {107703} (\bibinfo {year} {2019})}\BibitemShut {NoStop}%
\bibitem [{\citenamefont {Dvir}\ \emph {et~al.}(2019)\citenamefont {Dvir},
  \citenamefont {Aprili}, \citenamefont {Quay},\ and\ \citenamefont
  {Steinberg}}]{PhysRevLett.123.217003}%
  \BibitemOpen
  \bibfield  {author} {\bibinfo {author} {\bibfnamefont {T.}~\bibnamefont
  {Dvir}}, \bibinfo {author} {\bibfnamefont {M.}~\bibnamefont {Aprili}},
  \bibinfo {author} {\bibfnamefont {C.~H.~L.}\ \bibnamefont {Quay}},\ and\
  \bibinfo {author} {\bibfnamefont {H.}~\bibnamefont {Steinberg}},\ }\bibfield
  {title} {\bibinfo {title} {Zeeman tunability of {A}ndreev bound states in van
  der {W}aals tunnel barriers},\ }\href
  {https://doi.org/10.1103/PhysRevLett.123.217003} {\bibfield  {journal}
  {\bibinfo  {journal} {Phys. Rev. Lett.}\ }\textbf {\bibinfo {volume} {123}},\
  \bibinfo {pages} {217003} (\bibinfo {year} {2019})}\BibitemShut {NoStop}%
\bibitem [{\citenamefont {Vuik}\ \emph {et~al.}(2019)\citenamefont {Vuik},
  \citenamefont {Nijholt}, \citenamefont {Akhmerov},\ and\ \citenamefont
  {Wimmer}}]{10.21468/SciPostPhys.7.5.061}%
  \BibitemOpen
  \bibfield  {author} {\bibinfo {author} {\bibfnamefont {A.}~\bibnamefont
  {Vuik}}, \bibinfo {author} {\bibfnamefont {B.}~\bibnamefont {Nijholt}},
  \bibinfo {author} {\bibfnamefont {A.~R.}\ \bibnamefont {Akhmerov}},\ and\
  \bibinfo {author} {\bibfnamefont {M.}~\bibnamefont {Wimmer}},\ }\bibfield
  {title} {\bibinfo {title} {{Reproducing topological properties with
  quasi-{M}ajorana states}},\ }\href
  {https://doi.org/10.21468/SciPostPhys.7.5.061} {\bibfield  {journal}
  {\bibinfo  {journal} {SciPost Phys.}\ }\textbf {\bibinfo {volume} {7}},\
  \bibinfo {pages} {61} (\bibinfo {year} {2019})}\BibitemShut {NoStop}%
\bibitem [{\citenamefont {Zhang}\ \emph {et~al.}(2020)\citenamefont {Zhang},
  \citenamefont {Guo},\ and\ \citenamefont {Liu}}]{Zhang2020Transport}%
  \BibitemOpen
  \bibfield  {author} {\bibinfo {author} {\bibfnamefont {Y.}~\bibnamefont
  {Zhang}}, \bibinfo {author} {\bibfnamefont {K.}~\bibnamefont {Guo}},\ and\
  \bibinfo {author} {\bibfnamefont {J.}~\bibnamefont {Liu}},\ }\bibfield
  {title} {\bibinfo {title} {Transport characterization of topological
  superconductivity in a planar {J}osephson junction},\ }\href
  {https://doi.org/10.1103/PhysRevB.102.245403} {\bibfield  {journal} {\bibinfo
   {journal} {Phys. Rev. B}\ }\textbf {\bibinfo {volume} {102}},\ \bibinfo
  {pages} {245403} (\bibinfo {year} {2020})}\BibitemShut {NoStop}%
\bibitem [{\citenamefont {Valentini}\ \emph {et~al.}(2021)\citenamefont
  {Valentini}, \citenamefont {Pe\~naranda}, \citenamefont {Hofmann},
  \citenamefont {Brauns}, \citenamefont {Hauschild}, \citenamefont {Krogstrup},
  \citenamefont {San-Jose}, \citenamefont {Prada}, \citenamefont {Aguado},\
  and\ \citenamefont {Katsaros}}]{valentini2020nontopological}%
  \BibitemOpen
  \bibfield  {author} {\bibinfo {author} {\bibfnamefont {M.}~\bibnamefont
  {Valentini}}, \bibinfo {author} {\bibfnamefont {F.}~\bibnamefont
  {Pe\~naranda}}, \bibinfo {author} {\bibfnamefont {A.}~\bibnamefont
  {Hofmann}}, \bibinfo {author} {\bibfnamefont {M.}~\bibnamefont {Brauns}},
  \bibinfo {author} {\bibfnamefont {R.}~\bibnamefont {Hauschild}}, \bibinfo
  {author} {\bibfnamefont {P.}~\bibnamefont {Krogstrup}}, \bibinfo {author}
  {\bibfnamefont {P.}~\bibnamefont {San-Jose}}, \bibinfo {author}
  {\bibfnamefont {E.}~\bibnamefont {Prada}}, \bibinfo {author} {\bibfnamefont
  {R.}~\bibnamefont {Aguado}},\ and\ \bibinfo {author} {\bibfnamefont
  {G.}~\bibnamefont {Katsaros}},\ }\bibfield  {title} {\bibinfo {title}
  {Nontopological zero-bias peaks in full-shell nanowires induced by
  flux-tunable andreev states},\ }\href
  {https://doi.org/10.1126/science.abf1513} {\bibfield  {journal} {\bibinfo
  {journal} {Science}\ }\textbf {\bibinfo {volume} {373}},\ \bibinfo {pages}
  {82–88} (\bibinfo {year} {2021})}\BibitemShut {NoStop}%
\bibitem [{\citenamefont {J\"{u}nger}\ \emph {et~al.}(2020)\citenamefont
  {J\"{u}nger}, \citenamefont {Delagrange}, \citenamefont {Chevallier},
  \citenamefont {Lehmann}, \citenamefont {Dick}, \citenamefont {Thelander},
  \citenamefont {Klinovaja}, \citenamefont {Loss}, \citenamefont
  {Baumgartner},\ and\ \citenamefont
  {Sch\"{o}nenberger}}]{PhysRevLett.125.017701}%
  \BibitemOpen
  \bibfield  {author} {\bibinfo {author} {\bibfnamefont {C.}~\bibnamefont
  {J\"{u}nger}}, \bibinfo {author} {\bibfnamefont {R.}~\bibnamefont
  {Delagrange}}, \bibinfo {author} {\bibfnamefont {D.}~\bibnamefont
  {Chevallier}}, \bibinfo {author} {\bibfnamefont {S.}~\bibnamefont {Lehmann}},
  \bibinfo {author} {\bibfnamefont {K.~A.}\ \bibnamefont {Dick}}, \bibinfo
  {author} {\bibfnamefont {C.}~\bibnamefont {Thelander}}, \bibinfo {author}
  {\bibfnamefont {J.}~\bibnamefont {Klinovaja}}, \bibinfo {author}
  {\bibfnamefont {D.}~\bibnamefont {Loss}}, \bibinfo {author} {\bibfnamefont
  {A.}~\bibnamefont {Baumgartner}},\ and\ \bibinfo {author} {\bibfnamefont
  {C.}~\bibnamefont {Sch\"{o}nenberger}},\ }\bibfield  {title} {\bibinfo
  {title} {Magnetic-field-independent subgap states in hybrid {R}ashba
  nanowires},\ }\href {https://doi.org/10.1103/PhysRevLett.125.017701}
  {\bibfield  {journal} {\bibinfo  {journal} {Phys. Rev. Lett.}\ }\textbf
  {\bibinfo {volume} {125}},\ \bibinfo {pages} {017701} (\bibinfo {year}
  {2020})}\BibitemShut {NoStop}%
\bibitem [{\citenamefont {Razmadze}\ \emph {et~al.}(2020)\citenamefont
  {Razmadze}, \citenamefont {O'Farrell}, \citenamefont {Krogstrup},\ and\
  \citenamefont {Marcus}}]{PhysRevLett.125.116803}%
  \BibitemOpen
  \bibfield  {author} {\bibinfo {author} {\bibfnamefont {D.}~\bibnamefont
  {Razmadze}}, \bibinfo {author} {\bibfnamefont {E.~C.~T.}\ \bibnamefont
  {O'Farrell}}, \bibinfo {author} {\bibfnamefont {P.}~\bibnamefont
  {Krogstrup}},\ and\ \bibinfo {author} {\bibfnamefont {C.~M.}\ \bibnamefont
  {Marcus}},\ }\bibfield  {title} {\bibinfo {title} {Quantum dot parity effects
  in trivial and topological {J}osephson junctions},\ }\href
  {https://doi.org/10.1103/PhysRevLett.125.116803} {\bibfield  {journal}
  {\bibinfo  {journal} {Phys. Rev. Lett.}\ }\textbf {\bibinfo {volume} {125}},\
  \bibinfo {pages} {116803} (\bibinfo {year} {2020})}\BibitemShut {NoStop}%
\bibitem [{\citenamefont {Cayao}\ and\ \citenamefont
  {Burset}(2021)}]{PhysRevB.104.134507}%
  \BibitemOpen
  \bibfield  {author} {\bibinfo {author} {\bibfnamefont {J.}~\bibnamefont
  {Cayao}}\ and\ \bibinfo {author} {\bibfnamefont {P.}~\bibnamefont {Burset}},\
  }\bibfield  {title} {\bibinfo {title} {Confinement-induced zero-bias peaks in
  conventional superconductor hybrids},\ }\href
  {https://doi.org/10.1103/PhysRevB.104.134507} {\bibfield  {journal} {\bibinfo
   {journal} {Phys. Rev. B}\ }\textbf {\bibinfo {volume} {104}},\ \bibinfo
  {pages} {134507} (\bibinfo {year} {2021})}\BibitemShut {NoStop}%
\bibitem [{\citenamefont {Yu}\ \emph {et~al.}(2021)\citenamefont {Yu},
  \citenamefont {Chen}, \citenamefont {Gomanko}, \citenamefont {Badawy},
  \citenamefont {Bakkers}, \citenamefont {Zuo}, \citenamefont {Mourik},\ and\
  \citenamefont {Frolov}}]{yu2020non}%
  \BibitemOpen
  \bibfield  {author} {\bibinfo {author} {\bibfnamefont {P.}~\bibnamefont
  {Yu}}, \bibinfo {author} {\bibfnamefont {J.}~\bibnamefont {Chen}}, \bibinfo
  {author} {\bibfnamefont {M.}~\bibnamefont {Gomanko}}, \bibinfo {author}
  {\bibfnamefont {G.}~\bibnamefont {Badawy}}, \bibinfo {author} {\bibfnamefont
  {E.~P. A.~M.}\ \bibnamefont {Bakkers}}, \bibinfo {author} {\bibfnamefont
  {K.}~\bibnamefont {Zuo}}, \bibinfo {author} {\bibfnamefont {V.}~\bibnamefont
  {Mourik}},\ and\ \bibinfo {author} {\bibfnamefont {S.~M.}\ \bibnamefont
  {Frolov}},\ }\bibfield  {title} {\bibinfo {title} {Non-{M}ajorana states
  yield nearly quantized conductance in superconductor-semiconductor nanowire
  devices},\ }\href {https://doi.org/10.1038/s41567-020-01107-w} {\bibfield
  {journal} {\bibinfo  {journal} {Nat. Phys.}\ }\textbf {\bibinfo {volume}
  {17}},\ \bibinfo {pages} {482} (\bibinfo {year} {2021})}\BibitemShut
  {NoStop}%
\bibitem [{\citenamefont {Marra}\ and\ \citenamefont
  {Nigro}(2022)}]{marra2022majorana}%
  \BibitemOpen
  \bibfield  {author} {\bibinfo {author} {\bibfnamefont {P.}~\bibnamefont
  {Marra}}\ and\ \bibinfo {author} {\bibfnamefont {A.}~\bibnamefont {Nigro}},\
  }\bibfield  {title} {\bibinfo {title} {Majorana/andreev crossover and the
  fate of the topological phase transition in inhomogeneous nanowires},\
  }\href@noop {} {\bibfield  {journal} {\bibinfo  {journal} {Journal of
  Physics: Condensed Matter}\ }\textbf {\bibinfo {volume} {34}},\ \bibinfo
  {pages} {124001} (\bibinfo {year} {2022})}\BibitemShut {NoStop}%
\bibitem [{\citenamefont {Hess}\ \emph {et~al.}(2022)\citenamefont {Hess},
  \citenamefont {Legg}, \citenamefont {Loss},\ and\ \citenamefont
  {Klinovaja}}]{Hess2022trivial}%
  \BibitemOpen
  \bibfield  {author} {\bibinfo {author} {\bibfnamefont {R.}~\bibnamefont
  {Hess}}, \bibinfo {author} {\bibfnamefont {H.~F.}\ \bibnamefont {Legg}},
  \bibinfo {author} {\bibfnamefont {D.}~\bibnamefont {Loss}},\ and\ \bibinfo
  {author} {\bibfnamefont {J.}~\bibnamefont {Klinovaja}},\ }\bibfield  {title}
  {\bibinfo {title} {Trivial andreev band mimicking topological bulk gap
  reopening in the non-local conductance of long rashba nanowires},\ }\href
  {https://arxiv.org/abs/2210.03507} {\bibfield  {journal} {\bibinfo  {journal}
  {arXiv:2210:03507}\ } (\bibinfo {year} {2022})}\BibitemShut {NoStop}%
\bibitem [{\citenamefont {Chen}\ \emph {et~al.}(2022)\citenamefont {Chen},
  \citenamefont {Luo}, \citenamefont {Fang}, \citenamefont {Paltiel},
  \citenamefont {Millo}, \citenamefont {Guo},\ and\ \citenamefont
  {Sun}}]{chen2022topologically}%
  \BibitemOpen
  \bibfield  {author} {\bibinfo {author} {\bibfnamefont {X.-F.}\ \bibnamefont
  {Chen}}, \bibinfo {author} {\bibfnamefont {W.}~\bibnamefont {Luo}}, \bibinfo
  {author} {\bibfnamefont {T.-F.}\ \bibnamefont {Fang}}, \bibinfo {author}
  {\bibfnamefont {Y.}~\bibnamefont {Paltiel}}, \bibinfo {author} {\bibfnamefont
  {O.}~\bibnamefont {Millo}}, \bibinfo {author} {\bibfnamefont {A.-M.}\
  \bibnamefont {Guo}},\ and\ \bibinfo {author} {\bibfnamefont {Q.-F.}\
  \bibnamefont {Sun}},\ }\bibfield  {title} {\bibinfo {title} {Topologically
  nontrivial and trivial zero modes in chiral molecules},\ }\href@noop {}
  {\bibfield  {journal} {\bibinfo  {journal} {arXiv preprint arXiv:2208.13352}\
  } (\bibinfo {year} {2022})}\BibitemShut {NoStop}%
\bibitem [{\citenamefont {Sahu}\ \emph {et~al.}(2022)\citenamefont {Sahu},
  \citenamefont {Khade},\ and\ \citenamefont {Gangadharaiah}}]{sahu2022effect}%
  \BibitemOpen
  \bibfield  {author} {\bibinfo {author} {\bibfnamefont {D.}~\bibnamefont
  {Sahu}}, \bibinfo {author} {\bibfnamefont {V.}~\bibnamefont {Khade}},\ and\
  \bibinfo {author} {\bibfnamefont {S.}~\bibnamefont {Gangadharaiah}},\
  }\bibfield  {title} {\bibinfo {title} {Effect of topological length on bound
  states signatures in a topological nanowire},\ }\href@noop {} {\bibfield
  {journal} {\bibinfo  {journal} {arXiv preprint arXiv:2211.03045}\ } (\bibinfo
  {year} {2022})}\BibitemShut {NoStop}%
\bibitem [{\citenamefont {Kells}\ \emph {et~al.}(2012)\citenamefont {Kells},
  \citenamefont {Meidan},\ and\ \citenamefont {Brouwer}}]{kells2021Near}%
  \BibitemOpen
  \bibfield  {author} {\bibinfo {author} {\bibfnamefont {G.}~\bibnamefont
  {Kells}}, \bibinfo {author} {\bibfnamefont {D.}~\bibnamefont {Meidan}},\ and\
  \bibinfo {author} {\bibfnamefont {P.~W.}\ \bibnamefont {Brouwer}},\
  }\bibfield  {title} {\bibinfo {title} {Near-zero-energy end states in
  topologically trivial spin-orbit coupled superconducting nanowires with a
  smooth confinement},\ }\href {https://doi.org/10.1103/PhysRevB.86.100503}
  {\bibfield  {journal} {\bibinfo  {journal} {Phys. Rev. B}\ }\textbf {\bibinfo
  {volume} {86}},\ \bibinfo {pages} {100503} (\bibinfo {year}
  {2012})}\BibitemShut {NoStop}%
\bibitem [{\citenamefont {Cayao}\ \emph {et~al.}(2015)\citenamefont {Cayao},
  \citenamefont {Prada}, \citenamefont {San-Jose},\ and\ \citenamefont
  {Aguado}}]{PhysRevB.91.024514}%
  \BibitemOpen
  \bibfield  {author} {\bibinfo {author} {\bibfnamefont {J.}~\bibnamefont
  {Cayao}}, \bibinfo {author} {\bibfnamefont {E.}~\bibnamefont {Prada}},
  \bibinfo {author} {\bibfnamefont {P.}~\bibnamefont {San-Jose}},\ and\
  \bibinfo {author} {\bibfnamefont {R.}~\bibnamefont {Aguado}},\ }\bibfield
  {title} {\bibinfo {title} {{SNS} junctions in nanowires with spin-orbit
  coupling: {R}ole of confinement and helicity on the subgap spectrum},\ }\href
  {https://doi.org/10.1103/PhysRevB.91.024514} {\bibfield  {journal} {\bibinfo
  {journal} {Phys. Rev. B}\ }\textbf {\bibinfo {volume} {91}},\ \bibinfo
  {pages} {024514} (\bibinfo {year} {2015})}\BibitemShut {NoStop}%
\bibitem [{\citenamefont {de~Moor}\ \emph {et~al.}(2018)\citenamefont
  {de~Moor}, \citenamefont {Bommer}, \citenamefont {Xu}, \citenamefont
  {Winkler}, \citenamefont {Antipov}, \citenamefont {Bargerbos}, \citenamefont
  {Wang}, \citenamefont {van Loo}, \citenamefont {het Veld}, \citenamefont
  {Gazibegovic}, \citenamefont {Car}, \citenamefont {Logan}, \citenamefont
  {Pendharkar}, \citenamefont {Lee}, \citenamefont {Bakkers}, \citenamefont
  {Palmstr{\o}m}, \citenamefont {Lutchyn}, \citenamefont {Kouwenhoven},\ and\
  \citenamefont {Zhang}}]{deMoor2018Electric}%
  \BibitemOpen
  \bibfield  {author} {\bibinfo {author} {\bibfnamefont {M.~W.~A.}\
  \bibnamefont {de~Moor}}, \bibinfo {author} {\bibfnamefont {J.~D.~S.}\
  \bibnamefont {Bommer}}, \bibinfo {author} {\bibfnamefont {D.}~\bibnamefont
  {Xu}}, \bibinfo {author} {\bibfnamefont {G.~W.}\ \bibnamefont {Winkler}},
  \bibinfo {author} {\bibfnamefont {A.~E.}\ \bibnamefont {Antipov}}, \bibinfo
  {author} {\bibfnamefont {A.}~\bibnamefont {Bargerbos}}, \bibinfo {author}
  {\bibfnamefont {G.}~\bibnamefont {Wang}}, \bibinfo {author} {\bibfnamefont
  {N.}~\bibnamefont {van Loo}}, \bibinfo {author} {\bibfnamefont {R.~L. M.~O.}\
  \bibnamefont {het Veld}}, \bibinfo {author} {\bibfnamefont {S.}~\bibnamefont
  {Gazibegovic}}, \bibinfo {author} {\bibfnamefont {D.}~\bibnamefont {Car}},
  \bibinfo {author} {\bibfnamefont {J.~A.}\ \bibnamefont {Logan}}, \bibinfo
  {author} {\bibfnamefont {M.}~\bibnamefont {Pendharkar}}, \bibinfo {author}
  {\bibfnamefont {J.~S.}\ \bibnamefont {Lee}}, \bibinfo {author} {\bibfnamefont
  {E.~P. A.~M.}\ \bibnamefont {Bakkers}}, \bibinfo {author} {\bibfnamefont
  {C.~J.}\ \bibnamefont {Palmstr{\o}m}}, \bibinfo {author} {\bibfnamefont
  {R.~M.}\ \bibnamefont {Lutchyn}}, \bibinfo {author} {\bibfnamefont {L.~P.}\
  \bibnamefont {Kouwenhoven}},\ and\ \bibinfo {author} {\bibfnamefont
  {H.}~\bibnamefont {Zhang}},\ }\bibfield  {title} {\bibinfo {title} {Electric
  field tunable superconductor-semiconductor coupling in {M}ajorana
  nanowires},\ }\href {https://doi.org/10.1088/1367-2630/aae61d} {\bibfield
  {journal} {\bibinfo  {journal} {New J. Phys.}\ }\textbf {\bibinfo {volume}
  {20}},\ \bibinfo {pages} {103049} (\bibinfo {year} {2018})}\BibitemShut
  {NoStop}%
\bibitem [{\citenamefont {Awoga}\ \emph {et~al.}(2019)\citenamefont {Awoga},
  \citenamefont {Cayao},\ and\ \citenamefont
  {Black-Schaffer}}]{Awoga2019Supercurrent}%
  \BibitemOpen
  \bibfield  {author} {\bibinfo {author} {\bibfnamefont {O.~A.}\ \bibnamefont
  {Awoga}}, \bibinfo {author} {\bibfnamefont {J.}~\bibnamefont {Cayao}},\ and\
  \bibinfo {author} {\bibfnamefont {A.~M.}\ \bibnamefont {Black-Schaffer}},\
  }\bibfield  {title} {\bibinfo {title} {Supercurrent detection of
  topologically trivial zero-energy states in nanowire junctions},\ }\href
  {https://doi.org/10.1103/PhysRevLett.123.117001} {\bibfield  {journal}
  {\bibinfo  {journal} {Phys. Rev. Lett.}\ }\textbf {\bibinfo {volume} {123}},\
  \bibinfo {pages} {117001} (\bibinfo {year} {2019})}\BibitemShut {NoStop}%
\bibitem [{\citenamefont {Reeg}\ \emph
  {et~al.}(2018{\natexlab{a}})\citenamefont {Reeg}, \citenamefont {Dmytruk},
  \citenamefont {Chevallier}, \citenamefont {Loss},\ and\ \citenamefont
  {Klinovaja}}]{Reeg2018Zero}%
  \BibitemOpen
  \bibfield  {author} {\bibinfo {author} {\bibfnamefont {C.}~\bibnamefont
  {Reeg}}, \bibinfo {author} {\bibfnamefont {O.}~\bibnamefont {Dmytruk}},
  \bibinfo {author} {\bibfnamefont {D.}~\bibnamefont {Chevallier}}, \bibinfo
  {author} {\bibfnamefont {D.}~\bibnamefont {Loss}},\ and\ \bibinfo {author}
  {\bibfnamefont {J.}~\bibnamefont {Klinovaja}},\ }\bibfield  {title} {\bibinfo
  {title} {Zero-energy {A}ndreev bound states from quantum dots in proximitized
  {R}ashba nanowires},\ }\href {https://doi.org/10.1103/PhysRevB.98.245407}
  {\bibfield  {journal} {\bibinfo  {journal} {Phys. Rev. B}\ }\textbf {\bibinfo
  {volume} {98}},\ \bibinfo {pages} {245407} (\bibinfo {year}
  {2018}{\natexlab{a}})}\BibitemShut {NoStop}%
\bibitem [{\citenamefont {Awoga}\ \emph {et~al.}(2022)\citenamefont {Awoga},
  \citenamefont {Cayao},\ and\ \citenamefont
  {Black-Schaffer}}]{Awoga2022Robust}%
  \BibitemOpen
  \bibfield  {author} {\bibinfo {author} {\bibfnamefont {O.~A.}\ \bibnamefont
  {Awoga}}, \bibinfo {author} {\bibfnamefont {J.}~\bibnamefont {Cayao}},\ and\
  \bibinfo {author} {\bibfnamefont {A.~M.}\ \bibnamefont {Black-Schaffer}},\
  }\bibfield  {title} {\bibinfo {title} {Robust topological superconductivity
  in weakly coupled nanowire-superconductor hybrid structures},\ }\href
  {https://doi.org/10.1103/PhysRevB.105.144509} {\bibfield  {journal} {\bibinfo
   {journal} {Phys. Rev. B}\ }\textbf {\bibinfo {volume} {105}},\ \bibinfo
  {pages} {144509} (\bibinfo {year} {2022})}\BibitemShut {NoStop}%
\bibitem [{\citenamefont {Escribano}\ \emph {et~al.}(2022)\citenamefont
  {Escribano}, \citenamefont {Maiani}, \citenamefont {Leijnse}, \citenamefont
  {Flensberg}, \citenamefont {Oreg}, \citenamefont {Yeyati}, \citenamefont
  {Prada},\ and\ \citenamefont {Souto}}]{Escribano2022Proximity}%
  \BibitemOpen
  \bibfield  {author} {\bibinfo {author} {\bibfnamefont {S.~D.}\ \bibnamefont
  {Escribano}}, \bibinfo {author} {\bibfnamefont {A.}~\bibnamefont {Maiani}},
  \bibinfo {author} {\bibfnamefont {M.}~\bibnamefont {Leijnse}}, \bibinfo
  {author} {\bibfnamefont {K.}~\bibnamefont {Flensberg}}, \bibinfo {author}
  {\bibfnamefont {Y.}~\bibnamefont {Oreg}}, \bibinfo {author} {\bibfnamefont
  {A.~L.}\ \bibnamefont {Yeyati}}, \bibinfo {author} {\bibfnamefont
  {E.}~\bibnamefont {Prada}},\ and\ \bibinfo {author} {\bibfnamefont {R.~S.}\
  \bibnamefont {Souto}},\ }\bibfield  {title} {\bibinfo {title}
  {Semiconductor-ferromagnet-superconductor planar heterostructures for 1d
  topological superconductivity},\ }\href
  {https://doi.org/10.1038\%2Fs41535-022-00489-9} {\bibfield  {journal}
  {\bibinfo  {journal} {npj Quantum Mater.}\ }\textbf {\bibinfo {volume} {7}}
  (\bibinfo {year} {2022})}\BibitemShut {NoStop}%
\bibitem [{\citenamefont {Motrunich}\ \emph {et~al.}(2001)\citenamefont
  {Motrunich}, \citenamefont {Damle},\ and\ \citenamefont
  {Huse}}]{PhysRevB.63.224204}%
  \BibitemOpen
  \bibfield  {author} {\bibinfo {author} {\bibfnamefont {O.}~\bibnamefont
  {Motrunich}}, \bibinfo {author} {\bibfnamefont {K.}~\bibnamefont {Damle}},\
  and\ \bibinfo {author} {\bibfnamefont {D.~A.}\ \bibnamefont {Huse}},\
  }\bibfield  {title} {\bibinfo {title} {Griffiths effects and quantum critical
  points in dirty superconductors without spin-rotation invariance:
  One-dimensional examples},\ }\href
  {https://doi.org/10.1103/PhysRevB.63.224204} {\bibfield  {journal} {\bibinfo
  {journal} {Phys. Rev. B}\ }\textbf {\bibinfo {volume} {63}},\ \bibinfo
  {pages} {224204} (\bibinfo {year} {2001})}\BibitemShut {NoStop}%
\bibitem [{\citenamefont {Brouwer}\ \emph {et~al.}(2011)\citenamefont
  {Brouwer}, \citenamefont {Duckheim}, \citenamefont {Romito},\ and\
  \citenamefont {von Oppen}}]{PhysRevB.84.144526}%
  \BibitemOpen
  \bibfield  {author} {\bibinfo {author} {\bibfnamefont {P.~W.}\ \bibnamefont
  {Brouwer}}, \bibinfo {author} {\bibfnamefont {M.}~\bibnamefont {Duckheim}},
  \bibinfo {author} {\bibfnamefont {A.}~\bibnamefont {Romito}},\ and\ \bibinfo
  {author} {\bibfnamefont {F.}~\bibnamefont {von Oppen}},\ }\bibfield  {title}
  {\bibinfo {title} {Topological superconducting phases in disordered quantum
  wires with strong spin-orbit coupling},\ }\href
  {https://doi.org/10.1103/PhysRevB.84.144526} {\bibfield  {journal} {\bibinfo
  {journal} {Phys. Rev. B}\ }\textbf {\bibinfo {volume} {84}},\ \bibinfo
  {pages} {144526} (\bibinfo {year} {2011})}\BibitemShut {NoStop}%
\bibitem [{\citenamefont {Akhmerov}\ \emph {et~al.}(2011)\citenamefont
  {Akhmerov}, \citenamefont {Dahlhaus}, \citenamefont {Hassler}, \citenamefont
  {Wimmer},\ and\ \citenamefont {Beenakker}}]{Akhmerov2011Quantized}%
  \BibitemOpen
  \bibfield  {author} {\bibinfo {author} {\bibfnamefont {A.~R.}\ \bibnamefont
  {Akhmerov}}, \bibinfo {author} {\bibfnamefont {J.~P.}\ \bibnamefont
  {Dahlhaus}}, \bibinfo {author} {\bibfnamefont {F.}~\bibnamefont {Hassler}},
  \bibinfo {author} {\bibfnamefont {M.}~\bibnamefont {Wimmer}},\ and\ \bibinfo
  {author} {\bibfnamefont {C.~W.~J.}\ \bibnamefont {Beenakker}},\ }\bibfield
  {title} {\bibinfo {title} {Quantized conductance at the majorana phase
  transition in a disordered superconducting wire},\ }\href
  {https://doi.org/10.1103/PhysRevLett.106.057001} {\bibfield  {journal}
  {\bibinfo  {journal} {Phys. Rev. Lett.}\ }\textbf {\bibinfo {volume} {106}},\
  \bibinfo {pages} {057001} (\bibinfo {year} {2011})}\BibitemShut {NoStop}%
\bibitem [{\citenamefont {Potter}\ and\ \citenamefont
  {Lee}(2011)}]{Potter2011Enginering}%
  \BibitemOpen
  \bibfield  {author} {\bibinfo {author} {\bibfnamefont {A.~C.}\ \bibnamefont
  {Potter}}\ and\ \bibinfo {author} {\bibfnamefont {P.~A.}\ \bibnamefont
  {Lee}},\ }\bibfield  {title} {\bibinfo {title} {Engineering a $p+\mathit{ip}$
  superconductor: Comparison of topological insulator and {R}ashba
  spin-orbit-coupled materials},\ }\href
  {http://link.aps.org/doi/10.1103/PhysRevB.83.184520} {\bibfield  {journal}
  {\bibinfo  {journal} {Phys. Rev. B}\ }\textbf {\bibinfo {volume} {83}},\
  \bibinfo {pages} {184520} (\bibinfo {year} {2011})}\BibitemShut {NoStop}%
\bibitem [{\citenamefont {Bagrets}\ and\ \citenamefont
  {Altland}(2012)}]{Bagrets:PRL12}%
  \BibitemOpen
  \bibfield  {author} {\bibinfo {author} {\bibfnamefont {D.}~\bibnamefont
  {Bagrets}}\ and\ \bibinfo {author} {\bibfnamefont {A.}~\bibnamefont
  {Altland}},\ }\bibfield  {title} {\bibinfo {title} {Class {$D$} spectral peak
  in {M}ajorana quantum wires},\ }\href
  {https://link.aps.org/doi/10.1103/PhysRevLett.109.227005} {\bibfield
  {journal} {\bibinfo  {journal} {Phys. Rev. Lett.}\ }\textbf {\bibinfo
  {volume} {109}},\ \bibinfo {pages} {227005} (\bibinfo {year}
  {2012})}\BibitemShut {NoStop}%
\bibitem [{\citenamefont {Lutchyn}\ \emph {et~al.}(2012)\citenamefont
  {Lutchyn}, \citenamefont {Stanescu},\ and\ \citenamefont
  {Das~Sarma}}]{PhysRevB.85.140513}%
  \BibitemOpen
  \bibfield  {author} {\bibinfo {author} {\bibfnamefont {R.~M.}\ \bibnamefont
  {Lutchyn}}, \bibinfo {author} {\bibfnamefont {T.~D.}\ \bibnamefont
  {Stanescu}},\ and\ \bibinfo {author} {\bibfnamefont {S.}~\bibnamefont
  {Das~Sarma}},\ }\bibfield  {title} {\bibinfo {title} {Momentum relaxation in
  a semiconductor proximity-coupled to a disordered $s$-wave superconductor:
  Effect of scattering on topological superconductivity},\ }\href
  {https://doi.org/10.1103/PhysRevB.85.140513} {\bibfield  {journal} {\bibinfo
  {journal} {Phys. Rev. B}\ }\textbf {\bibinfo {volume} {85}},\ \bibinfo
  {pages} {140513} (\bibinfo {year} {2012})}\BibitemShut {NoStop}%
\bibitem [{\citenamefont {Pientka}\ \emph {et~al.}(2012)\citenamefont
  {Pientka}, \citenamefont {Kells}, \citenamefont {Romito}, \citenamefont
  {Brouwer},\ and\ \citenamefont {von Oppen}}]{Pientka2012Enhanced}%
  \BibitemOpen
  \bibfield  {author} {\bibinfo {author} {\bibfnamefont {F.}~\bibnamefont
  {Pientka}}, \bibinfo {author} {\bibfnamefont {G.}~\bibnamefont {Kells}},
  \bibinfo {author} {\bibfnamefont {A.}~\bibnamefont {Romito}}, \bibinfo
  {author} {\bibfnamefont {P.~W.}\ \bibnamefont {Brouwer}},\ and\ \bibinfo
  {author} {\bibfnamefont {F.}~\bibnamefont {von Oppen}},\ }\bibfield  {title}
  {\bibinfo {title} {Enhanced zero-bias {M}ajorana peak in the differential
  tunneling conductance of disordered multisubband quantum-wire/superconductor
  junctions},\ }\href {https://doi.org/10.1103/PhysRevLett.109.227006}
  {\bibfield  {journal} {\bibinfo  {journal} {Phys. Rev. Lett.}\ }\textbf
  {\bibinfo {volume} {109}},\ \bibinfo {pages} {227006} (\bibinfo {year}
  {2012})}\BibitemShut {NoStop}%
\bibitem [{\citenamefont {Pikulin}\ \emph {et~al.}(2012)\citenamefont
  {Pikulin}, \citenamefont {Dahlhaus}, \citenamefont {Wimmer}, \citenamefont
  {Schomerus},\ and\ \citenamefont {Beenakker}}]{Pikulin2012A}%
  \BibitemOpen
  \bibfield  {author} {\bibinfo {author} {\bibfnamefont {D.~I.}\ \bibnamefont
  {Pikulin}}, \bibinfo {author} {\bibfnamefont {J.~P.}\ \bibnamefont
  {Dahlhaus}}, \bibinfo {author} {\bibfnamefont {M.}~\bibnamefont {Wimmer}},
  \bibinfo {author} {\bibfnamefont {H.}~\bibnamefont {Schomerus}},\ and\
  \bibinfo {author} {\bibfnamefont {C.~W.~J.}\ \bibnamefont {Beenakker}},\
  }\bibfield  {title} {\bibinfo {title} {A zero-voltage conductance peak from
  weak antilocalization in a {M}ajorana nanowire},\ }\href
  {https://doi.org/10.1088/1367-2630/14/12/125011} {\bibfield  {journal}
  {\bibinfo  {journal} {New J. Phys.}\ }\textbf {\bibinfo {volume} {14}},\
  \bibinfo {pages} {125011} (\bibinfo {year} {2012})}\BibitemShut {NoStop}%
\bibitem [{\citenamefont {Sau}\ and\ \citenamefont {Das~Sarma}(2013)}]{Sau:13}%
  \BibitemOpen
  \bibfield  {author} {\bibinfo {author} {\bibfnamefont {J.~D.}\ \bibnamefont
  {Sau}}\ and\ \bibinfo {author} {\bibfnamefont {S.}~\bibnamefont
  {Das~Sarma}},\ }\bibfield  {title} {\bibinfo {title} {Density of states of
  disordered topological superconductor-semiconductor hybrid nanowires},\
  }\href {http://link.aps.org/doi/10.1103/PhysRevB.88.064506} {\bibfield
  {journal} {\bibinfo  {journal} {Phys. Rev. B}\ }\textbf {\bibinfo {volume}
  {88}},\ \bibinfo {pages} {064506} (\bibinfo {year} {2013})}\BibitemShut
  {NoStop}%
\bibitem [{\citenamefont {Cole}\ \emph {et~al.}(2016)\citenamefont {Cole},
  \citenamefont {Sau},\ and\ \citenamefont {Das~Sarma}}]{DasCole2016Proximity}%
  \BibitemOpen
  \bibfield  {author} {\bibinfo {author} {\bibfnamefont {W.~S.}\ \bibnamefont
  {Cole}}, \bibinfo {author} {\bibfnamefont {J.~D.}\ \bibnamefont {Sau}},\ and\
  \bibinfo {author} {\bibfnamefont {S.}~\bibnamefont {Das~Sarma}},\ }\bibfield
  {title} {\bibinfo {title} {Proximity effect and majorana bound states in
  clean semiconductor nanowires coupled to disordered superconductors},\ }\href
  {https://doi.org/10.1103/PhysRevB.94.140505} {\bibfield  {journal} {\bibinfo
  {journal} {Phys. Rev. B}\ }\textbf {\bibinfo {volume} {94}},\ \bibinfo
  {pages} {140505} (\bibinfo {year} {2016})}\BibitemShut {NoStop}%
\bibitem [{\citenamefont {Pan}\ and\ \citenamefont
  {Das~Sarma}(2020)}]{PhysRevResearch.2.013377}%
  \BibitemOpen
  \bibfield  {author} {\bibinfo {author} {\bibfnamefont {H.}~\bibnamefont
  {Pan}}\ and\ \bibinfo {author} {\bibfnamefont {S.}~\bibnamefont
  {Das~Sarma}},\ }\bibfield  {title} {\bibinfo {title} {Physical mechanisms for
  zero-bias conductance peaks in {M}ajorana nanowires},\ }\href
  {https://doi.org/10.1103/PhysRevResearch.2.013377} {\bibfield  {journal}
  {\bibinfo  {journal} {Phys. Rev. Res.}\ }\textbf {\bibinfo {volume} {2}},\
  \bibinfo {pages} {013377} (\bibinfo {year} {2020})}\BibitemShut {NoStop}%
\bibitem [{\citenamefont {Thamm}\ and\ \citenamefont
  {Rosenow}(2022)}]{Thamm2022Machine}%
  \BibitemOpen
  \bibfield  {author} {\bibinfo {author} {\bibfnamefont {M.}~\bibnamefont
  {Thamm}}\ and\ \bibinfo {author} {\bibfnamefont {B.}~\bibnamefont
  {Rosenow}},\ }\bibfield  {title} {\bibinfo {title} {Machine learning
  optimization of {M}ajorana hybrid nanowires},\ }\href
  {https://arxiv.org/abs/2208.02182} {\bibfield  {journal} {\bibinfo  {journal}
  {arXiv:2208.02182}\ } (\bibinfo {year} {2022})}\BibitemShut {NoStop}%
\bibitem [{\citenamefont {Zhang}\ \emph {et~al.}(2021)\citenamefont {Zhang},
  \citenamefont {de~Moor}, \citenamefont {Bommer}, \citenamefont {Xu},
  \citenamefont {Wang}, \citenamefont {van Loo}, \citenamefont {Liu},
  \citenamefont {Gazibegovic}, \citenamefont {Logan}, \citenamefont {Car},
  \citenamefont {Veld}, \citenamefont {van Veldhoven}, \citenamefont
  {Koelling}, \citenamefont {Verheijen}, \citenamefont {Pendharkar},
  \citenamefont {Pennachio}, \citenamefont {Shojaei}, \citenamefont {Lee},
  \citenamefont {Palmstrøm}, \citenamefont {Bakkers}, \citenamefont {Sarma},\
  and\ \citenamefont {Kouwenhoven}}]{Zhang2021large}%
  \BibitemOpen
  \bibfield  {author} {\bibinfo {author} {\bibfnamefont {H.}~\bibnamefont
  {Zhang}}, \bibinfo {author} {\bibfnamefont {M.~W.~A.}\ \bibnamefont
  {de~Moor}}, \bibinfo {author} {\bibfnamefont {J.~D.~S.}\ \bibnamefont
  {Bommer}}, \bibinfo {author} {\bibfnamefont {D.}~\bibnamefont {Xu}}, \bibinfo
  {author} {\bibfnamefont {G.}~\bibnamefont {Wang}}, \bibinfo {author}
  {\bibfnamefont {N.}~\bibnamefont {van Loo}}, \bibinfo {author} {\bibfnamefont
  {C.-X.}\ \bibnamefont {Liu}}, \bibinfo {author} {\bibfnamefont
  {S.}~\bibnamefont {Gazibegovic}}, \bibinfo {author} {\bibfnamefont {J.~A.}\
  \bibnamefont {Logan}}, \bibinfo {author} {\bibfnamefont {D.}~\bibnamefont
  {Car}}, \bibinfo {author} {\bibfnamefont {R.~L. M. O.~h.}\ \bibnamefont
  {Veld}}, \bibinfo {author} {\bibfnamefont {P.~J.}\ \bibnamefont {van
  Veldhoven}}, \bibinfo {author} {\bibfnamefont {S.}~\bibnamefont {Koelling}},
  \bibinfo {author} {\bibfnamefont {M.~A.}\ \bibnamefont {Verheijen}}, \bibinfo
  {author} {\bibfnamefont {M.}~\bibnamefont {Pendharkar}}, \bibinfo {author}
  {\bibfnamefont {D.~J.}\ \bibnamefont {Pennachio}}, \bibinfo {author}
  {\bibfnamefont {B.}~\bibnamefont {Shojaei}}, \bibinfo {author} {\bibfnamefont
  {J.~S.}\ \bibnamefont {Lee}}, \bibinfo {author} {\bibfnamefont {C.~J.}\
  \bibnamefont {Palmstrøm}}, \bibinfo {author} {\bibfnamefont {E.~P. A.~M.}\
  \bibnamefont {Bakkers}}, \bibinfo {author} {\bibfnamefont {S.~D.}\
  \bibnamefont {Sarma}},\ and\ \bibinfo {author} {\bibfnamefont {L.~P.}\
  \bibnamefont {Kouwenhoven}},\ }\bibfield  {title} {\bibinfo {title} {Large
  zero-bias peaks in insb-al hybrid semiconductor-superconductor nanowire
  devices},\ }\href {https://arxiv.org/abs/2101.11456} {\bibfield  {journal}
  {\bibinfo  {journal} {arXiv:2101.11456}\ } (\bibinfo {year}
  {2021})}\BibitemShut {NoStop}%
\bibitem [{\citenamefont {Ahn}\ \emph {et~al.}(2021)\citenamefont {Ahn},
  \citenamefont {Pan}, \citenamefont {Woods}, \citenamefont {Stanescu},\ and\
  \citenamefont {Das~Sarma}}]{Ahn2021Estimating}%
  \BibitemOpen
  \bibfield  {author} {\bibinfo {author} {\bibfnamefont {S.}~\bibnamefont
  {Ahn}}, \bibinfo {author} {\bibfnamefont {H.}~\bibnamefont {Pan}}, \bibinfo
  {author} {\bibfnamefont {B.}~\bibnamefont {Woods}}, \bibinfo {author}
  {\bibfnamefont {T.~D.}\ \bibnamefont {Stanescu}},\ and\ \bibinfo {author}
  {\bibfnamefont {S.}~\bibnamefont {Das~Sarma}},\ }\bibfield  {title} {\bibinfo
  {title} {Estimating disorder and its adverse effects in semiconductor
  majorana nanowires},\ }\href
  {https://doi.org/10.1103/PhysRevMaterials.5.124602} {\bibfield  {journal}
  {\bibinfo  {journal} {Phys. Rev. Mater.}\ }\textbf {\bibinfo {volume} {5}},\
  \bibinfo {pages} {124602} (\bibinfo {year} {2021})}\BibitemShut {NoStop}%
\bibitem [{\citenamefont {Das~Sarma}\ and\ \citenamefont
  {Pan}(2021)}]{DasSarma2021Disorder}%
  \BibitemOpen
  \bibfield  {author} {\bibinfo {author} {\bibfnamefont {S.}~\bibnamefont
  {Das~Sarma}}\ and\ \bibinfo {author} {\bibfnamefont {H.}~\bibnamefont
  {Pan}},\ }\bibfield  {title} {\bibinfo {title} {Disorder-induced zero-bias
  peaks in {M}ajorana nanowires},\ }\href
  {https://doi.org/10.1103/PhysRevB.103.195158} {\bibfield  {journal} {\bibinfo
   {journal} {Phys. Rev. B}\ }\textbf {\bibinfo {volume} {103}},\ \bibinfo
  {pages} {195158} (\bibinfo {year} {2021})}\BibitemShut {NoStop}%
\bibitem [{\citenamefont {Kuzmanovski}\ \emph {et~al.}(2020)\citenamefont
  {Kuzmanovski}, \citenamefont {Black-Schaffer},\ and\ \citenamefont
  {Cayao}}]{PhysRevB.101.094506}%
  \BibitemOpen
  \bibfield  {author} {\bibinfo {author} {\bibfnamefont {D.}~\bibnamefont
  {Kuzmanovski}}, \bibinfo {author} {\bibfnamefont {A.~M.}\ \bibnamefont
  {Black-Schaffer}},\ and\ \bibinfo {author} {\bibfnamefont {J.}~\bibnamefont
  {Cayao}},\ }\bibfield  {title} {\bibinfo {title} {Suppression of
  odd-frequency pairing by phase disorder in a nanowire coupled to majorana
  zero modes},\ }\href {https://doi.org/10.1103/PhysRevB.101.094506} {\bibfield
   {journal} {\bibinfo  {journal} {Phys. Rev. B}\ }\textbf {\bibinfo {volume}
  {101}},\ \bibinfo {pages} {094506} (\bibinfo {year} {2020})}\BibitemShut
  {NoStop}%
\bibitem [{\citenamefont {Aghaee}\ \emph {et~al.}(2022)\citenamefont {Aghaee},
  \citenamefont {Akkala}, \citenamefont {Alam}, \citenamefont {Ali},
  \citenamefont {Ramirez}, \citenamefont {Andrzejczuk}, \citenamefont
  {Antipov}, \citenamefont {Aseev}, \citenamefont {Astafev}, \citenamefont
  {Bauer}, \citenamefont {Becker}, \citenamefont {Boddapati}, \citenamefont
  {Boekhout}, \citenamefont {Bommer}, \citenamefont {Hansen}, \citenamefont
  {Bosma}, \citenamefont {Bourdet}, \citenamefont {Boutin}, \citenamefont
  {Caroff}, \citenamefont {Casparis}, \citenamefont {Cassidy}, \citenamefont
  {Christensen}, \citenamefont {Clay}, \citenamefont {Cole}, \citenamefont
  {Corsetti}, \citenamefont {Cui}, \citenamefont {Dalampiras}, \citenamefont
  {Dokania}, \citenamefont {de~Lange}, \citenamefont {de~Moor}, \citenamefont
  {Saldaña}, \citenamefont {Fallahi}, \citenamefont {Fathabad}, \citenamefont
  {Gamble}, \citenamefont {Gardner}, \citenamefont {Govender}, \citenamefont
  {Griggio}, \citenamefont {Grigoryan}, \citenamefont {Gronin}, \citenamefont
  {Gukelberger}, \citenamefont {Heedt}, \citenamefont {Zamorano}, \citenamefont
  {Ho}, \citenamefont {Holgaard}, \citenamefont {Nielsen}, \citenamefont
  {Ingerslev}, \citenamefont {Krogstrup}, \citenamefont {Johansson},
  \citenamefont {Jones}, \citenamefont {Kallaher}, \citenamefont {Karimi},
  \citenamefont {Karzig}, \citenamefont {King}, \citenamefont {Kloster},
  \citenamefont {Knapp}, \citenamefont {Kocon}, \citenamefont {Koski},
  \citenamefont {Kostamo}, \citenamefont {Kumar}, \citenamefont {Laeven},
  \citenamefont {Larsen}, \citenamefont {Li}, \citenamefont {Lindemann},
  \citenamefont {Love}, \citenamefont {Lutchyn}, \citenamefont {Manfra},
  \citenamefont {Memisevic}, \citenamefont {Nayak}, \citenamefont {Nijholt},
  \citenamefont {Madsen}, \citenamefont {Markussen}, \citenamefont {Martinez},
  \citenamefont {McNeil}, \citenamefont {Mullally}, \citenamefont {Nielsen},
  \citenamefont {Nurmohamed}, \citenamefont {O'Farrell}, \citenamefont {Otani},
  \citenamefont {Pauka}, \citenamefont {Petersson}, \citenamefont {Petit},
  \citenamefont {Pikulin}, \citenamefont {Preiss}, \citenamefont {Perez},
  \citenamefont {Rasmussen}, \citenamefont {Rajpalke}, \citenamefont
  {Razmadze}, \citenamefont {Reentila}, \citenamefont {Reilly}, \citenamefont
  {Rouse}, \citenamefont {Sadovskyy}, \citenamefont {Sainiemi}, \citenamefont
  {Schreppler}, \citenamefont {Sidorkin}, \citenamefont {Singh}, \citenamefont
  {Singh}, \citenamefont {Sinha}, \citenamefont {Sohr}, \citenamefont
  {Stankevič}, \citenamefont {Stek}, \citenamefont {Suominen}, \citenamefont
  {Suter}, \citenamefont {Svidenko}, \citenamefont {Teicher}, \citenamefont
  {Temuerhan}, \citenamefont {Thiyagarajah}, \citenamefont {Tholapi},
  \citenamefont {Thomas}, \citenamefont {Toomey}, \citenamefont {Upadhyay},
  \citenamefont {Urban}, \citenamefont {Vaitiekėnas}, \citenamefont
  {Van~Hoogdalem}, \citenamefont {Viazmitinov}, \citenamefont {Waddy},
  \citenamefont {Van~Woerkom}, \citenamefont {Vogel}, \citenamefont {Watson},
  \citenamefont {Weston}, \citenamefont {Winkler}, \citenamefont {Yang},
  \citenamefont {Yau}, \citenamefont {Yi}, \citenamefont {Yucelen},
  \citenamefont {Webster}, \citenamefont {Zeisel},\ and\ \citenamefont
  {Zhao}}]{aghaee2022inas1}%
  \BibitemOpen
  \bibfield  {author} {\bibinfo {author} {\bibfnamefont {M.}~\bibnamefont
  {Aghaee}}, \bibinfo {author} {\bibfnamefont {A.}~\bibnamefont {Akkala}},
  \bibinfo {author} {\bibfnamefont {Z.}~\bibnamefont {Alam}}, \bibinfo {author}
  {\bibfnamefont {R.}~\bibnamefont {Ali}}, \bibinfo {author} {\bibfnamefont
  {A.~A.}\ \bibnamefont {Ramirez}}, \bibinfo {author} {\bibfnamefont
  {M.}~\bibnamefont {Andrzejczuk}}, \bibinfo {author} {\bibfnamefont {A.~E.}\
  \bibnamefont {Antipov}}, \bibinfo {author} {\bibfnamefont {P.}~\bibnamefont
  {Aseev}}, \bibinfo {author} {\bibfnamefont {M.}~\bibnamefont {Astafev}},
  \bibinfo {author} {\bibfnamefont {B.}~\bibnamefont {Bauer}}, \bibinfo
  {author} {\bibfnamefont {J.}~\bibnamefont {Becker}}, \bibinfo {author}
  {\bibfnamefont {S.}~\bibnamefont {Boddapati}}, \bibinfo {author}
  {\bibfnamefont {F.}~\bibnamefont {Boekhout}}, \bibinfo {author}
  {\bibfnamefont {J.}~\bibnamefont {Bommer}}, \bibinfo {author} {\bibfnamefont
  {E.~B.}\ \bibnamefont {Hansen}}, \bibinfo {author} {\bibfnamefont
  {T.}~\bibnamefont {Bosma}}, \bibinfo {author} {\bibfnamefont
  {L.}~\bibnamefont {Bourdet}}, \bibinfo {author} {\bibfnamefont
  {S.}~\bibnamefont {Boutin}}, \bibinfo {author} {\bibfnamefont
  {P.}~\bibnamefont {Caroff}}, \bibinfo {author} {\bibfnamefont
  {L.}~\bibnamefont {Casparis}}, \bibinfo {author} {\bibfnamefont
  {M.}~\bibnamefont {Cassidy}}, \bibinfo {author} {\bibfnamefont {A.~W.}\
  \bibnamefont {Christensen}}, \bibinfo {author} {\bibfnamefont
  {N.}~\bibnamefont {Clay}}, \bibinfo {author} {\bibfnamefont {W.~S.}\
  \bibnamefont {Cole}}, \bibinfo {author} {\bibfnamefont {F.}~\bibnamefont
  {Corsetti}}, \bibinfo {author} {\bibfnamefont {A.}~\bibnamefont {Cui}},
  \bibinfo {author} {\bibfnamefont {P.}~\bibnamefont {Dalampiras}}, \bibinfo
  {author} {\bibfnamefont {A.}~\bibnamefont {Dokania}}, \bibinfo {author}
  {\bibfnamefont {G.}~\bibnamefont {de~Lange}}, \bibinfo {author}
  {\bibfnamefont {M.}~\bibnamefont {de~Moor}}, \bibinfo {author} {\bibfnamefont
  {J.~C.~E.}\ \bibnamefont {Saldaña}}, \bibinfo {author} {\bibfnamefont
  {S.}~\bibnamefont {Fallahi}}, \bibinfo {author} {\bibfnamefont {Z.~H.}\
  \bibnamefont {Fathabad}}, \bibinfo {author} {\bibfnamefont {J.}~\bibnamefont
  {Gamble}}, \bibinfo {author} {\bibfnamefont {G.}~\bibnamefont {Gardner}},
  \bibinfo {author} {\bibfnamefont {D.}~\bibnamefont {Govender}}, \bibinfo
  {author} {\bibfnamefont {F.}~\bibnamefont {Griggio}}, \bibinfo {author}
  {\bibfnamefont {R.}~\bibnamefont {Grigoryan}}, \bibinfo {author}
  {\bibfnamefont {S.}~\bibnamefont {Gronin}}, \bibinfo {author} {\bibfnamefont
  {J.}~\bibnamefont {Gukelberger}}, \bibinfo {author} {\bibfnamefont
  {S.}~\bibnamefont {Heedt}}, \bibinfo {author} {\bibfnamefont {J.~H.}\
  \bibnamefont {Zamorano}}, \bibinfo {author} {\bibfnamefont {S.}~\bibnamefont
  {Ho}}, \bibinfo {author} {\bibfnamefont {U.~L.}\ \bibnamefont {Holgaard}},
  \bibinfo {author} {\bibfnamefont {W.~H.~P.}\ \bibnamefont {Nielsen}},
  \bibinfo {author} {\bibfnamefont {H.}~\bibnamefont {Ingerslev}}, \bibinfo
  {author} {\bibfnamefont {P.~J.}\ \bibnamefont {Krogstrup}}, \bibinfo {author}
  {\bibfnamefont {L.}~\bibnamefont {Johansson}}, \bibinfo {author}
  {\bibfnamefont {J.}~\bibnamefont {Jones}}, \bibinfo {author} {\bibfnamefont
  {R.}~\bibnamefont {Kallaher}}, \bibinfo {author} {\bibfnamefont
  {F.}~\bibnamefont {Karimi}}, \bibinfo {author} {\bibfnamefont
  {T.}~\bibnamefont {Karzig}}, \bibinfo {author} {\bibfnamefont
  {C.}~\bibnamefont {King}}, \bibinfo {author} {\bibfnamefont {M.~E.}\
  \bibnamefont {Kloster}}, \bibinfo {author} {\bibfnamefont {C.}~\bibnamefont
  {Knapp}}, \bibinfo {author} {\bibfnamefont {D.}~\bibnamefont {Kocon}},
  \bibinfo {author} {\bibfnamefont {J.}~\bibnamefont {Koski}}, \bibinfo
  {author} {\bibfnamefont {P.}~\bibnamefont {Kostamo}}, \bibinfo {author}
  {\bibfnamefont {M.}~\bibnamefont {Kumar}}, \bibinfo {author} {\bibfnamefont
  {T.}~\bibnamefont {Laeven}}, \bibinfo {author} {\bibfnamefont
  {T.}~\bibnamefont {Larsen}}, \bibinfo {author} {\bibfnamefont
  {K.}~\bibnamefont {Li}}, \bibinfo {author} {\bibfnamefont {T.}~\bibnamefont
  {Lindemann}}, \bibinfo {author} {\bibfnamefont {J.}~\bibnamefont {Love}},
  \bibinfo {author} {\bibfnamefont {R.}~\bibnamefont {Lutchyn}}, \bibinfo
  {author} {\bibfnamefont {M.}~\bibnamefont {Manfra}}, \bibinfo {author}
  {\bibfnamefont {E.}~\bibnamefont {Memisevic}}, \bibinfo {author}
  {\bibfnamefont {C.}~\bibnamefont {Nayak}}, \bibinfo {author} {\bibfnamefont
  {B.}~\bibnamefont {Nijholt}}, \bibinfo {author} {\bibfnamefont {M.~H.}\
  \bibnamefont {Madsen}}, \bibinfo {author} {\bibfnamefont {S.}~\bibnamefont
  {Markussen}}, \bibinfo {author} {\bibfnamefont {E.}~\bibnamefont {Martinez}},
  \bibinfo {author} {\bibfnamefont {R.}~\bibnamefont {McNeil}}, \bibinfo
  {author} {\bibfnamefont {A.}~\bibnamefont {Mullally}}, \bibinfo {author}
  {\bibfnamefont {J.}~\bibnamefont {Nielsen}}, \bibinfo {author} {\bibfnamefont
  {A.}~\bibnamefont {Nurmohamed}}, \bibinfo {author} {\bibfnamefont
  {E.}~\bibnamefont {O'Farrell}}, \bibinfo {author} {\bibfnamefont
  {K.}~\bibnamefont {Otani}}, \bibinfo {author} {\bibfnamefont
  {S.}~\bibnamefont {Pauka}}, \bibinfo {author} {\bibfnamefont
  {K.}~\bibnamefont {Petersson}}, \bibinfo {author} {\bibfnamefont
  {L.}~\bibnamefont {Petit}}, \bibinfo {author} {\bibfnamefont
  {D.}~\bibnamefont {Pikulin}}, \bibinfo {author} {\bibfnamefont
  {F.}~\bibnamefont {Preiss}}, \bibinfo {author} {\bibfnamefont {M.~Q.}\
  \bibnamefont {Perez}}, \bibinfo {author} {\bibfnamefont {K.}~\bibnamefont
  {Rasmussen}}, \bibinfo {author} {\bibfnamefont {M.}~\bibnamefont {Rajpalke}},
  \bibinfo {author} {\bibfnamefont {D.}~\bibnamefont {Razmadze}}, \bibinfo
  {author} {\bibfnamefont {O.}~\bibnamefont {Reentila}}, \bibinfo {author}
  {\bibfnamefont {D.}~\bibnamefont {Reilly}}, \bibinfo {author} {\bibfnamefont
  {R.}~\bibnamefont {Rouse}}, \bibinfo {author} {\bibfnamefont
  {I.}~\bibnamefont {Sadovskyy}}, \bibinfo {author} {\bibfnamefont
  {L.}~\bibnamefont {Sainiemi}}, \bibinfo {author} {\bibfnamefont
  {S.}~\bibnamefont {Schreppler}}, \bibinfo {author} {\bibfnamefont
  {V.}~\bibnamefont {Sidorkin}}, \bibinfo {author} {\bibfnamefont
  {A.}~\bibnamefont {Singh}}, \bibinfo {author} {\bibfnamefont
  {S.}~\bibnamefont {Singh}}, \bibinfo {author} {\bibfnamefont
  {S.}~\bibnamefont {Sinha}}, \bibinfo {author} {\bibfnamefont
  {P.}~\bibnamefont {Sohr}}, \bibinfo {author} {\bibfnamefont {T.}~\bibnamefont
  {Stankevič}}, \bibinfo {author} {\bibfnamefont {L.}~\bibnamefont {Stek}},
  \bibinfo {author} {\bibfnamefont {H.}~\bibnamefont {Suominen}}, \bibinfo
  {author} {\bibfnamefont {J.}~\bibnamefont {Suter}}, \bibinfo {author}
  {\bibfnamefont {V.}~\bibnamefont {Svidenko}}, \bibinfo {author}
  {\bibfnamefont {S.}~\bibnamefont {Teicher}}, \bibinfo {author} {\bibfnamefont
  {M.}~\bibnamefont {Temuerhan}}, \bibinfo {author} {\bibfnamefont
  {N.}~\bibnamefont {Thiyagarajah}}, \bibinfo {author} {\bibfnamefont
  {R.}~\bibnamefont {Tholapi}}, \bibinfo {author} {\bibfnamefont
  {M.}~\bibnamefont {Thomas}}, \bibinfo {author} {\bibfnamefont
  {E.}~\bibnamefont {Toomey}}, \bibinfo {author} {\bibfnamefont
  {S.}~\bibnamefont {Upadhyay}}, \bibinfo {author} {\bibfnamefont
  {I.}~\bibnamefont {Urban}}, \bibinfo {author} {\bibfnamefont
  {S.}~\bibnamefont {Vaitiekėnas}}, \bibinfo {author} {\bibfnamefont
  {K.}~\bibnamefont {Van~Hoogdalem}}, \bibinfo {author} {\bibfnamefont {D.~V.}\
  \bibnamefont {Viazmitinov}}, \bibinfo {author} {\bibfnamefont
  {S.}~\bibnamefont {Waddy}}, \bibinfo {author} {\bibfnamefont
  {D.}~\bibnamefont {Van~Woerkom}}, \bibinfo {author} {\bibfnamefont
  {D.}~\bibnamefont {Vogel}}, \bibinfo {author} {\bibfnamefont
  {J.}~\bibnamefont {Watson}}, \bibinfo {author} {\bibfnamefont
  {J.}~\bibnamefont {Weston}}, \bibinfo {author} {\bibfnamefont {G.~W.}\
  \bibnamefont {Winkler}}, \bibinfo {author} {\bibfnamefont {C.~K.}\
  \bibnamefont {Yang}}, \bibinfo {author} {\bibfnamefont {S.}~\bibnamefont
  {Yau}}, \bibinfo {author} {\bibfnamefont {D.}~\bibnamefont {Yi}}, \bibinfo
  {author} {\bibfnamefont {E.}~\bibnamefont {Yucelen}}, \bibinfo {author}
  {\bibfnamefont {A.}~\bibnamefont {Webster}}, \bibinfo {author} {\bibfnamefont
  {R.}~\bibnamefont {Zeisel}},\ and\ \bibinfo {author} {\bibfnamefont
  {R.}~\bibnamefont {Zhao}},\ }\bibfield  {title} {\bibinfo {title}
  {{I}n{A}s-{A}l hybrid devices passing the topological gap protocol},\ }\href
  {https://arxiv.org/abs/2207.02472} {\bibfield  {journal} {\bibinfo  {journal}
  {arXiv:2207.02472}\ } (\bibinfo {year} {2022})}\BibitemShut {NoStop}%
\bibitem [{\citenamefont {Sarma}(2022)}]{sarma2022search}%
  \BibitemOpen
  \bibfield  {author} {\bibinfo {author} {\bibfnamefont {S.~D.}\ \bibnamefont
  {Sarma}},\ }\bibfield  {title} {\bibinfo {title} {In search of {M}ajorana},\
  }\href {https://arxiv.org/abs/2210.17365} {\bibfield  {journal} {\bibinfo
  {journal} {arXiv:2210.17365}\ } (\bibinfo {year} {2022})}\BibitemShut
  {NoStop}%
\bibitem [{\citenamefont {Chang}\ \emph {et~al.}(2015)\citenamefont {Chang},
  \citenamefont {Albrecht}, \citenamefont {Jespersen}, \citenamefont
  {Kuemmeth}, \citenamefont {Krogstrup}, \citenamefont {Nyg{\aa}rd},\ and\
  \citenamefont {Marcus}}]{Chang2015Hard}%
  \BibitemOpen
  \bibfield  {author} {\bibinfo {author} {\bibfnamefont {W.}~\bibnamefont
  {Chang}}, \bibinfo {author} {\bibfnamefont {S.~M.}\ \bibnamefont {Albrecht}},
  \bibinfo {author} {\bibfnamefont {T.~S.}\ \bibnamefont {Jespersen}}, \bibinfo
  {author} {\bibfnamefont {F.}~\bibnamefont {Kuemmeth}}, \bibinfo {author}
  {\bibfnamefont {P.}~\bibnamefont {Krogstrup}}, \bibinfo {author}
  {\bibfnamefont {J.}~\bibnamefont {Nyg{\aa}rd}},\ and\ \bibinfo {author}
  {\bibfnamefont {C.~M.}\ \bibnamefont {Marcus}},\ }\bibfield  {title}
  {\bibinfo {title} {Hard gap in epitaxial
  semiconductor{\textendash}superconductor nanowires},\ }\href
  {https://doi.org/10.1038/nnano.2014.306} {\bibfield  {journal} {\bibinfo
  {journal} {Nature Nanotech.}\ }\textbf {\bibinfo {volume} {10}},\ \bibinfo
  {pages} {232} (\bibinfo {year} {2015})}\BibitemShut {NoStop}%
\bibitem [{\citenamefont {Lee}\ \emph {et~al.}(2017)\citenamefont {Lee},
  \citenamefont {Jiang}, \citenamefont {\ifmmode~\check{Z}\else
  \v{Z}\fi{}itko}, \citenamefont {Aguado}, \citenamefont {Lieber},\ and\
  \citenamefont {De~Franceschi}}]{Lee2017Scaling}%
  \BibitemOpen
  \bibfield  {author} {\bibinfo {author} {\bibfnamefont {E.~J.~H.}\
  \bibnamefont {Lee}}, \bibinfo {author} {\bibfnamefont {X.}~\bibnamefont
  {Jiang}}, \bibinfo {author} {\bibfnamefont {R.}~\bibnamefont
  {\ifmmode~\check{Z}\else \v{Z}\fi{}itko}}, \bibinfo {author} {\bibfnamefont
  {R.}~\bibnamefont {Aguado}}, \bibinfo {author} {\bibfnamefont {C.~M.}\
  \bibnamefont {Lieber}},\ and\ \bibinfo {author} {\bibfnamefont
  {S.}~\bibnamefont {De~Franceschi}},\ }\bibfield  {title} {\bibinfo {title}
  {Scaling of subgap excitations in a superconductor-semiconductor nanowire
  quantum dot},\ }\href {https://doi.org/10.1103/PhysRevB.95.180502} {\bibfield
   {journal} {\bibinfo  {journal} {Phys. Rev. B}\ }\textbf {\bibinfo {volume}
  {95}},\ \bibinfo {pages} {180502} (\bibinfo {year} {2017})}\BibitemShut
  {NoStop}%
\bibitem [{\citenamefont {Reeg}\ \emph
  {et~al.}(2018{\natexlab{b}})\citenamefont {Reeg}, \citenamefont {Loss},\ and\
  \citenamefont {Klinovaja}}]{Reeg2018Metallization}%
  \BibitemOpen
  \bibfield  {author} {\bibinfo {author} {\bibfnamefont {C.}~\bibnamefont
  {Reeg}}, \bibinfo {author} {\bibfnamefont {D.}~\bibnamefont {Loss}},\ and\
  \bibinfo {author} {\bibfnamefont {J.}~\bibnamefont {Klinovaja}},\ }\bibfield
  {title} {\bibinfo {title} {Metallization of a {R}ashba wire by a
  superconducting layer in the strong-proximity regime},\ }\href
  {https://doi.org/10.1103/PhysRevB.97.165425} {\bibfield  {journal} {\bibinfo
  {journal} {Phys. Rev. B}\ }\textbf {\bibinfo {volume} {97}},\ \bibinfo
  {pages} {165425} (\bibinfo {year} {2018}{\natexlab{b}})}\BibitemShut
  {NoStop}%
\bibitem [{\citenamefont {Reeg}\ \emph {et~al.}(2017)\citenamefont {Reeg},
  \citenamefont {Loss},\ and\ \citenamefont {Klinovaja}}]{Reeg2017Finite}%
  \BibitemOpen
  \bibfield  {author} {\bibinfo {author} {\bibfnamefont {C.}~\bibnamefont
  {Reeg}}, \bibinfo {author} {\bibfnamefont {D.}~\bibnamefont {Loss}},\ and\
  \bibinfo {author} {\bibfnamefont {J.}~\bibnamefont {Klinovaja}},\ }\bibfield
  {title} {\bibinfo {title} {Finite-size effects in a nanowire strongly coupled
  to a thin superconducting shell},\ }\href
  {https://doi.org/10.1103/PhysRevB.96.125426} {\bibfield  {journal} {\bibinfo
  {journal} {Phys. Rev. B}\ }\textbf {\bibinfo {volume} {96}},\ \bibinfo
  {pages} {125426} (\bibinfo {year} {2017})}\BibitemShut {NoStop}%
\bibitem [{\citenamefont {Stanescu}\ and\ \citenamefont
  {Das~Sarma}(2017)}]{Stanescu2017Proximity}%
  \BibitemOpen
  \bibfield  {author} {\bibinfo {author} {\bibfnamefont {T.~D.}\ \bibnamefont
  {Stanescu}}\ and\ \bibinfo {author} {\bibfnamefont {S.}~\bibnamefont
  {Das~Sarma}},\ }\bibfield  {title} {\bibinfo {title} {Proximity-induced
  low-energy renormalization in hybrid semiconductor-superconductor {M}ajorana
  structures},\ }\href {https://doi.org/10.1103/PhysRevB.96.014510} {\bibfield
  {journal} {\bibinfo  {journal} {Phys. Rev. B}\ }\textbf {\bibinfo {volume}
  {96}},\ \bibinfo {pages} {014510} (\bibinfo {year} {2017})}\BibitemShut
  {NoStop}%
\bibitem [{\citenamefont {Anderson}(1958)}]{PhysRev.109.1492}%
  \BibitemOpen
  \bibfield  {author} {\bibinfo {author} {\bibfnamefont {P.~W.}\ \bibnamefont
  {Anderson}},\ }\bibfield  {title} {\bibinfo {title} {Absence of diffusion in
  certain random lattices},\ }\href {https://doi.org/10.1103/PhysRev.109.1492}
  {\bibfield  {journal} {\bibinfo  {journal} {Phys. Rev.}\ }\textbf {\bibinfo
  {volume} {109}},\ \bibinfo {pages} {1492} (\bibinfo {year}
  {1958})}\BibitemShut {NoStop}%
\bibitem [{\citenamefont {L\"othman}\ \emph {et~al.}(2021)\citenamefont
  {L\"othman}, \citenamefont {Triola}, \citenamefont {Cayao},\ and\
  \citenamefont {Black-Schaffer}}]{PhysRevB.104.094503}%
  \BibitemOpen
  \bibfield  {author} {\bibinfo {author} {\bibfnamefont {T.}~\bibnamefont
  {L\"othman}}, \bibinfo {author} {\bibfnamefont {C.}~\bibnamefont {Triola}},
  \bibinfo {author} {\bibfnamefont {J.}~\bibnamefont {Cayao}},\ and\ \bibinfo
  {author} {\bibfnamefont {A.~M.}\ \bibnamefont {Black-Schaffer}},\ }\bibfield
  {title} {\bibinfo {title} {Disorder-robust $p$-wave pairing with
  odd-frequency dependence in normal metal--conventional superconductor
  junctions},\ }\href {https://doi.org/10.1103/PhysRevB.104.094503} {\bibfield
  {journal} {\bibinfo  {journal} {Phys. Rev. B}\ }\textbf {\bibinfo {volume}
  {104}},\ \bibinfo {pages} {094503} (\bibinfo {year} {2021})}\BibitemShut
  {NoStop}%
\bibitem [{\citenamefont {Mashkoori}\ \emph {et~al.}(2023)\citenamefont
  {Mashkoori}, \citenamefont {Parhizgar}, \citenamefont {Rachel},\ and\
  \citenamefont {Black-Schaffer}}]{Mashkoori2023}%
  \BibitemOpen
  \bibfield  {author} {\bibinfo {author} {\bibfnamefont {M.}~\bibnamefont
  {Mashkoori}}, \bibinfo {author} {\bibfnamefont {F.}~\bibnamefont
  {Parhizgar}}, \bibinfo {author} {\bibfnamefont {S.}~\bibnamefont {Rachel}},\
  and\ \bibinfo {author} {\bibfnamefont {A.~M.}\ \bibnamefont
  {Black-Schaffer}},\ }\bibfield  {title} {\bibinfo {title} {Detrimental
  effects of disorder in two-dimensional time-reversal invariant topological
  superconductors},\ }\href {https://doi.org/10.1103/PhysRevB.107.014512}
  {\bibfield  {journal} {\bibinfo  {journal} {Phys. Rev. B}\ }\textbf {\bibinfo
  {volume} {107}},\ \bibinfo {pages} {014512} (\bibinfo {year}
  {2023})}\BibitemShut {NoStop}%
\bibitem [{Note1()}]{Note1}%
  \BibitemOpen
  \bibinfo {note} {For a N region with a vanishing length, $L_{\protect \rm N}
  = 0$, we have verified that our results do not change}\BibitemShut {NoStop}%
\bibitem [{\citenamefont {Arnoldi}(1951)}]{arnoldi1951principle}%
  \BibitemOpen
  \bibfield  {author} {\bibinfo {author} {\bibfnamefont {W.~E.}\ \bibnamefont
  {Arnoldi}},\ }\bibfield  {title} {\bibinfo {title} {The principle of
  minimized iterations in the solution of the matrix eigenvalue problem},\
  }\href {https://doi.org/10.1090/qam/42792} {\bibfield  {journal} {\bibinfo
  {journal} {Quart. Appl. Math.}\ }\textbf {\bibinfo {volume} {9}},\ \bibinfo
  {pages} {17} (\bibinfo {year} {1951})}\BibitemShut {NoStop}%
\bibitem [{\citenamefont {Maiani}\ \emph {et~al.}(2021)\citenamefont {Maiani},
  \citenamefont {Seoane~Souto}, \citenamefont {Leijnse},\ and\ \citenamefont
  {Flensberg}}]{Maiani2021Topological}%
  \BibitemOpen
  \bibfield  {author} {\bibinfo {author} {\bibfnamefont {A.}~\bibnamefont
  {Maiani}}, \bibinfo {author} {\bibfnamefont {R.}~\bibnamefont
  {Seoane~Souto}}, \bibinfo {author} {\bibfnamefont {M.}~\bibnamefont
  {Leijnse}},\ and\ \bibinfo {author} {\bibfnamefont {K.}~\bibnamefont
  {Flensberg}},\ }\bibfield  {title} {\bibinfo {title} {Topological
  superconductivity in semiconductor-superconductor-magnetic-insulator
  heterostructures},\ }\href {https://doi.org/10.1103/PhysRevB.103.104508}
  {\bibfield  {journal} {\bibinfo  {journal} {Phys. Rev. B}\ }\textbf {\bibinfo
  {volume} {103}},\ \bibinfo {pages} {104508} (\bibinfo {year}
  {2021})}\BibitemShut {NoStop}%
\bibitem [{Note2()}]{Note2}%
  \BibitemOpen
  \bibinfo {note} {The vanishing of the gap depends on system size: short SMs
  develop a sharp profile at the TPT but $E_{0}$ does not usually reach zero,
  while for long systems $E_{0}$ reaches zero.}\BibitemShut {Stop}%
\bibitem [{\citenamefont {Sarma}\ \emph {et~al.}(2015)\citenamefont {Sarma},
  \citenamefont {Freedman},\ and\ \citenamefont {Nayak}}]{Sarma:16}%
  \BibitemOpen
  \bibfield  {author} {\bibinfo {author} {\bibfnamefont {S.~D.}\ \bibnamefont
  {Sarma}}, \bibinfo {author} {\bibfnamefont {M.}~\bibnamefont {Freedman}},\
  and\ \bibinfo {author} {\bibfnamefont {C.}~\bibnamefont {Nayak}},\ }\bibfield
   {title} {\bibinfo {title} {Majorana zero modes and topological quantum
  computation},\ }\href {https://doi.org/10.1038/npjqi.2015.1} {\bibfield
  {journal} {\bibinfo  {journal} {npj Quantum Inf.}\ }\textbf {\bibinfo
  {volume} {1}},\ \bibinfo {pages} {15001} (\bibinfo {year}
  {2015})}\BibitemShut {NoStop}%
\bibitem [{Note3()}]{Note3}%
  \BibitemOpen
  \bibinfo {note} {We have checked that other random disorder realizations
  support our claims.}\BibitemShut {Stop}%
\bibitem [{Note4()}]{Note4}%
  \BibitemOpen
  \bibinfo {note} {To be precise, in the strong coupling regime, while most of
  the wavefunctions of the TZESs are mostly located in the SC, part of them are
  also in the SM but that is much smaller.}\BibitemShut {Stop}%
\bibitem [{\citenamefont {Anderson}(1959)}]{ANDERSON195926}%
  \BibitemOpen
  \bibfield  {author} {\bibinfo {author} {\bibfnamefont {P.}~\bibnamefont
  {Anderson}},\ }\bibfield  {title} {\bibinfo {title} {Theory of dirty
  superconductors},\ }\href
  {https://doi.org/https://doi.org/10.1016/0022-3697(59)90036-8} {\bibfield
  {journal} {\bibinfo  {journal} {J. Phys. Chem. Solids}\ }\textbf {\bibinfo
  {volume} {11}},\ \bibinfo {pages} {26 } (\bibinfo {year} {1959})}\BibitemShut
  {NoStop}%
\bibitem [{\citenamefont {Gor'kov}\ and\ \citenamefont
  {Rashba}(2001)}]{PhysRevLett.87.037004}%
  \BibitemOpen
  \bibfield  {author} {\bibinfo {author} {\bibfnamefont {L.~P.}\ \bibnamefont
  {Gor'kov}}\ and\ \bibinfo {author} {\bibfnamefont {E.~I.}\ \bibnamefont
  {Rashba}},\ }\bibfield  {title} {\bibinfo {title} {Superconducting 2d system
  with lifted spin degeneracy: Mixed singlet-triplet state},\ }\href
  {https://doi.org/10.1103/PhysRevLett.87.037004} {\bibfield  {journal}
  {\bibinfo  {journal} {Phys. Rev. Lett.}\ }\textbf {\bibinfo {volume} {87}},\
  \bibinfo {pages} {037004} (\bibinfo {year} {2001})}\BibitemShut {NoStop}%
\bibitem [{\citenamefont {Bergeret}\ \emph {et~al.}(2005)\citenamefont
  {Bergeret}, \citenamefont {Volkov},\ and\ \citenamefont
  {Efetov}}]{RevModPhys.77.1321}%
  \BibitemOpen
  \bibfield  {author} {\bibinfo {author} {\bibfnamefont {F.~S.}\ \bibnamefont
  {Bergeret}}, \bibinfo {author} {\bibfnamefont {A.~F.}\ \bibnamefont
  {Volkov}},\ and\ \bibinfo {author} {\bibfnamefont {K.~B.}\ \bibnamefont
  {Efetov}},\ }\bibfield  {title} {\bibinfo {title} {Odd triplet
  superconductivity and related phenomena in superconductor-ferromagnet
  structures},\ }\href {https://doi.org/10.1103/RevModPhys.77.1321} {\bibfield
  {journal} {\bibinfo  {journal} {Rev. Mod. Phys.}\ }\textbf {\bibinfo {volume}
  {77}},\ \bibinfo {pages} {1321} (\bibinfo {year} {2005})}\BibitemShut
  {NoStop}%
\bibitem [{\citenamefont {Linder}\ and\ \citenamefont
  {Robinson}(2015)}]{linder2015superconducting}%
  \BibitemOpen
  \bibfield  {author} {\bibinfo {author} {\bibfnamefont {J.}~\bibnamefont
  {Linder}}\ and\ \bibinfo {author} {\bibfnamefont {J.~W.}\ \bibnamefont
  {Robinson}},\ }\bibfield  {title} {\bibinfo {title} {Superconducting
  spintronics},\ }\href@noop {} {\bibfield  {journal} {\bibinfo  {journal}
  {Nature Physics}\ }\textbf {\bibinfo {volume} {11}},\ \bibinfo {pages} {307}
  (\bibinfo {year} {2015})}\BibitemShut {NoStop}%
\bibitem [{\citenamefont {Liu}\ \emph {et~al.}(2015)\citenamefont {Liu},
  \citenamefont {Sau},\ and\ \citenamefont {Das~Sarma}}]{PhysRevB.92.014513}%
  \BibitemOpen
  \bibfield  {author} {\bibinfo {author} {\bibfnamefont {X.}~\bibnamefont
  {Liu}}, \bibinfo {author} {\bibfnamefont {J.~D.}\ \bibnamefont {Sau}},\ and\
  \bibinfo {author} {\bibfnamefont {S.}~\bibnamefont {Das~Sarma}},\ }\bibfield
  {title} {\bibinfo {title} {Universal spin-triplet superconducting
  correlations of {M}ajorana fermions},\ }\href
  {https://doi.org/10.1103/PhysRevB.92.014513} {\bibfield  {journal} {\bibinfo
  {journal} {Phys. Rev. B}\ }\textbf {\bibinfo {volume} {92}},\ \bibinfo
  {pages} {014513} (\bibinfo {year} {2015})}\BibitemShut {NoStop}%
\bibitem [{\citenamefont {Jacobsen}\ \emph {et~al.}(2016)\citenamefont
  {Jacobsen}, \citenamefont {Ouassou},\ and\ \citenamefont {Linder}}]{solbook}%
  \BibitemOpen
  \bibfield  {author} {\bibinfo {author} {\bibfnamefont {S.~H.}\ \bibnamefont
  {Jacobsen}}, \bibinfo {author} {\bibfnamefont {J.~A.}\ \bibnamefont
  {Ouassou}},\ and\ \bibinfo {author} {\bibfnamefont {J.}~\bibnamefont
  {Linder}},\ }\bibinfo {title} {Superconducting order in magnetic
  heterostructures},\ in\ \href
  {https://doi.org/https://doi.org/10.1002/9781119241966.ch1} {\emph {\bibinfo
  {booktitle} {Advanced Magnetic and Optical Materials}}}\ (\bibinfo
  {publisher} {John Wiley \& Sons, Ltd},\ \bibinfo {year} {2016})\
  Chap.~\bibinfo {chapter} {1}, pp.\ \bibinfo {pages} {1--46}\BibitemShut
  {NoStop}%
\bibitem [{\citenamefont {Cayao}\ and\ \citenamefont
  {Black-Schaffer}(2018)}]{PhysRevB.98.075425}%
  \BibitemOpen
  \bibfield  {author} {\bibinfo {author} {\bibfnamefont {J.}~\bibnamefont
  {Cayao}}\ and\ \bibinfo {author} {\bibfnamefont {A.~M.}\ \bibnamefont
  {Black-Schaffer}},\ }\bibfield  {title} {\bibinfo {title} {Odd-frequency
  superconducting pairing in junctions with rashba spin-orbit coupling},\
  }\href {https://doi.org/10.1103/PhysRevB.98.075425} {\bibfield  {journal}
  {\bibinfo  {journal} {Phys. Rev. B}\ }\textbf {\bibinfo {volume} {98}},\
  \bibinfo {pages} {075425} (\bibinfo {year} {2018})}\BibitemShut {NoStop}%
\bibitem [{\citenamefont {Tamura}\ \emph {et~al.}(2019)\citenamefont {Tamura},
  \citenamefont {Hoshino},\ and\ \citenamefont {Tanaka}}]{PhysRevB.99.184512}%
  \BibitemOpen
  \bibfield  {author} {\bibinfo {author} {\bibfnamefont {S.}~\bibnamefont
  {Tamura}}, \bibinfo {author} {\bibfnamefont {S.}~\bibnamefont {Hoshino}},\
  and\ \bibinfo {author} {\bibfnamefont {Y.}~\bibnamefont {Tanaka}},\
  }\bibfield  {title} {\bibinfo {title} {Odd-frequency pairs in chiral
  symmetric systems: Spectral bulk-boundary correspondence and topological
  criticality},\ }\href {https://doi.org/10.1103/PhysRevB.99.184512} {\bibfield
   {journal} {\bibinfo  {journal} {Phys. Rev. B}\ }\textbf {\bibinfo {volume}
  {99}},\ \bibinfo {pages} {184512} (\bibinfo {year} {2019})}\BibitemShut
  {NoStop}%
\bibitem [{\citenamefont {Tsintzis}\ \emph {et~al.}(2019)\citenamefont
  {Tsintzis}, \citenamefont {Black-Schaffer},\ and\ \citenamefont
  {Cayao}}]{PhysRevB.100.115433}%
  \BibitemOpen
  \bibfield  {author} {\bibinfo {author} {\bibfnamefont {A.}~\bibnamefont
  {Tsintzis}}, \bibinfo {author} {\bibfnamefont {A.~M.}\ \bibnamefont
  {Black-Schaffer}},\ and\ \bibinfo {author} {\bibfnamefont {J.}~\bibnamefont
  {Cayao}},\ }\bibfield  {title} {\bibinfo {title} {Odd-frequency
  superconducting pairing in kitaev-based junctions},\ }\href
  {https://doi.org/10.1103/PhysRevB.100.115433} {\bibfield  {journal} {\bibinfo
   {journal} {Phys. Rev. B}\ }\textbf {\bibinfo {volume} {100}},\ \bibinfo
  {pages} {115433} (\bibinfo {year} {2019})}\BibitemShut {NoStop}%
\bibitem [{\citenamefont {Cayao}\ \emph {et~al.}(2020)\citenamefont {Cayao},
  \citenamefont {Triola},\ and\ \citenamefont {Black-Schaffer}}]{cayao2019odd}%
  \BibitemOpen
  \bibfield  {author} {\bibinfo {author} {\bibfnamefont {J.}~\bibnamefont
  {Cayao}}, \bibinfo {author} {\bibfnamefont {C.}~\bibnamefont {Triola}},\ and\
  \bibinfo {author} {\bibfnamefont {A.~M.}\ \bibnamefont {Black-Schaffer}},\
  }\bibfield  {title} {\bibinfo {title} {Odd-frequency superconducting pairing
  in one-dimensional systems},\ }\href@noop {} {\bibfield  {journal} {\bibinfo
  {journal} {Eur. Phys. J. Spec. Top.}\ }\textbf {\bibinfo {volume} {229}},\
  \bibinfo {pages} {545} (\bibinfo {year} {2020})}\BibitemShut {NoStop}%
\bibitem [{Note5()}]{Note5}%
  \BibitemOpen
  \bibinfo {note} {Apart from the chemical potential, other parameters such as
  spin-orbit coupling and $g$-factor also suffer a renormalization in the
  strong coupling regime}\BibitemShut {NoStop}%
\bibitem [{\citenamefont {Awoga}\ \emph {et~al.}(2017)\citenamefont {Awoga},
  \citenamefont {Bj\"{o}rnson},\ and\ \citenamefont
  {Black-Schaffer}}]{Awoga2017disorder}%
  \BibitemOpen
  \bibfield  {author} {\bibinfo {author} {\bibfnamefont {O.~A.}\ \bibnamefont
  {Awoga}}, \bibinfo {author} {\bibfnamefont {K.}~\bibnamefont
  {Bj\"{o}rnson}},\ and\ \bibinfo {author} {\bibfnamefont {A.~M.}\ \bibnamefont
  {Black-Schaffer}},\ }\bibfield  {title} {\bibinfo {title} {Disorder
  robustness and protection of {M}ajorana bound states in ferromagnetic chains
  on conventional superconductors},\ }\href
  {https://link.aps.org/doi/10.1103/PhysRevB.95.184511} {\bibfield  {journal}
  {\bibinfo  {journal} {Phys. Rev. B}\ }\textbf {\bibinfo {volume} {95}},\
  \bibinfo {pages} {184511} (\bibinfo {year} {2017})}\BibitemShut {NoStop}%
\bibitem [{\citenamefont {Haim}\ and\ \citenamefont
  {Stern}(2019)}]{Haim2019Benefits}%
  \BibitemOpen
  \bibfield  {author} {\bibinfo {author} {\bibfnamefont {A.}~\bibnamefont
  {Haim}}\ and\ \bibinfo {author} {\bibfnamefont {A.}~\bibnamefont {Stern}},\
  }\bibfield  {title} {\bibinfo {title} {Benefits of weak disorder in
  one-dimensional topological superconductors},\ }\href
  {https://doi.org/10.1103/PhysRevLett.122.126801} {\bibfield  {journal}
  {\bibinfo  {journal} {Phys. Rev. Lett.}\ }\textbf {\bibinfo {volume} {122}},\
  \bibinfo {pages} {126801} (\bibinfo {year} {2019})}\BibitemShut {NoStop}%
\bibitem [{\citenamefont {Dom{\'\i}nguez}\ \emph {et~al.}(2017)\citenamefont
  {Dom{\'\i}nguez}, \citenamefont {Cayao}, \citenamefont {San-Jose},
  \citenamefont {Aguado}, \citenamefont {Yeyati},\ and\ \citenamefont
  {Prada}}]{dominguez2017zero}%
  \BibitemOpen
  \bibfield  {author} {\bibinfo {author} {\bibfnamefont {F.}~\bibnamefont
  {Dom{\'\i}nguez}}, \bibinfo {author} {\bibfnamefont {J.}~\bibnamefont
  {Cayao}}, \bibinfo {author} {\bibfnamefont {P.}~\bibnamefont {San-Jose}},
  \bibinfo {author} {\bibfnamefont {R.}~\bibnamefont {Aguado}}, \bibinfo
  {author} {\bibfnamefont {A.~L.}\ \bibnamefont {Yeyati}},\ and\ \bibinfo
  {author} {\bibfnamefont {E.}~\bibnamefont {Prada}},\ }\bibfield  {title}
  {\bibinfo {title} {Zero-energy pinning from interactions in majorana
  nanowires},\ }\href@noop {} {\bibfield  {journal} {\bibinfo  {journal} {npj
  Quantum Mater.}\ }\textbf {\bibinfo {volume} {2}},\ \bibinfo {pages} {13}
  (\bibinfo {year} {2017})}\BibitemShut {NoStop}%
\bibitem [{\citenamefont {Woods}\ \emph {et~al.}(2018)\citenamefont {Woods},
  \citenamefont {Stanescu},\ and\ \citenamefont
  {Das~Sarma}}]{PhysRevB.98.035428}%
  \BibitemOpen
  \bibfield  {author} {\bibinfo {author} {\bibfnamefont {B.~D.}\ \bibnamefont
  {Woods}}, \bibinfo {author} {\bibfnamefont {T.~D.}\ \bibnamefont
  {Stanescu}},\ and\ \bibinfo {author} {\bibfnamefont {S.}~\bibnamefont
  {Das~Sarma}},\ }\bibfield  {title} {\bibinfo {title} {Effective theory
  approach to the schr\"odinger-poisson problem in semiconductor majorana
  devices},\ }\href {https://doi.org/10.1103/PhysRevB.98.035428} {\bibfield
  {journal} {\bibinfo  {journal} {Phys. Rev. B}\ }\textbf {\bibinfo {volume}
  {98}},\ \bibinfo {pages} {035428} (\bibinfo {year} {2018})}\BibitemShut
  {NoStop}%
\bibitem [{\citenamefont {Escribano}\ \emph {et~al.}(2018)\citenamefont
  {Escribano}, \citenamefont {Yeyati},\ and\ \citenamefont
  {Prada}}]{escribano2018interaction}%
  \BibitemOpen
  \bibfield  {author} {\bibinfo {author} {\bibfnamefont {S.~D.}\ \bibnamefont
  {Escribano}}, \bibinfo {author} {\bibfnamefont {A.~L.}\ \bibnamefont
  {Yeyati}},\ and\ \bibinfo {author} {\bibfnamefont {E.}~\bibnamefont
  {Prada}},\ }\bibfield  {title} {\bibinfo {title} {Interaction-induced
  zero-energy pinning and quantum dot formation in majorana nanowires},\
  }\href@noop {} {\bibfield  {journal} {\bibinfo  {journal} {Beilstein J.
  Nanotechnol.}\ }\textbf {\bibinfo {volume} {9}},\ \bibinfo {pages} {2171}
  (\bibinfo {year} {2018})}\BibitemShut {NoStop}%
\bibitem [{\citenamefont {Mikkelsen}\ \emph {et~al.}(2018)\citenamefont
  {Mikkelsen}, \citenamefont {Kotetes}, \citenamefont {Krogstrup},\ and\
  \citenamefont {Flensberg}}]{PhysRevX.8.031040}%
  \BibitemOpen
  \bibfield  {author} {\bibinfo {author} {\bibfnamefont {A.~E.~G.}\
  \bibnamefont {Mikkelsen}}, \bibinfo {author} {\bibfnamefont {P.}~\bibnamefont
  {Kotetes}}, \bibinfo {author} {\bibfnamefont {P.}~\bibnamefont {Krogstrup}},\
  and\ \bibinfo {author} {\bibfnamefont {K.}~\bibnamefont {Flensberg}},\
  }\bibfield  {title} {\bibinfo {title} {Hybridization at
  superconductor-semiconductor interfaces},\ }\href
  {https://doi.org/10.1103/PhysRevX.8.031040} {\bibfield  {journal} {\bibinfo
  {journal} {Phys. Rev. X}\ }\textbf {\bibinfo {volume} {8}},\ \bibinfo {pages}
  {031040} (\bibinfo {year} {2018})}\BibitemShut {NoStop}%
\bibitem [{\citenamefont {Antipov}\ \emph {et~al.}(2018)\citenamefont
  {Antipov}, \citenamefont {Bargerbos}, \citenamefont {Winkler}, \citenamefont
  {Bauer}, \citenamefont {Rossi},\ and\ \citenamefont
  {Lutchyn}}]{PhysRevX.8.031041}%
  \BibitemOpen
  \bibfield  {author} {\bibinfo {author} {\bibfnamefont {A.~E.}\ \bibnamefont
  {Antipov}}, \bibinfo {author} {\bibfnamefont {A.}~\bibnamefont {Bargerbos}},
  \bibinfo {author} {\bibfnamefont {G.~W.}\ \bibnamefont {Winkler}}, \bibinfo
  {author} {\bibfnamefont {B.}~\bibnamefont {Bauer}}, \bibinfo {author}
  {\bibfnamefont {E.}~\bibnamefont {Rossi}},\ and\ \bibinfo {author}
  {\bibfnamefont {R.~M.}\ \bibnamefont {Lutchyn}},\ }\bibfield  {title}
  {\bibinfo {title} {Effects of gate-induced electric fields on semiconductor
  majorana nanowires},\ }\href {https://doi.org/10.1103/PhysRevX.8.031041}
  {\bibfield  {journal} {\bibinfo  {journal} {Phys. Rev. X}\ }\textbf {\bibinfo
  {volume} {8}},\ \bibinfo {pages} {031041} (\bibinfo {year}
  {2018})}\BibitemShut {NoStop}%
\bibitem [{\citenamefont {Lutchyn}\ \emph {et~al.}(2011)\citenamefont
  {Lutchyn}, \citenamefont {Stanescu},\ and\ \citenamefont
  {Das~Sarma}}]{PhysRevLett.106.127001}%
  \BibitemOpen
  \bibfield  {author} {\bibinfo {author} {\bibfnamefont {R.~M.}\ \bibnamefont
  {Lutchyn}}, \bibinfo {author} {\bibfnamefont {T.~D.}\ \bibnamefont
  {Stanescu}},\ and\ \bibinfo {author} {\bibfnamefont {S.}~\bibnamefont
  {Das~Sarma}},\ }\bibfield  {title} {\bibinfo {title} {Search for majorana
  fermions in multiband semiconducting nanowires},\ }\href
  {https://doi.org/10.1103/PhysRevLett.106.127001} {\bibfield  {journal}
  {\bibinfo  {journal} {Phys. Rev. Lett.}\ }\textbf {\bibinfo {volume} {106}},\
  \bibinfo {pages} {127001} (\bibinfo {year} {2011})}\BibitemShut {NoStop}%
\bibitem [{\citenamefont {Stanescu}\ \emph {et~al.}(2011)\citenamefont
  {Stanescu}, \citenamefont {Lutchyn},\ and\ \citenamefont
  {Das~Sarma}}]{Stanescu2011Majorana}%
  \BibitemOpen
  \bibfield  {author} {\bibinfo {author} {\bibfnamefont {T.~D.}\ \bibnamefont
  {Stanescu}}, \bibinfo {author} {\bibfnamefont {R.~M.}\ \bibnamefont
  {Lutchyn}},\ and\ \bibinfo {author} {\bibfnamefont {S.}~\bibnamefont
  {Das~Sarma}},\ }\bibfield  {title} {\bibinfo {title} {Majorana fermions in
  semiconductor nanowires},\ }\href
  {https://doi.org/10.1103/PhysRevB.84.144522} {\bibfield  {journal} {\bibinfo
  {journal} {Phys. Rev. B}\ }\textbf {\bibinfo {volume} {84}},\ \bibinfo
  {pages} {144522} (\bibinfo {year} {2011})}\BibitemShut {NoStop}%
\bibitem [{\citenamefont {San-Jose}\ \emph {et~al.}(2014)\citenamefont
  {San-Jose}, \citenamefont {Prada},\ and\ \citenamefont
  {Aguado}}]{PhysRevLett.112.137001}%
  \BibitemOpen
  \bibfield  {author} {\bibinfo {author} {\bibfnamefont {P.}~\bibnamefont
  {San-Jose}}, \bibinfo {author} {\bibfnamefont {E.}~\bibnamefont {Prada}},\
  and\ \bibinfo {author} {\bibfnamefont {R.}~\bibnamefont {Aguado}},\
  }\bibfield  {title} {\bibinfo {title} {Mapping the topological phase diagram
  of multiband semiconductors with supercurrents},\ }\href
  {http://journals.aps.org/prl/abstract/10.1103/PhysRevLett.112.137001}
  {\bibfield  {journal} {\bibinfo  {journal} {Phys. Rev. Lett.}\ }\textbf
  {\bibinfo {volume} {112}},\ \bibinfo {pages} {137001} (\bibinfo {year}
  {2014})}\BibitemShut {NoStop}%
\bibitem [{\citenamefont {Lim}\ \emph {et~al.}(2012)\citenamefont {Lim},
  \citenamefont {Serra}, \citenamefont {L\'opez},\ and\ \citenamefont
  {Aguado}}]{PhysRevB.86.121103}%
  \BibitemOpen
  \bibfield  {author} {\bibinfo {author} {\bibfnamefont {J.~S.}\ \bibnamefont
  {Lim}}, \bibinfo {author} {\bibfnamefont {L.~m.~c.}\ \bibnamefont {Serra}},
  \bibinfo {author} {\bibfnamefont {R.}~\bibnamefont {L\'opez}},\ and\ \bibinfo
  {author} {\bibfnamefont {R.}~\bibnamefont {Aguado}},\ }\bibfield  {title}
  {\bibinfo {title} {Magnetic-field instability of majorana modes in multiband
  semiconductor wires},\ }\href {https://doi.org/10.1103/PhysRevB.86.121103}
  {\bibfield  {journal} {\bibinfo  {journal} {Phys. Rev. B}\ }\textbf {\bibinfo
  {volume} {86}},\ \bibinfo {pages} {121103} (\bibinfo {year}
  {2012})}\BibitemShut {NoStop}%
\bibitem [{\citenamefont {Potter}\ and\ \citenamefont
  {Lee}(2010)}]{PhysRevLett.105.227003}%
  \BibitemOpen
  \bibfield  {author} {\bibinfo {author} {\bibfnamefont {A.~C.}\ \bibnamefont
  {Potter}}\ and\ \bibinfo {author} {\bibfnamefont {P.~A.}\ \bibnamefont
  {Lee}},\ }\bibfield  {title} {\bibinfo {title} {Multichannel generalization
  of kitaev's majorana end states and a practical route to realize them in thin
  films},\ }\href {https://doi.org/10.1103/PhysRevLett.105.227003} {\bibfield
  {journal} {\bibinfo  {journal} {Phys. Rev. Lett.}\ }\textbf {\bibinfo
  {volume} {105}},\ \bibinfo {pages} {227003} (\bibinfo {year}
  {2010})}\BibitemShut {NoStop}%
\bibitem [{\citenamefont {Datta}(1997)}]{Datta:97}%
  \BibitemOpen
  \bibfield  {author} {\bibinfo {author} {\bibfnamefont {S.}~\bibnamefont
  {Datta}},\ }\href@noop {} {\emph {\bibinfo {title} {Electronic transport in
  mesoscopic systems}}}\ (\bibinfo  {publisher} {Cambridge university press},\
  \bibinfo {year} {1997})\BibitemShut {NoStop}%
\end{thebibliography}%

\end{document}